\documentclass[aps,pre,twocolumn,showpacs,showkeys,amssymb,floatfix,reprint]{revtex4-1}
\usepackage{graphicx}
\usepackage{amssymb}
\usepackage{amsmath}
\usepackage[caption=false]{subfig}
\usepackage{booktabs}
\usepackage{float}

\bibpunct{[}{]}{,}{n}{,}{,}

\begin{document}

\title{ Self assembled linear polymeric chains with tuneable semiflexibility using isotropic interactions}

\author{Alex Abraham$^{1}$, Apratim Chatterji$^{1,2}$}
\email{apratim@iiserpune.ac.in}
\affiliation{
$^1$ Dept. of Physics, IISER-Pune, Dr. Homi Bhaba Road,  Pune-411008, India.\\
$^2$ Center for Energy Science, IISER-Pune,  Dr. Homi Bhaba Road,  Pune-411008, India.
}

\date{\today}
\begin{abstract}
We propose a two-body spherically symmetric (isotropic) potential such that particles  interacting by
the potential self assemble into linear semiflexible polymeric chains without branching. By suitable control of the
potential parameters we can control the persistence length of the polymer,
and can even introduce a controlled number of branches.  Thus we  show how to achieve effective directional
interactions starting from spherically symmetric potentials.
The self assembled polymers have a exponential distribution of chain lengths akin to what is observed
for worm-like micellar systems. On increasing  particle density the polymeric chains self-organize to
a ordered line-hexagonal phase where every chain is surround by six  parallel chains, the transition is first order.
On further increase in monomer density, the order is destroyed and we get a branched gel like phase.
This potential can be used to model  semi-flexible equilibrium polymers with tunable semiflexibility and excluded
volume.  The use of the potential is computationally cheap and  hence,  can be used to  simulate and
probe complex micellar dynamics with long chains.  The potential also gives a plausible method of tuning
colloidal interactions in experiments such that one can obtain self-assembling polymeric chains made up of
colloids and probe polymer dynamics using an optical microscope.
\end{abstract}

\maketitle

\section{Introduction}

The self-assembly of microscopic particles by tuning the interactions between them to obtain well-defined 
target  structures has a well established direction of research in soft matter physics. 
The more recent  focus in this area has evolved  to heirarchical self assembly \citep{geissler} as well as
self-assembly with very many different constituent particles which will result in self-assembled 
structures of much higher complexity \cite{frenkel,frenkel1,luijten,schoot,mubeena,sa_grason,sa_mcmanus,sa_frenkel,
sa_bruinsma,sa_dellago,sa_whitelam}. 
A line of investigation has been the self assembly of spherical particles with directional interactions (patchy colloids)
and self assembly of particles with anisotropic shapes to obtain a zoo of different target structures 
and organizations \cite{glotzer}. Another line of studies has been  to  
obtain directed interactions starting out from spherically symmetric interactions 
between suitably dressed isotropic particles \cite{geissler,sanat,mossa,zacarelli}.

In particular, Mossa et.al. and others \cite{mossa, zacarelli, shibananda, snigdha} 
investigated the self assembly of spherical particles 
interacting with radially symmetric potentials with a short range attractive part and long range repulsive 
interactions (modelled by screened Coulomb interactions). 
On changing the relative contributions (range and strength) of these two parts of the interacting
potential, they obtained  clusters of particles with different 
organizations of particles within cluster, e.g., crystalline clusters as well as planar and linear
extended aggregates  of particles at temperature $T=0$ using energy minimization techniques. 
The organization also depended on the number of particles considered as it affects the average number of 
neighbours per particle in cluster. The range of attractive interaction was changed by changing the integer value of
$n$ in the potential of the form $ V_{att} = 4 \epsilon [(\sigma/r)^{2n} - (\sigma/r)^{n}]$,
following the previous work of Vliegenhart et. al. \cite{vliegenthart}. Vliegenthart et. al. had various 
different ranges of the attractive potential of the form $V_{att}$ to investigate the role of the range 
of attractive minima to obtain liquid phases in the fluid-solid phase diagram. 
   
Our aim is to develop a spherically symmetric model potential such that particles interacting by the potential
self-assemble to linear equilibrium polymeric chains which are semiflexible; real life examples of such self
 assembled polymeric chains is worm-like micelles \cite{berret,dreiss,dreiss1}.
There exists quite a few coarse-grained models which describe self-assembling micellar chains
\cite{cates_candau,milchev,ryck1,ryck2,ryck3,padding,pa1,pa2,pa3,mubeena,pandit,sunilkumar,sunilkumar1}.
Some use suitable rate constants
to model joining and breaking of bonds between  effective bead-spring monomers where only 2 bonds are allowed 
per monomer \cite{ryck1,ryck2,ryck3,padding,pa1,pa2,pa3}.
Other models have effective potentials for self assembly of particles into polymeric chains, where
semiflexibility is incorporated by suitable angle dependent potentials \cite{milchev,pandit}.
Branching or cluster formation
is prevented by suitable choice of parameters of 3-body or 4 body potentials \cite{mubeena,pandit,sunilkumar,sunilkumar1}. 
The use of 3-body or 4-body potentials
is cumbersome and computationally expensive, espcially when one wants to model systems of long chains to study
interesting phenomenon such as shear banding \cite{sood,sood1,ganapathy,chakrabarti}. A simpler potential with just
two-body spherically symmetric interaction potential would greatly help in modelling 
systems of long self assembled polymer chains, moreover, we would like to avoid branching or introduce
branches in a controlled manner.

The spherically symmetric potential that we developed has three parts ($i$) a repulsive potential at very short distances
of the form $(\sigma/r)^{24})$ which takes care of excluded volume interactions between self-assembling particles, 
where $r$ is the distance between 2 particles and $\sigma$ is the diameter ($ii$) a short range attractive  
potential $ -(\sigma/r)^{12}$ at distances just beyond $\sigma$, and ($iii$) a screened Coulomb interaction 
of the form $\sim \frac{1}{r} e^{-r/R}$ with a suitable
strength $\epsilon^*$ and range of interaction decided by $R$. We have two kinds of particles, A and B in  equal 
ratios. A-A interactions and B-B are purely repulsive, modelled by screened Coulomb interactions ($iii$).
The interaction between A-B particles is a combination of ($i$),($ii$) and ($iii$) such that 
A-B particles can attract each other when distances between the two
are just above $\sigma$ and repel each other at longer distances. 
The form of the potentials are reminiscent of interaction between colloidal particles 
and can be used to make colloidal-polymers, even without the use of DNA-linkers \cite{sa_uberti_frenkel}.
This would
open up possiblities to observe complex micellar dynamics e.g. rheochaos or shear banding to direct 
optical observation. In addition, because of the large size of the particles the dynamics would be much slower 
and can be tracked using standard optical microscopy techniques. 

The rest of the manuscript is as follows. We describe the model potential and computational details 
in the next section: Methods.
Then in the next Results-section we describe the various phases that we obtain as we change the number 
density of particles and temperature; we also describe how we control the branching properties and 
persistence length of the self assembled polymers.  We conclude with discussions in the fourth and final section.

\section{Model}

\begin{figure}[H]
\begin{center}
\includegraphics[width=0.5\columnwidth]{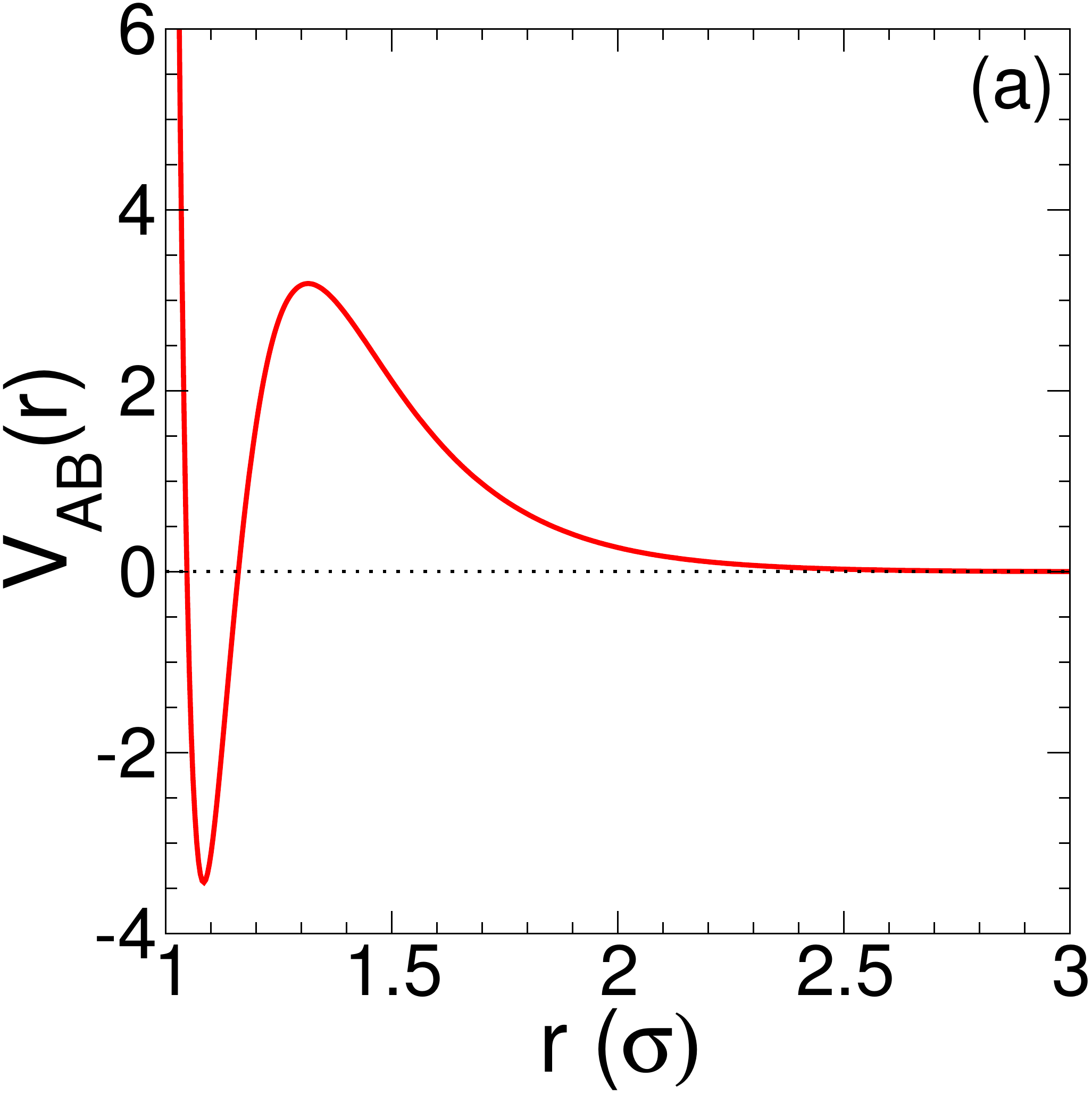}\\
\includegraphics[width=0.5\columnwidth]{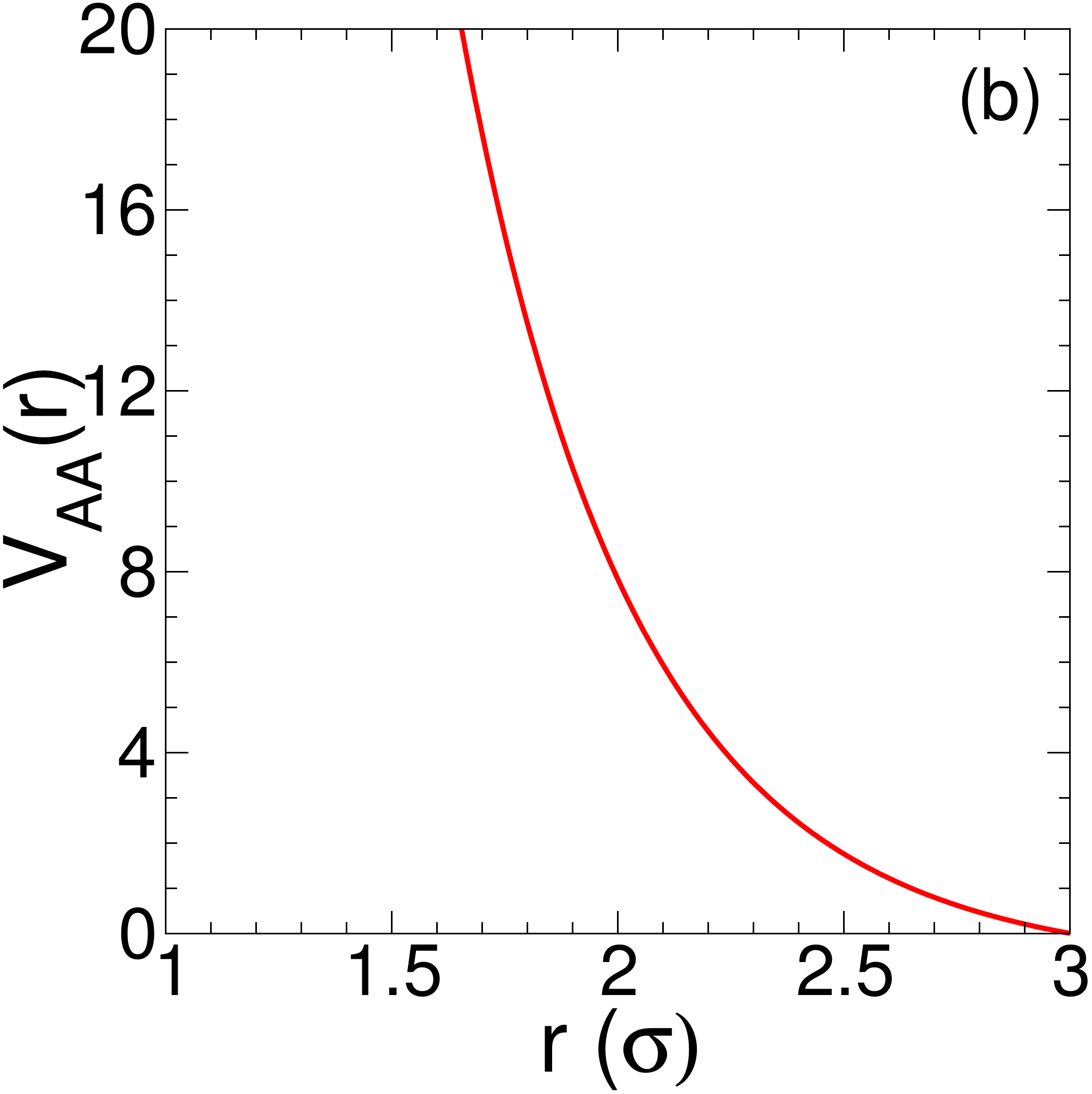}\\
\includegraphics[width=0.5\columnwidth]{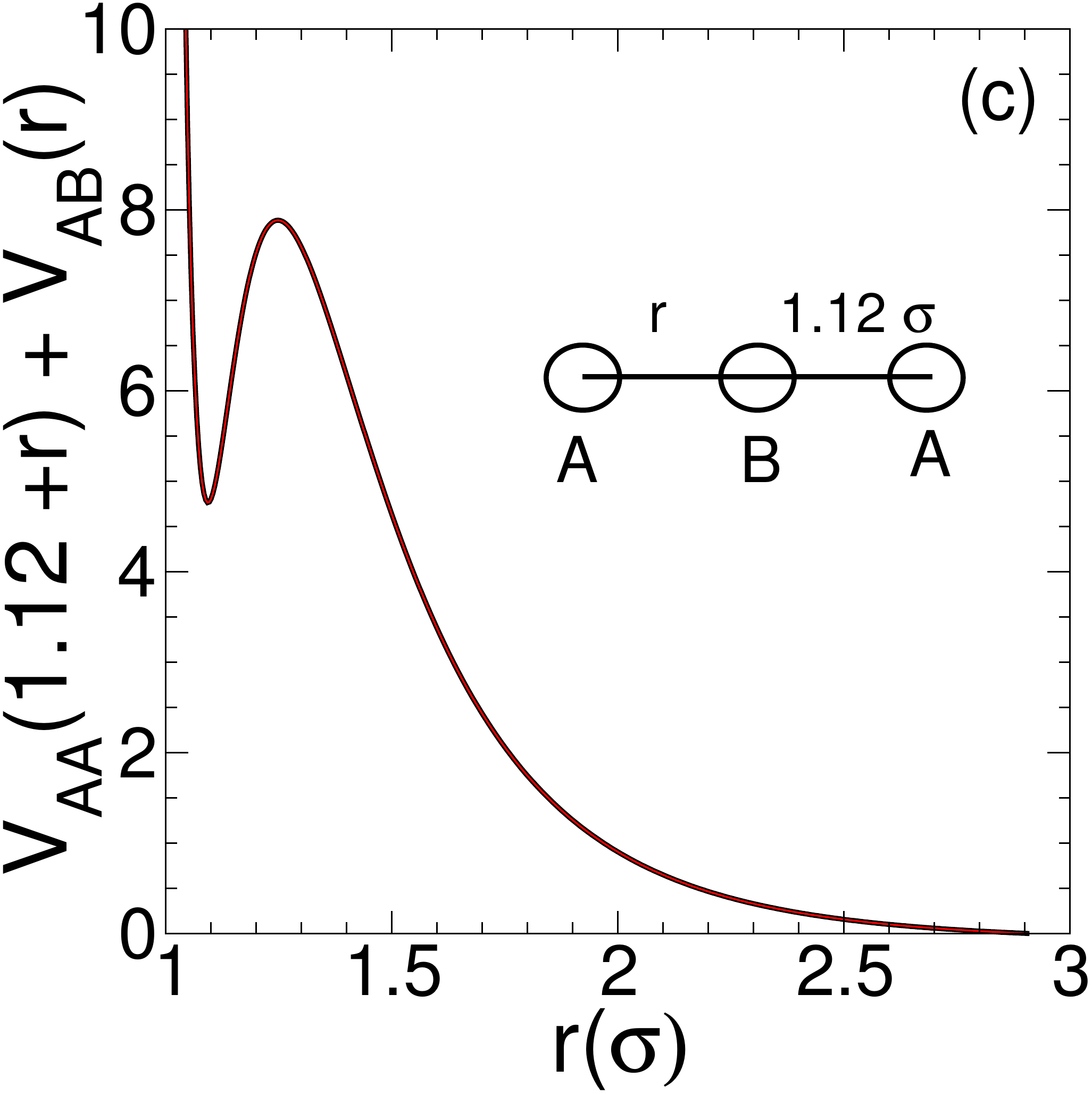}
\end{center}
\caption{\label{fig1} Representative plots of the potentials  (a) $V_{ab}(r)$ acting between centers of $A$ and $B$ 
(b)  $V_{aa}(r)$ acting between pair of $A-A$ or $B-B$ particles (c) $ V_{AA}(1.12 +r) + V_{AB}(r)$ for a linear triplet 
of $A-B-A$ (or $B-A-B$) particles, where the potential is plotted as a function of $r$, the distance of the third
particle from central B particle. The distance between the first particle and central particle is kept fixed at 
$1.12 \sigma$ (refer inset schematic diagram).     
}
\end{figure}

To develop an interaction potential between effective-monomers which self assemble to form a linear polymeric 
micellar chains, we avoid bead-spring potentials with suitable rate constants for bonds to join and break 
along the chain. Instead, we focus on developing a suitably modified Lennard Jones (LJ) type potential 
which encourages assembly of particles at short distances but  also with a potential maxima 
at an appropriate longer distance which should discourage spherical clusters or branching. Moreover,
right angles between bond-vectors in a triplet of monomers should be penalized. 
The monomers we have  in mind could also be colloidal
particles for which effective interaction can be tuned using changing surface properties or altering the
counter-ion or salt densities around a charged colloid. To that end, we modify the LJ-type interaction
by adding a Debye Huckel type screening potential between particles such that the total potential $V_{tot}$ is
of the form $ V_{tot} = \epsilon [(\sigma/r)^{2n} - (\sigma/r)^{n} + \epsilon^* e^{-r/R}/r]$ with exponent $n=12$, 
such that the potential looks akin to the potential shown in Fig.\ref{fig1}(a).  The $n=12$ potential have 
a sharper minima as compared to the usual LJ ($n=6$) potential, so that the peak due to the repulsive term
can be shifted to lower values of $r$ just beyond the position of potential minima as shown in Fig.\ref{fig1}a. 
The monomers self-assemble to form polymeric chains, however, contrary to our expectations, a large 
number of branches  get inadvertently formed  as the number density of monomers increases.

To prevent branching during the self assembly we introduce $2$ kinds of particles, 
(say) A and B; we maintain the  interaction potential between A-A, A-B and B-B to be spherically symmetric.
The interaction between a pair of A-B particles are kept to be of the same form as mentioned before, viz.,
\begin{multline}
        V_{AB}(r)= \epsilon_{AB}\left[ \left(\frac{\sigma}{r}\right)^{24}- \left(\frac{\sigma}{r}\right)^{12} 
 + \epsilon^{*}_{AB}  \frac{e^{-(r/R_{AB})}}{r} \right], \\ ..\forall r< r_C
\end{multline}
\label{eqn2}
where $\epsilon_{AB} = 105\,k_BT$, $\epsilon^{*}_{AB}=16.81$, and $R_{AB}=0.25 \sigma$. 
We model the interaction between similar particles (A-A or B-B) by screened Coulomb purely repulsive potential 
as given by  Eqn. \ref{eqn1}
\begin{equation}
        V_{\alpha \alpha}(r)= \epsilon_{\alpha \alpha} \frac{e^{-(r/R_{\alpha \alpha})}}{r}, \,\, \forall r< r_C
\label{eqn1}
\end{equation}

where $\alpha=A$ or $\alpha=B$; we choose $\epsilon_{AA} = \epsilon_{BB}$ and $r_C=3 \sigma$.  
The parameter values are $\epsilon_{AA} = 939.33\,k_BT$, 
 $R_{AA} = R_{BB} = 0.5 \sigma$. This choice of parameters prevents identical particles 
from coming  close to each other (energy at $r=2 \sigma$ is $ \approx 8 k_BT$, refer Fig.\ref{fig2}b), whereas 
A-B/B-A bound-pairs (effective-bonds) are easily formed  due to the existence of the
the potential minima in the interaction potential. We define A and B to be bonded  if the distance 
between the particles is $< r_b =1.3 \sigma$, where $r_b$  is the position of the potential maxima.
If the A-B distance becomes greater than $r_b$ due thermal fluctuations, we  define the effective bond to be broken.  
The A-B pairs in turn join up to form -A-B-A- chains.  Right angled A-B-A configurations are discouraged 
due to strong repulsive energy between like particles at distances $\sqrt{2}\sigma$ 
(refer Eqn.\ref{eqn1}) resulting in semiflexible chains. 
Furthermore, a third $A$ particle cannot bond at right angles to $B$ to form a branch in a existing A-B-A configiuration 
due to the combined repulsion from the two $A$ particles.
In a configuration where A-B-A forms a straight line, if a pair of A-B particles are kept at a fixed distance 
of $r_{12} =1.12 \sigma$ the potential felt by the third $A$-particle
as a combination of $V_{AA}$ and $V_{AB}$ as a function of distance $r$ 
($r$ is measured from the particle $B$ at the center) is given in 
Fig.\ref{fig1}$c$. As we show later, we can play with the parameters to allow limited amount of branching.

The excluded volume distance $\sigma$ of the $(\sigma/r)^{24} - (\sigma/r)^{12}$ potential sets the unit of length 
in our simulations,  we use $\sigma=1$. All  energies are measured in units of the thermal energy $k_BT$ ($k_BT=1$).
When we report the chain length distribution at different temperatures $T^*$, we maintain
the $\epsilon_{AA},\epsilon_{BB},\epsilon_{AB}$ fixed as mentioned above, and specify $T^*/T$.
The potential cutoff is at a distance of $r_C=3 \sigma$.  The simulation box size is chosen to be 
$30\times30\times 50 \sigma^3$ or $(20 \sigma)^3$, unless otherwise mentioned.  For equilibration, we use at least 
$3 \times 10^5$  Monte Carlo steps (MCS) for volume fractions less than $0.09$, 
at higher densities one needs longer runs to equilibrate.
The statistical quantities are calculated over at least over $5 \times 10^4$ (at times $10^5$)
independent snapshots (microstates), data to calculate statistical averages is collected every $10$ MCS. 
The  maximum value of the trial step-size is  $0.125 \sigma$ in each of $x,y,z$ directions
in a Monte Carlo displacement attempt of each monomer.

\section{Results}

We perform equilibrium Monte Carlo simulations (Metropolis algorithm) at a fixed temperature $T$ 
with equal number of $A$ and $B$ monomers in a simulation box with the potential described in the previous section. 
We have characterized the properties  the self-assembled linear polymeric chains as a fucntion of the volume 
fraction of monomers as we increase the number density.
The monomers are primarily found as monomers or bonded A-B pairs at low densities, but we observe that the self-assembled 
polymers self-organize  into line hexagonal phase of longer chains at medium densities.
At even higher monomer densities, long range order is destroyed and at very high monomer volume fractions the
self-assembled polymers form a gel phase with branching. We can  
modify the potential suitably to introduce branching at lower number densities, though our primary focus
is in obtaining linear polymer chains  without branching at medium densities,
with which we would like study more interesting problems in future.

\begin{figure*}[tbh]
\includegraphics[width=0.48\columnwidth]{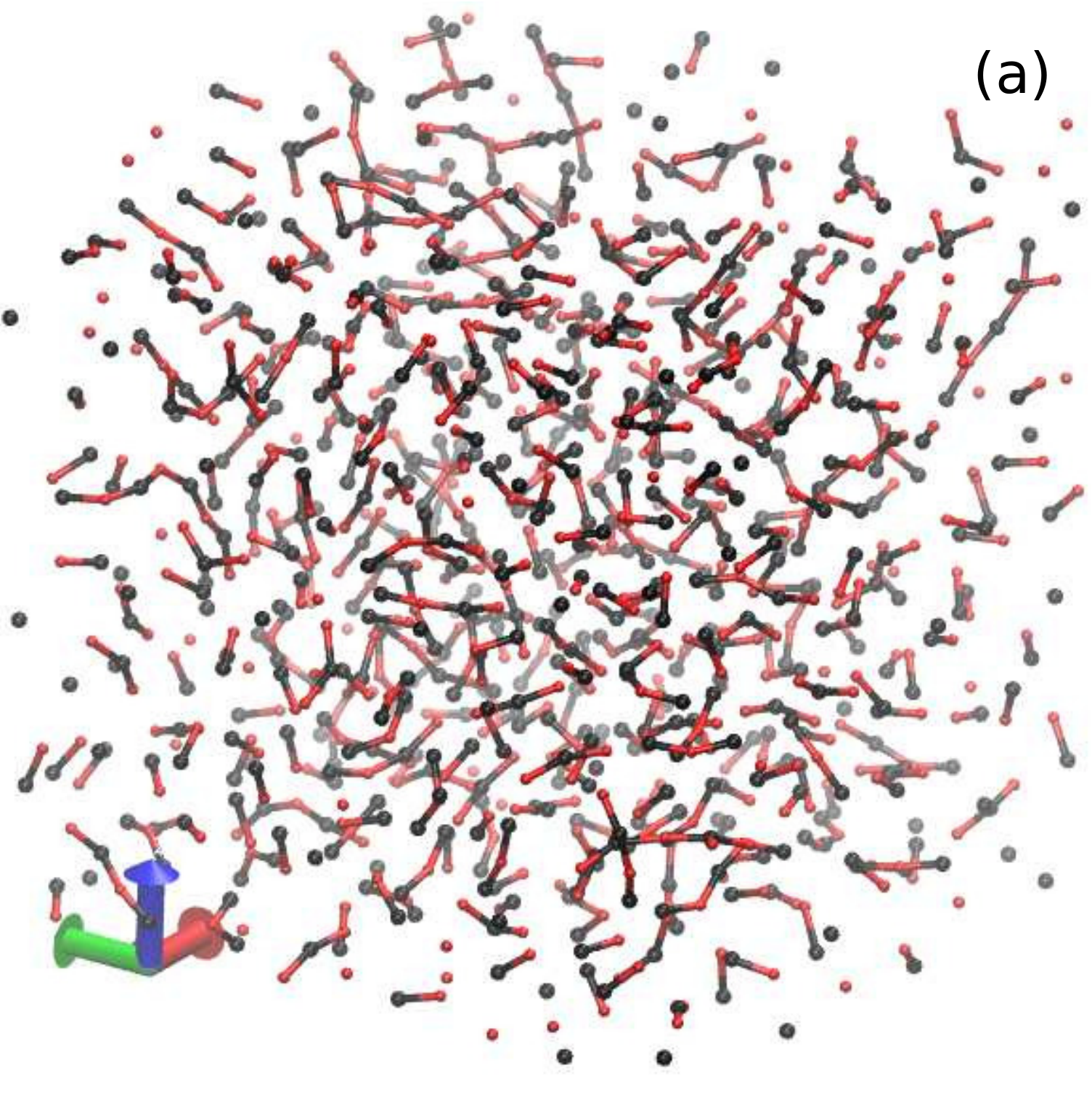}
\includegraphics[width=0.48\columnwidth]{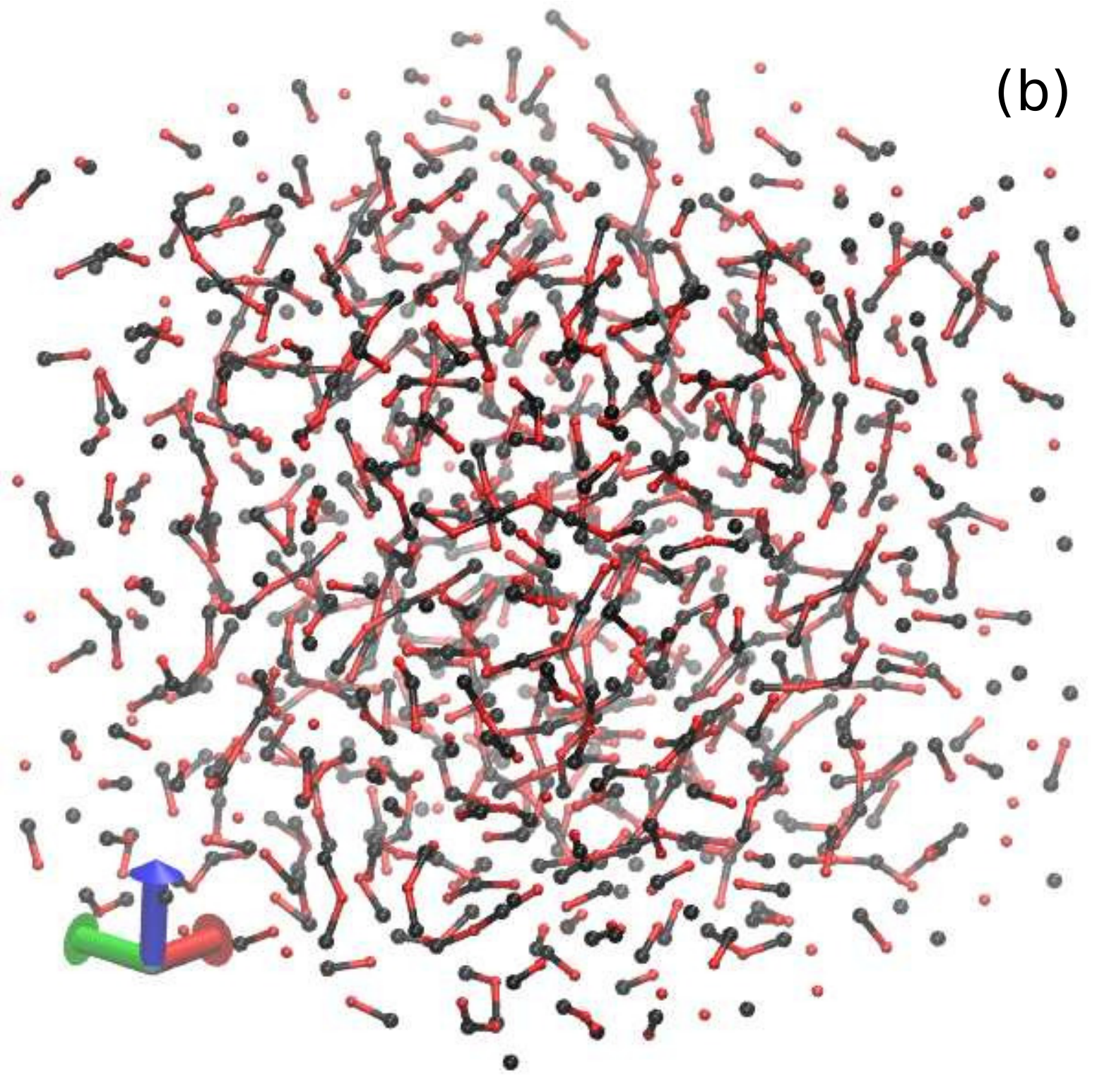}
\includegraphics[width=0.48\columnwidth]{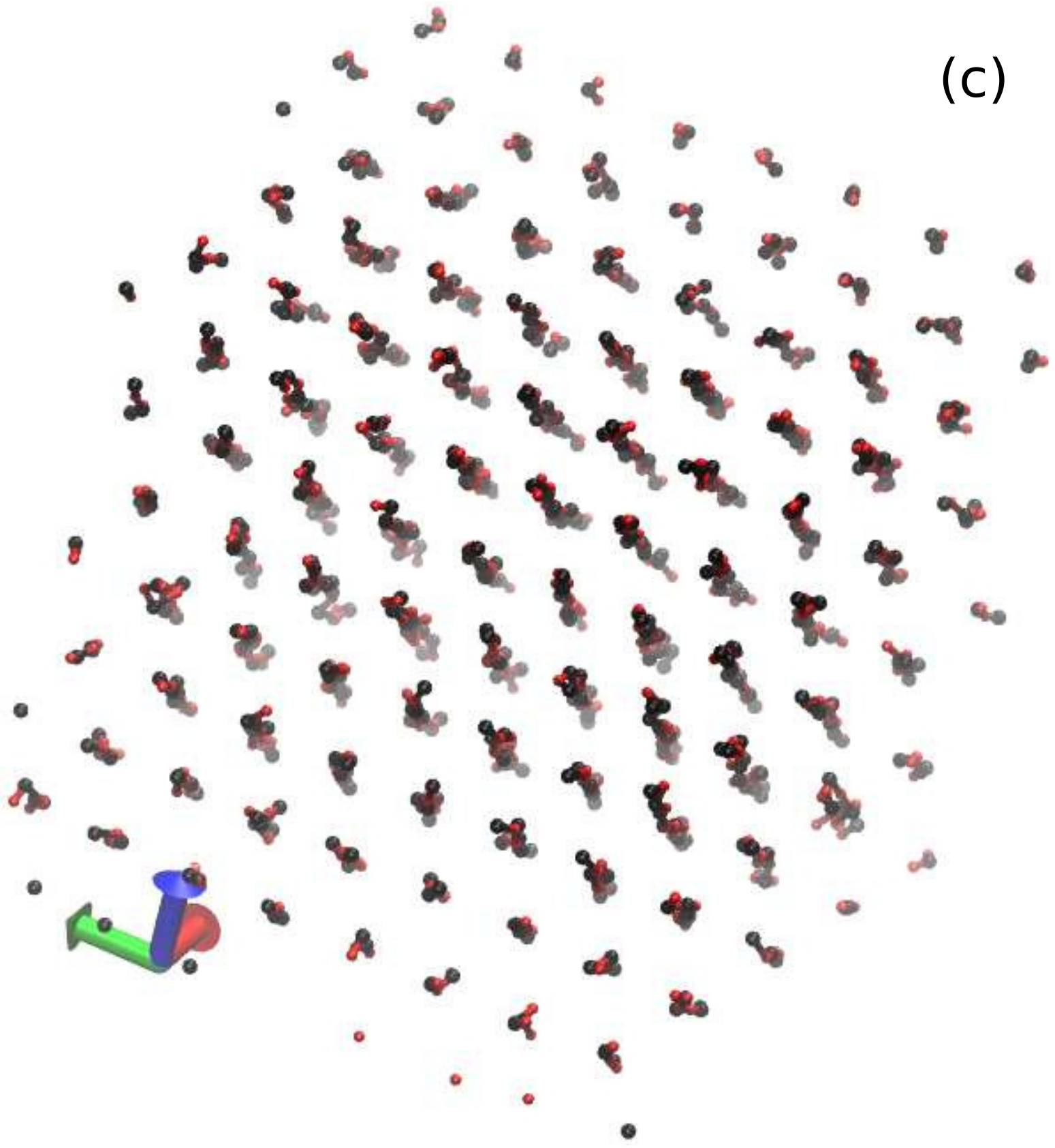}
\includegraphics[width=0.48\columnwidth]{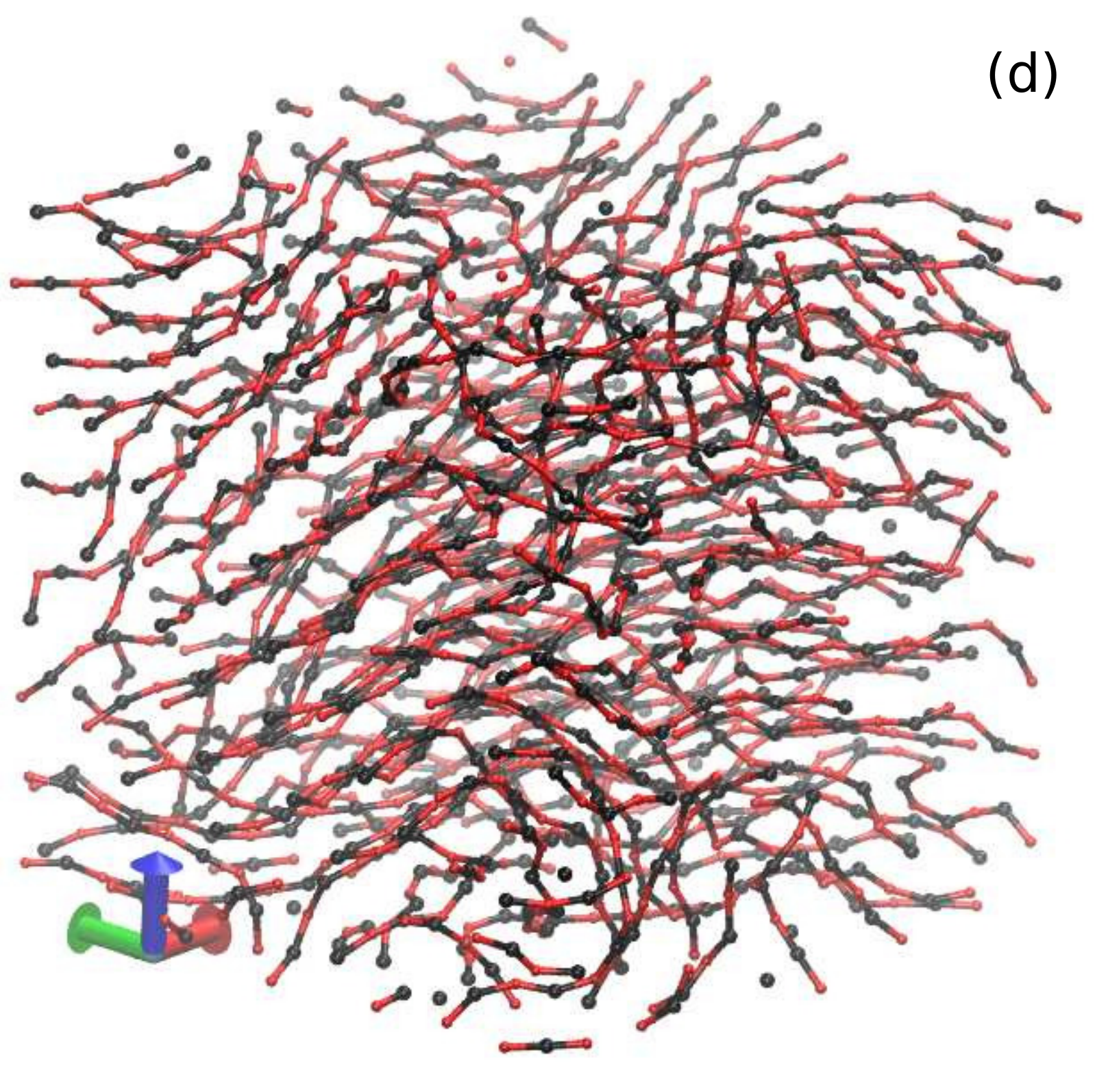}
\caption{\label{fig2} 
Representative snapshots of the assembled monomers at volume fractions 
$\phi_p$ of (a) $\phi_p = 0.08$ with short chains and many monomers and dimer (b) $\phi_p=0.09$ with 
larger fraction of long chains
(c) $\phi_p=0.095$ with long assembled
chains parallel to each other with line-hexagonal order (d) $\phi_p=0.12$ with long  polymers where hexagonal order is destroyed. 
There are equal number of A and B particles, each of diameter $\sigma$. 
The A-particles are coloured red and the $B-$ particles are black. 
The snapshots are for $20 \times 20 \times 20 \sigma^3$ box (for easier viewing),  corresponding snapshots for 
a bigger box size of  $30 \times 30 \times 50 \sigma^3$ are given in Figs S1, S2.
In the snapshot (c) 
the polymer chains are orientationally ordered and are all parallel to each other.
The view direction is parallel to the polymer chains and the orientation direction is into the plane of the paper, 
we choose this to enable the reader to clearly decipher the line-hexagonal order. Plots of same system as seen
from other directions are in Supplementary section (Fig. S3).
}
\end{figure*}

In Fig.\ref{fig2}, we show equilibrium snapshots of self assembled stuctures at low, medium and high volume 
fraction $\rho_p$ of particles in a box of volume $V_{box}=20\times20\times20 \sigma^3$.
Corresponding snapshots for a bigger simulation box, $30 \times 30 \times 50 \sigma^3$ are shown in the supplementary 
section, refer Figs. S1 and S2.
The volume fraction $\phi_p$ calculated as 
$(N \pi \sigma^3)/(6 V_{box})$, where $N$ is total number A and B particles.
The number of A particles in the box is $N/2$.  For volume fractions of $\phi_p=0.08$, one observes 
a large number of unassociated monomers, many dimers, as well as chains with $3$ or more monomers. 
There are very few side-branches emanating out from a linear polymer chain, we  have later quantified the 
number of branches in the system.  The chains become significantly longer at $\phi_p=0.09$ and
at $\phi_p=0.091$  (at $\phi_p=0.99$ for the bigger box) they span the length of the simulation box and arrange themselves in 
a line-hexagonal structure, refer Fig.\ref{fig2}(b) and (c), respectively.  The polymeric chains start developing branches 
at volume fraction of $\phi_p=0.12$ (Fig.\ref{fig2}d) and become a branched
gel at even higher densities.  Note that any particle 
feels a repulsive potential from other particles upto a distance of 
$R_c =3 \sigma$, and if we use $2 \sigma$ as an estimate of the diameter 
of a soft particle (the repulsive energy of a pair of A-A or A-B particle $\sim k_BT$ at $2 \sigma$), then
then an estimate of the effective volume fractions become $\rho_{eff}=0.64, \, 0.72, \, 0.76, \, 0.96$, 
respectively. As for any system of soft particles, the volume fraction $\rho$ or $\rho_{eff}$ 
has to be interpreted with care.  

\begin{figure}[H]
\includegraphics[width=0.46\columnwidth]{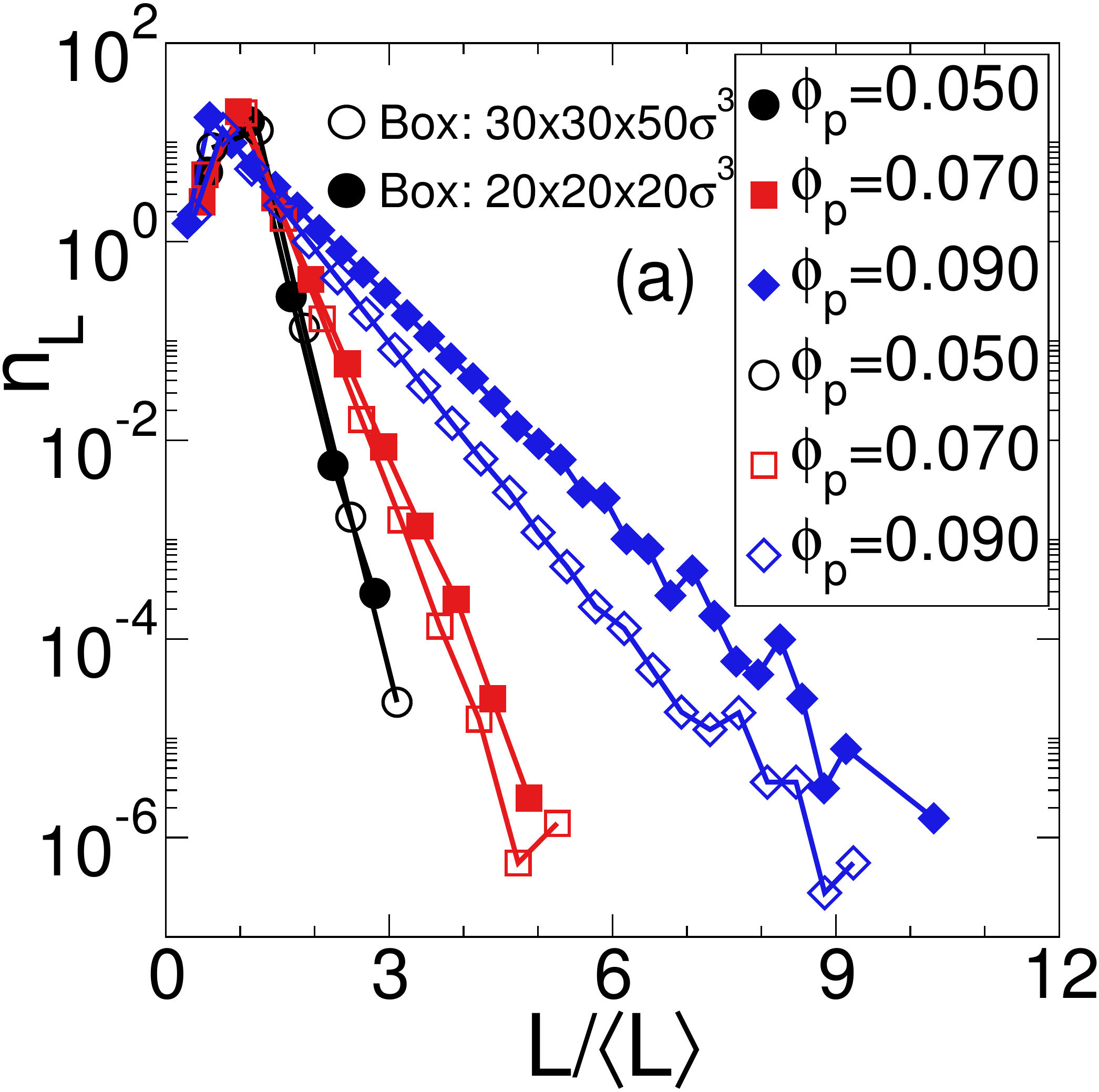}
\includegraphics[width=0.46\columnwidth]{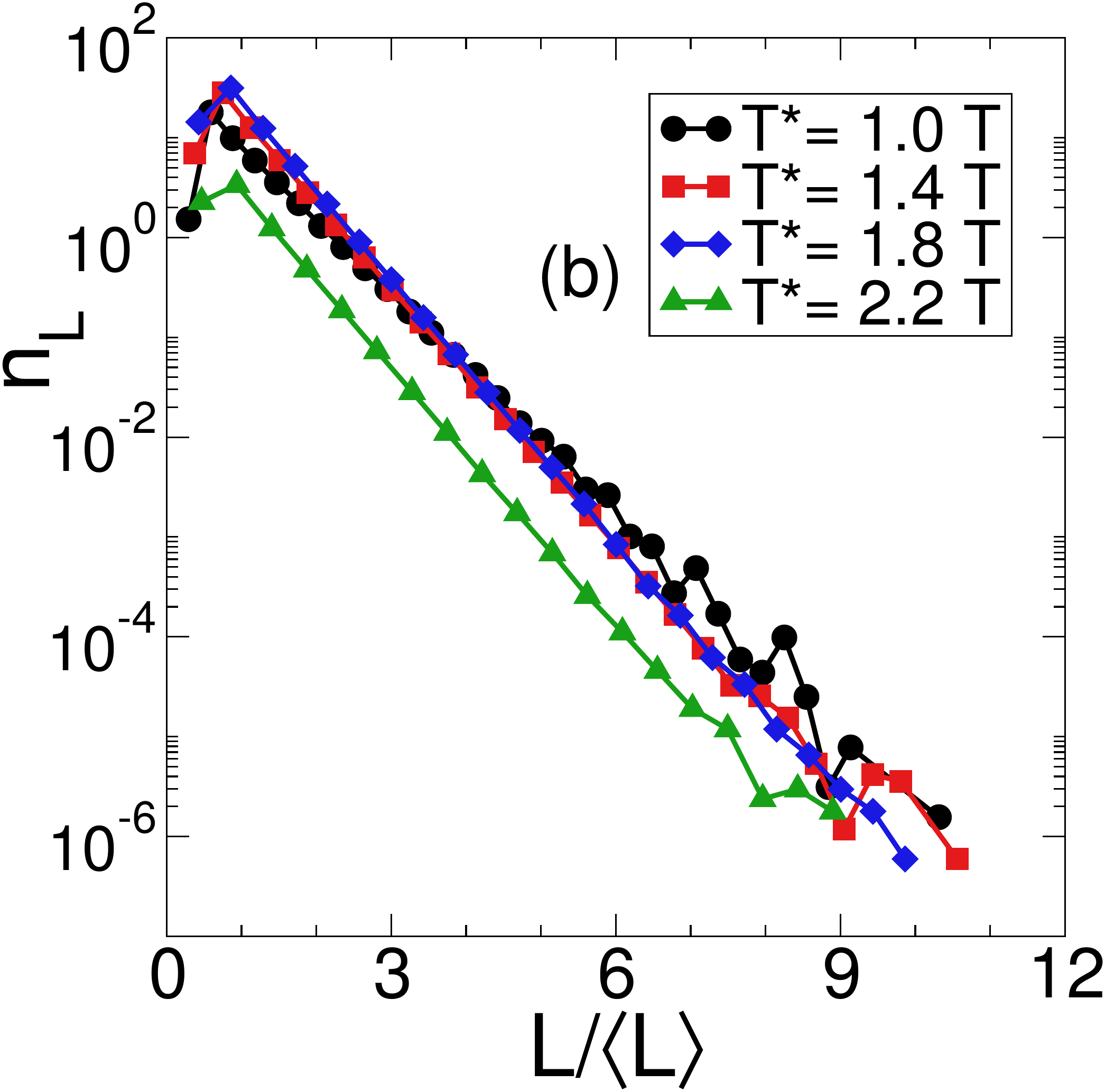}\\
\vskip0.05cm
\includegraphics[width=0.46\columnwidth]{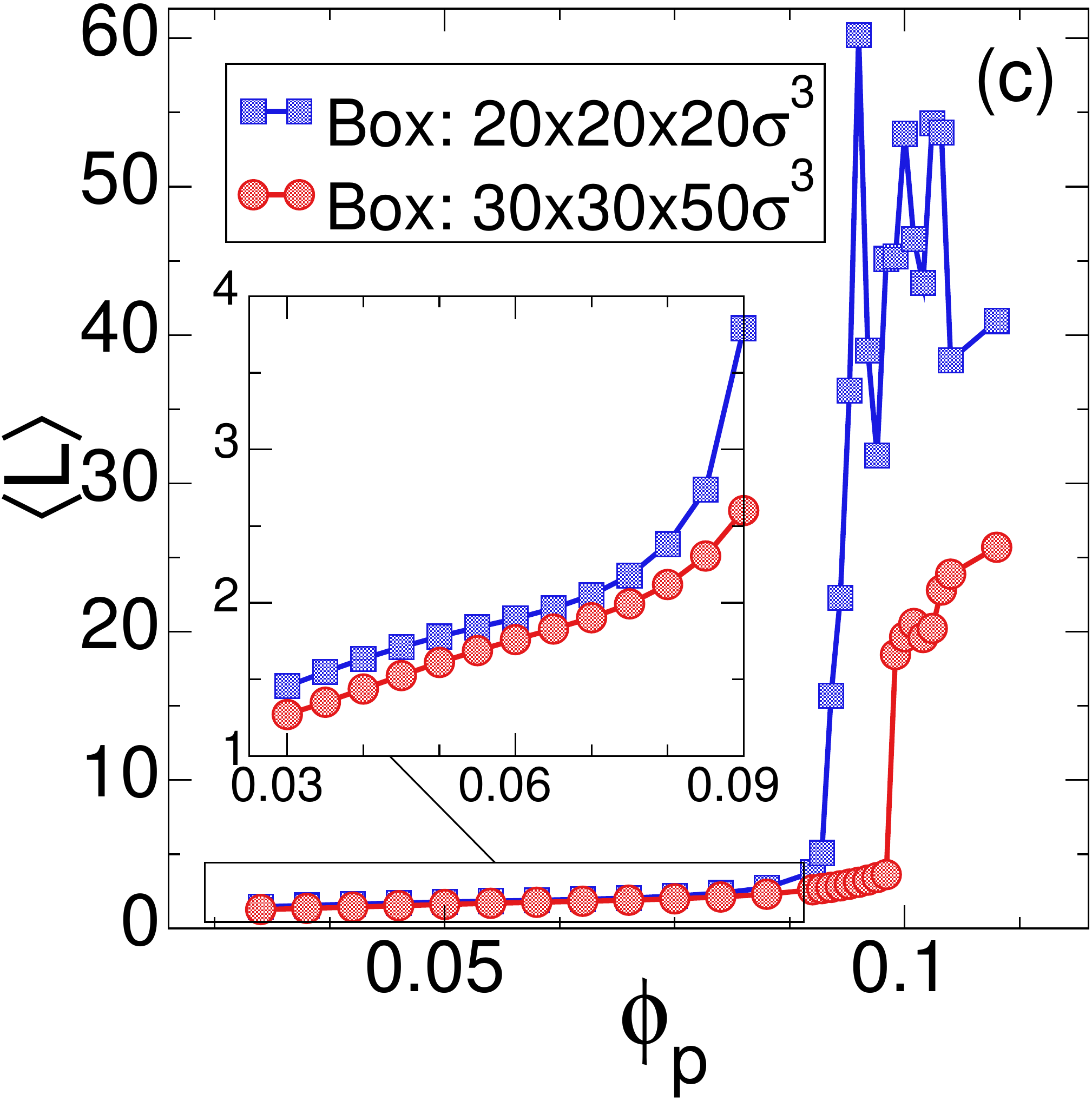}
\includegraphics[width=0.46\columnwidth]{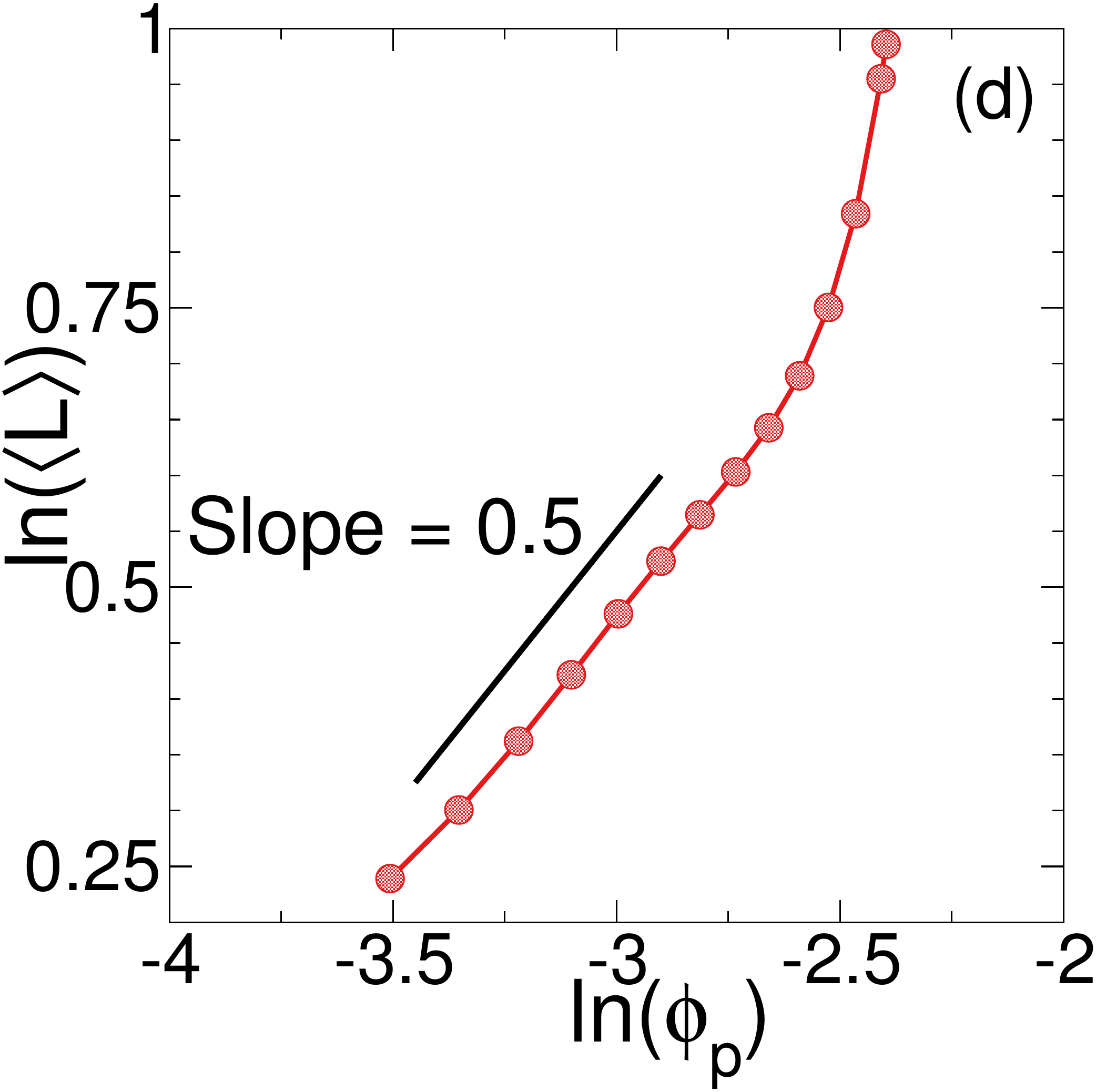}
\caption{\label{fig3} (a) The length distribution of assembled polymeric chains is  shown for 
different volume fractions of monomers. The y-axis shows the average number of chains $n_L$
suitably normalized for a box of size $(10\sigma)^3$ and the $x$-axis is the length $L$ of the chains, 
scaled by $\langle L \rangle$: the average length of chain
for a particular volume fraction $\phi_p$.  The number of chains of 
length $L$ is normalized to the volume of a $(10\sigma)^3$ box to obtain $n_L$.
Thermal energy $k_BT=1$. Data for $(20 \sigma)^3$ is shown with filled symbols; open symbols show data for 
simulation box $30 \times 30 \times 50 \sigma^3$.
(b) The length distibution of self-assembled chains at $\phi_p=0.09$ at different temperatures $T^*/T$ 
keeping the potential values fixed.
(c) The average length of the polymer chains $\langle L \rangle$ with plotted versus the volume fraction 
of monomers $\phi_p$. It shows a jump just above $\phi_p=0.091$ in simulation box and at $\phi_p=0.099$ in bigger box. The inset shows in detail the data 
shown in the boxed region of the main graph. (d) The average length $\langle L \rangle$
versus $\phi_p$ in a  log-log scale to identify the exponent $\alpha$ in $\langle L \rangle 
\sim \phi_p^{\alpha}$. 
}
\end{figure}

This system of particles behaves like a self-assembling system of equilibrium polymers with suitable length
distributions. We plot the chain-length distribution in Figure \ref{fig3}(a), 
i.e. the number of chains $n_L$ with $L$ number of monomers in a chain normalized for 
a box size of volume $(10 \sigma)^3$. Since  both the number of chains and the length of chains
keep fluctuating in the simulation box, we choose not to  
normalize by the total number of chains in box. But since we show data for two different box sizes,
we normalize  chain length distribution data by using the factor $(10\sigma)^3/V_{box}$; i.e. $n_L$ is the number of
chains one would find in a volume of $(10 \sigma)^3$.   
We show the change in the distribution  as we
vary  the volume fraction $\phi_p$ and the temperature $T^*$ ($k_B=1$). The particles start assembling above a 
 volume fraction $\phi_p \sim 0.05$ and thereafter maintain an exponential distribution of chain lengths
with increasing $\phi_p$, as expected for a worm-like micellar system. With increasing density the length 
of the longest chains present in the system also increases as can be confirmed from Fig.\ref{fig3}(a). 
Data for $\phi_p=0.09$ shows significant finite size effects, as the chains are longer and bigger than 
half the size of box for $(20\sigma)^3$ box.

Figure \ref{fig3}(b) shows that the the chains break into smaller assemblies
when the temperature $T^*$ is increased but the chain length distribution continues to remain linear 
in a semi-log plot, as expected for self-assembling equilibrium-polymers. 
In \ref{fig3}(c), we observe that  the average length of chains $\langle L \rangle$ gradually increases 
with $\phi_p$, but for $\phi_p >0.091$ it  shows a jump for a $(20 \sigma)^3$ box. For the bigger
box size of $ 30 \times 30 \times 50 \sigma^3$ box, the jump in $\langle L \rangle$ 
moves to a slightly higher value of $\phi_p$.
This happens because the chain arrange themselves 
in a line-haxagonal order. Simultaneously, they become long enough to span the simulation box to form
ring polymers using the periodic boundary condition (PBC) to minimize energy of a polymer chain due to presence of an
end-cap.  The formation of ring polymers invoking PBC is clearly an artefact of finite size of the simulation box. 
Snapshots for the self-assembly of particles at $\phi_p=0.095$ in different box sizes are presented 
in the supplementary section (Fig. S4 
 as well as Fig. S1. 
 From these figures we can conclude that 
in the thermodynamic limit one can expect the system to form domains 
of hexagonally ordered chains with different orientation of  chains in different domains. 
An individual long polymer can continue to meander from one domain to the other by suitable 
bending and looping. It may also possibly form rings (even without invoking PBC). 
Figure \ref{fig3}(d) shows that the mean length of the self-assembled
chains  $\langle L \rangle$ increases as $\phi_p^{0.5}$
for dilute systems, as predicted by Cates and Candau \cite{cates_candau} for worm-like micelles.

\begin{figure}[H]
\includegraphics[width=0.47\columnwidth]{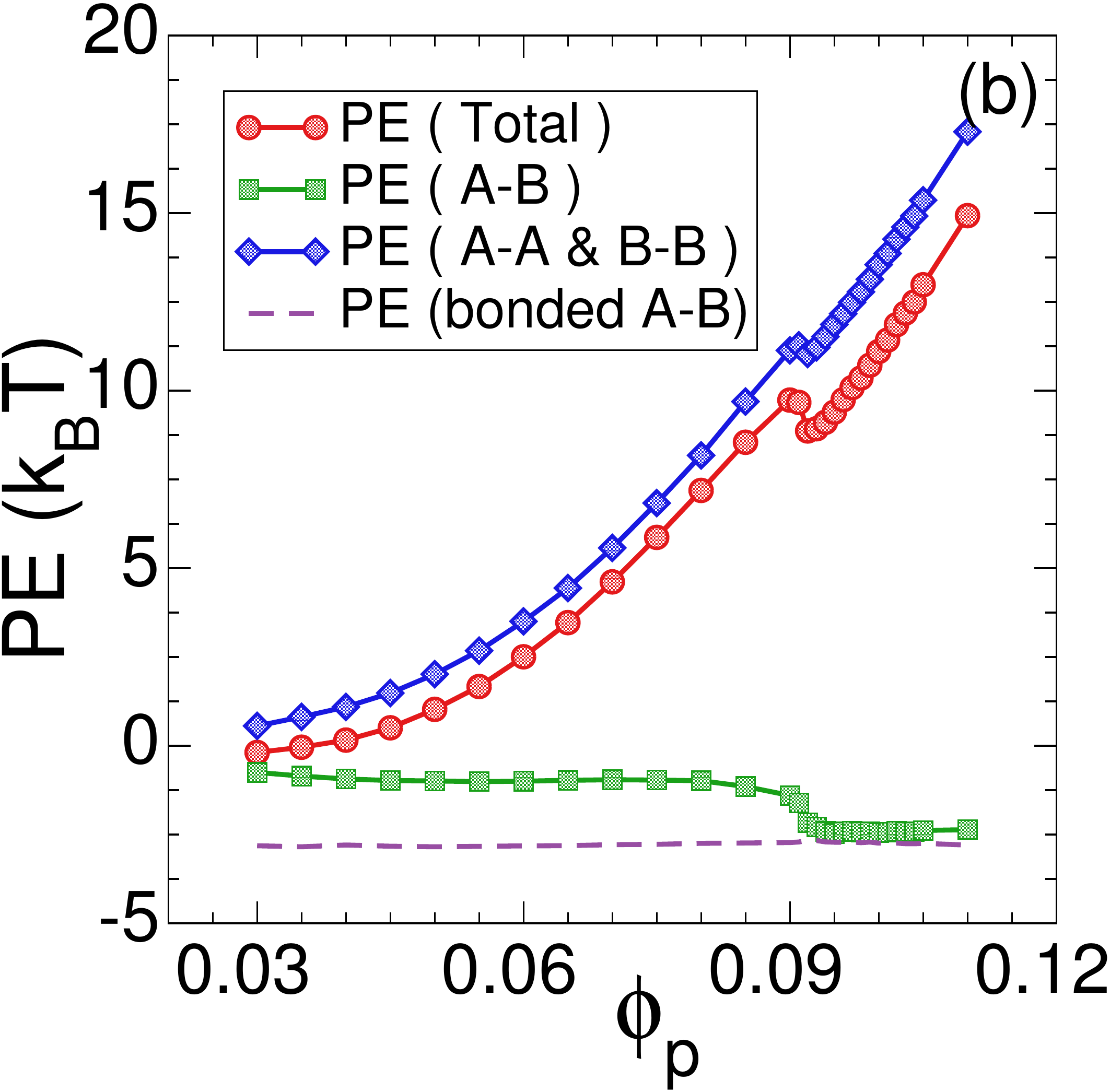}
\includegraphics[width=0.47\columnwidth]{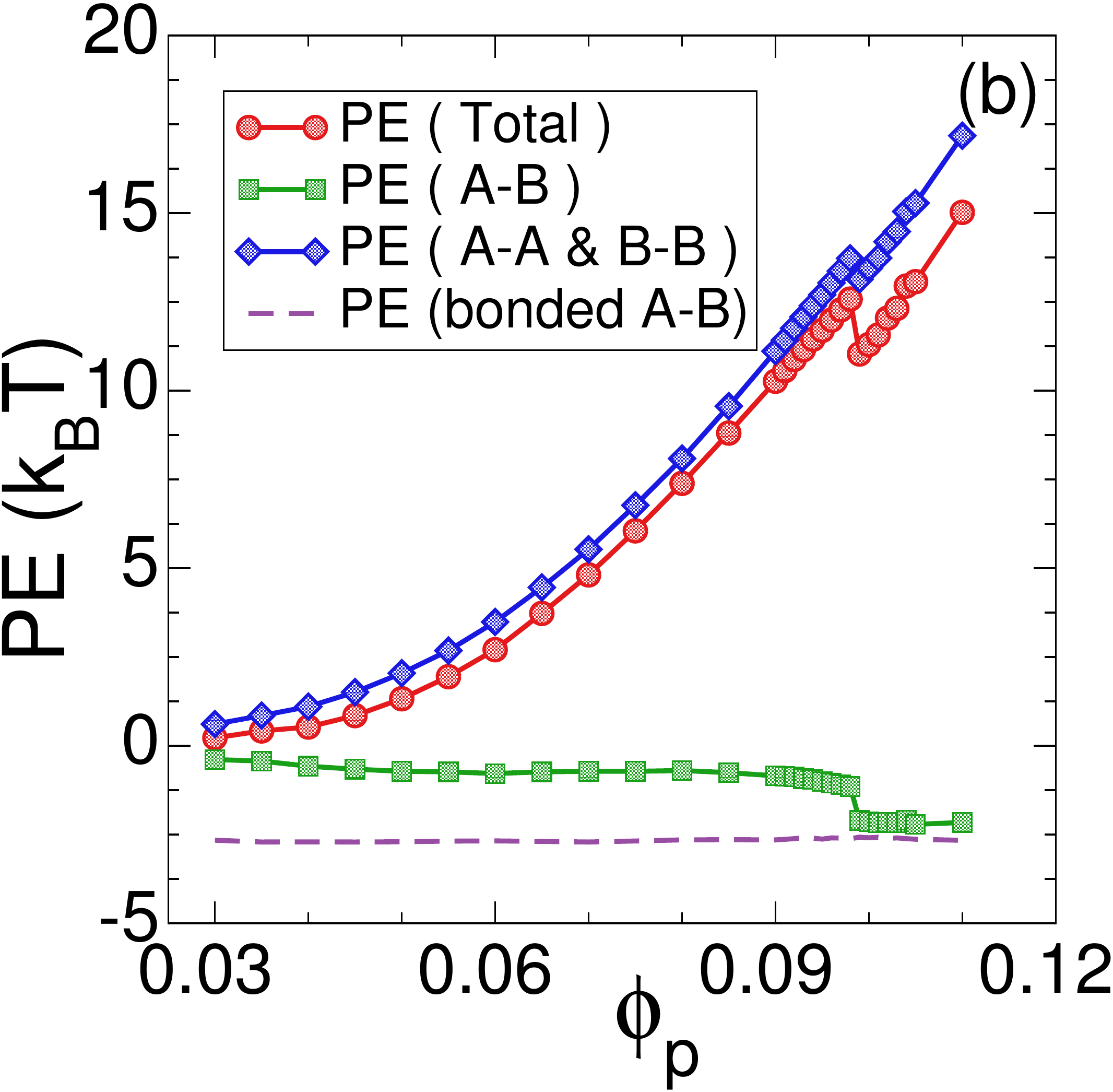}
\vskip0.2cm
\includegraphics[width=0.47\columnwidth]{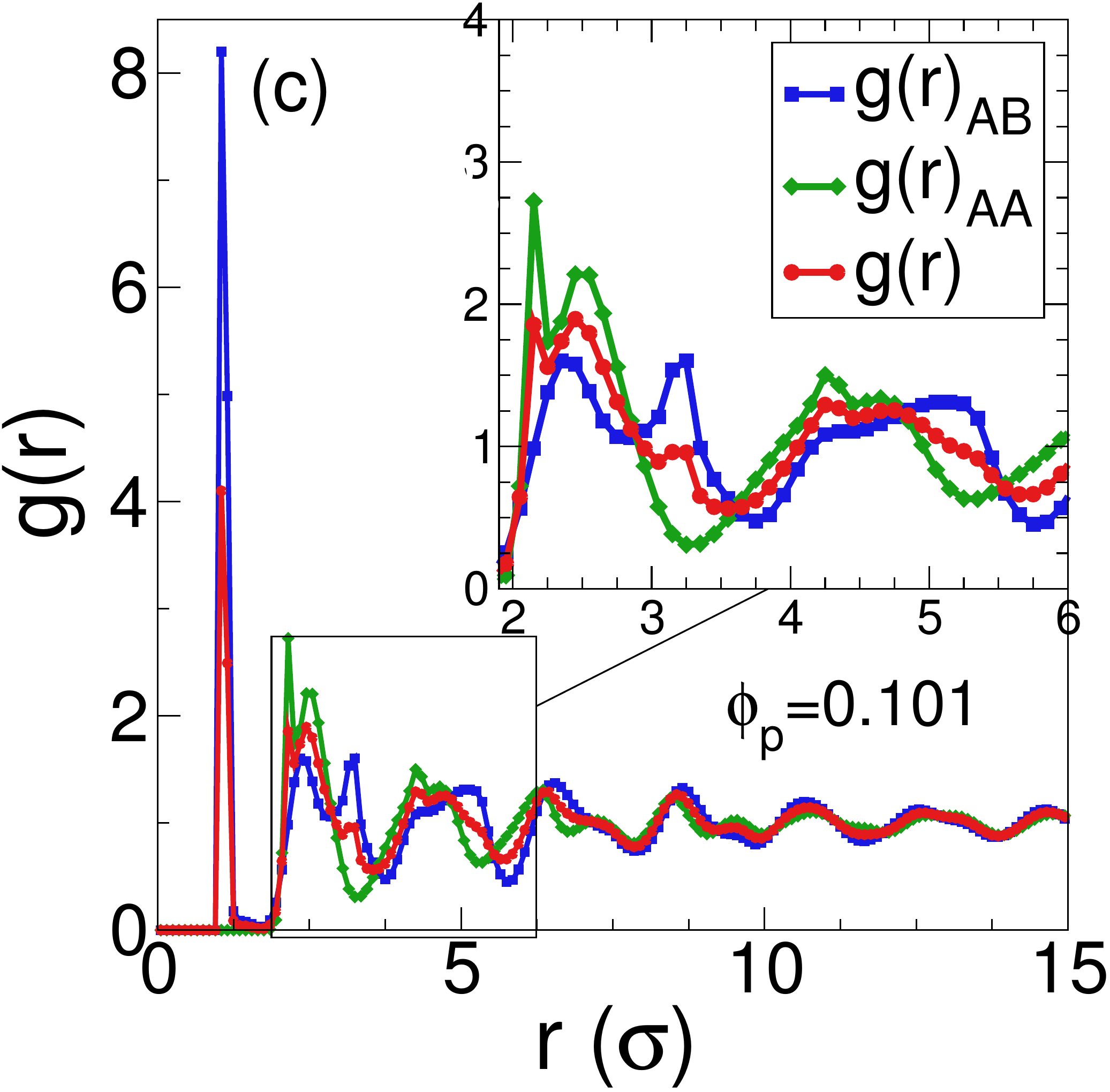}
\includegraphics[width=0.47\columnwidth]{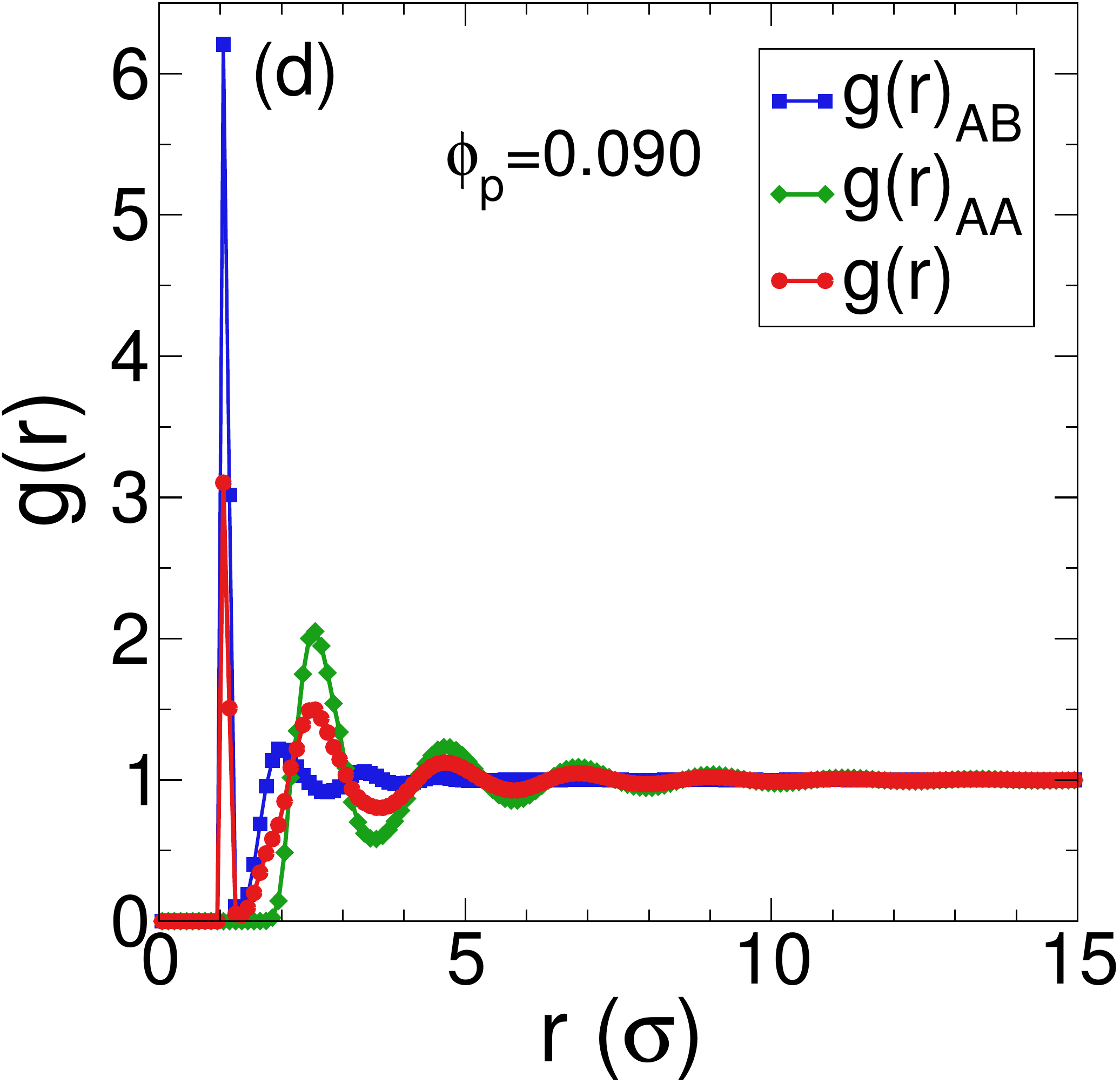}
\caption{\label{fig3a} 
Plots of energy contribution from different pair interaction of A and B monomers
for box sizes of (a) $(20\sigma)^3$ and (b) $30\times 30 \times 50 \sigma^3$
Different symbols indicate the total PE, PE due to A-B interactions (sum of attractive and repulsive interactions),
sum of purely repulsive A-A \& B-B interactions, and attractive ponetial energy between A-B 
which are at distances $r < 1.3 \sigma$ such that they are bonded. 
The average potential energy $PE$ in units of $k_BT$ of the system versus $\phi_p$ show a discontinuity 
at the same $\phi_p$ where $\langle L \rangle$ shows the jump.
Pair correlation functions $g(r)$ at volume fractions (c) $\phi_p =0.09$ and (d) $\phi_p=0.101$ [line-hexagonal
phase] for box size $30 \times 30 \times 50 \sigma^3 $ for all monomers, A-A/B-B pair of monomers and
A-B monomer pairs.
}
\end{figure}

To understand the energetics in comparison to thermal energy scales ($k_BT$) 
of the self-assembling systems of polymers we note that at very low densities 
($\phi_p= 0.03$ or $\rho_{eff}=8\phi_p$), a pair of bonded A and B monomers gain an average potential energy 
(PE) per particle of $\approx -2.5 k_BT$ (refer Fig.\ref{fig1}a for graph of potential and Fig.\ref{fig3a}(a,b) 
showing PE for two different box sizes); the dashed line  represents potential energy (per particle) 
contribution from bonded A-B interactions.  
This is offset by the repulsive interactions between particles (A-A,A-B and B-B) which act at larger 
distances between the particles,  such that the total PE combining all attractive and repulsive
contributions is nearly zero. 
But as the density increases the $A-A$ and $B-B$ repulsion starts to play a more significant role than 
the attractive A-B bond-energy ($\approx -2.5 k_BT$), the A-A or B-B PE contribution increases  and 
correspondingly the total PE of the system takes increasingly higher positive values, refer Fig. \ref{fig3a}a,b.
At higher densities, the particles bond to form long polymers (trimers or longer) by using the minima of the 
potential (refer Fig.\ref{fig1}c).  
Simultaneously we observe that the contribution of the non-bonded (repulsive) A-B interaction becomes nearly zero, as 
the value of the total PE of A-B interactions is equal that PE contribution due to bonds between A-B particles.

For $\phi_p >0.09$, the monomers form the line-hexagonal 
phase and moreover, A and B monomers from {\em adjacent} chains adjust their relative positions such that 
they are nearest each other and A-A (or B-B) repulsive interaction energy between monomers of adjacent chains 
is minimized.  This is clearly seen in the representative snapshot shown supplementary section Fig. S6
and can
also be confirmed from the pair correlation function $g(r)$ shown in Fig. \ref{fig3a}c.
Due to this ordering of chains and re-adjustment of the relative positions of 
A and B monomers from adjacent chains, we see a drop 
of total PE of A-B interaction [square symbols in Fig.\ref{fig3a}(a,b)] near $\phi_p=0.091$ ($\phi_p=0.099$ for
bigger box), however,
the mean negative energy (dashed line) contribution from intra-chain bonded $A-B$ monomers
 remains nearly unchanged across this transition.  The total PE of the system shows a discontinuity 
near $\phi_p=0.093$, indicating a first order transition to a orientationally ordered state.  The result is robust,
the jump in total PE is seen in both the box sizes, only the corresponding $\phi_p$ value changes slightly as one changes
the box size which is likely to be a finite size effect.

We calculate the pair correlation function  $g(r)$ between monomers, the $g(r)$ data is 
 shown in Figs.\ref{fig3a}(c,d). When the monomers 
form domains of parallel polymers ordered in line-hexagonal phase at $\phi_p=0.10$ for the bigger boxsize, 
the first peak of $g(r)_{AB}$ between A and B monomers shows up at a distance just beyond $1\sigma$ corresponding to the 
distance between adjacent A and B bonded monomers in a chain. The next peak in $g(r)_{AB}$ at distance 
just beyond $2 \sigma$ is due to the arrangment between A and B monomers from adjacent chains, as discussed
in the previous paragraph. The third peak just beyond a distance of $3 \sigma$ is a consequence of the 
arrangement of monomers belonging to the same chain: the second nearest neighbour A-monomer from a B-monomer
along a chain will be at a distance of about $3 \sigma$. The $g(r)_{AA}$, the pair correlation fucntion
between A-monomers (equivalently B monomers) shows the first peak at $2 \sigma$ corresponding to the smallest distance
between A-monomers along the contour length of a self-assembled chain. The two A-monomers have a B-monomers in between them.
The next peak corresponds to A-monomers from adjacent chain.
The quantity $g(r)$ is the pair correlation function between all monomers without discriminating
between A and B monomers, it shows the  peaks of both $g(r)_{AA}$ and $g(r)_{AB}$.
The total $g(r)$ shows regular peaks upto half the length of the box, which corresponds to long-range ordering
of chains and thereby monomer positions. In contrast $g(r)$ for lower volume fractions (e.g. $\phi_p=0.09$,
refer Figs.\ref{fig3a}(d) shows the first two peaks for $g(r)_{AA}$ and $g(r)_{BB}$ but there is no
long range order to be seen.

\begin{figure}[H]
\includegraphics[width=0.49\columnwidth]{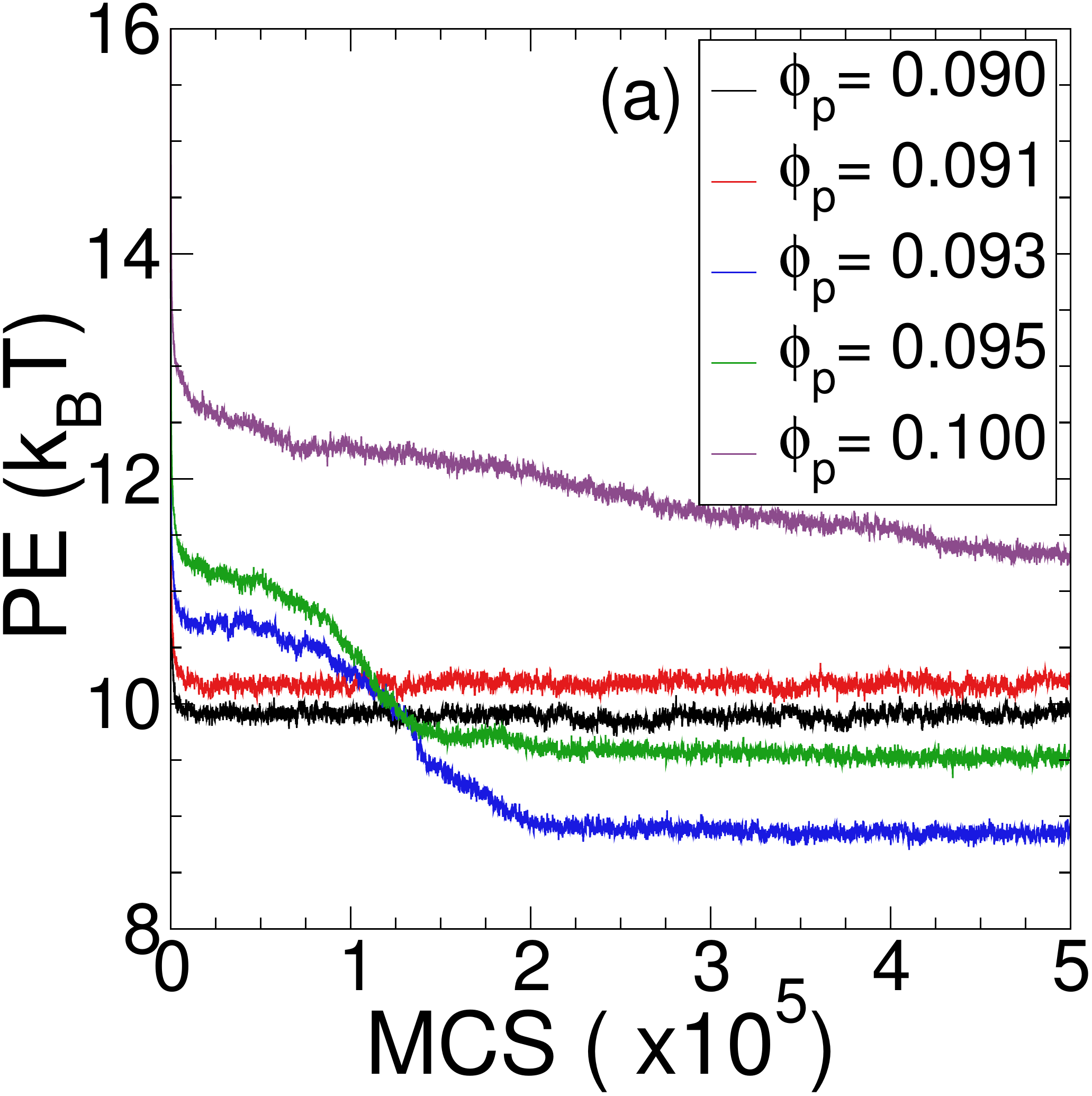}
\includegraphics[width=0.49\columnwidth]{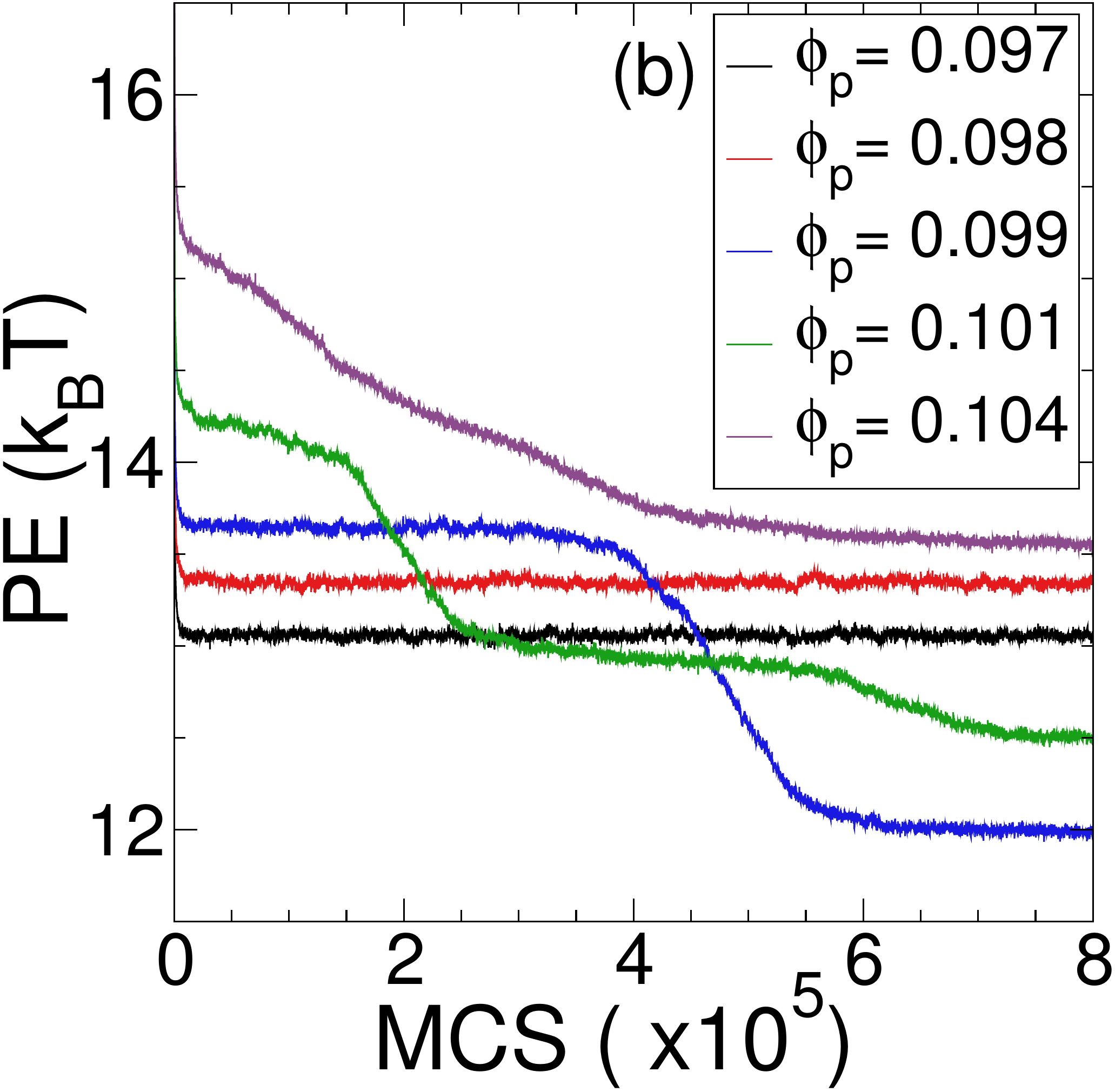}
\caption{\label{fig3b} 
 The total potential energy (PE) vs number of Monte Carlo steps (MCS) for different values of $\phi_p$ for box sizes 
(a) $(20\sigma)^3$ and (b) $30\times 30 \times 50 \sigma^3$ to check the  approach to equilibrium of
the different systems.
}
\end{figure}

Figures \ref{fig3b}(a,b) show the potential energy $PE$ of the system plotted every $100$ Monte 
Carlo steps (MCS)  for different values of $\phi_p$ to enable us to analyze the approach to equilibrium 
of the self-assembled chains, especially when domains with line-hexagonal phases are formed. 
For relatively lower values $\phi_p$ ($\phi_p=0.09$),
the system reaches equilibrium within $0.2 \times 10^5$ MCS for the box size of $(20\sigma)^3$
starting from a random initial configuration of monomers and then the energy  fluctuates about 
the average value of energy.  But  for $\phi_p=0.093$ and $\phi_p= 0.095$, the energy first shows 
a small plateau till around $1 \times 10^5$ MCS (for small box), and then settles down to a lower 
mean value of PE after $2 \times 10^5$ MCS. 
The mutual repulsion between self-assembled chains arising from the 
repulsion between A-A and A-B particles (for $r>1.3 \sigma$) keeps the polymer chains straight and 
parallel to each other. 

We see similar relaxation behaviour at $\phi_p=0.099$ and $\phi_p=0.101$ for simulations in the bigger box size
where there are steps in PE in at $2 \times 10^5$ and $4 \times 10^5$ MCS, respectively; longer relaxation times
for bigger box sizes are expected. 
For the runs with $\phi_p=0.101$,
 snapshots of the configuration before $2 \times 10^5$ MCS,  between $ 2 \times 10^5$ and  $6 \times 10^5$ MCS,
and after $6\times 10^5$ MCS are shown in the supplementary section for the bigger box size (Fig. S5).
After the initial transient relaxation (at $0.1 \times 10^5$ MCS)  long self-assembled 
chains are formed from the monomers  till $2 \times 10^5$ MCS, but after the PE dip at $
2 \times 10^5$ MCS, the polymers   spontaneously self-organize themselves  into small
 domains with hexagonal line order.  After $6 \times 10^5$ iterations, the small domains reorganize
into much larger domains, and one sees a corresponding second dip in the energy plot of \ref{fig3b}(b) 
for $\phi_p=0.101$. 
At densities of $\phi_p=0.1$ (or more), the system has not equilibrated within the 
simulation time scales, but we do not present statiscally  averaged quantities for the high density regime.

\begin{figure}[!tbh]
\includegraphics[width=0.49\columnwidth]{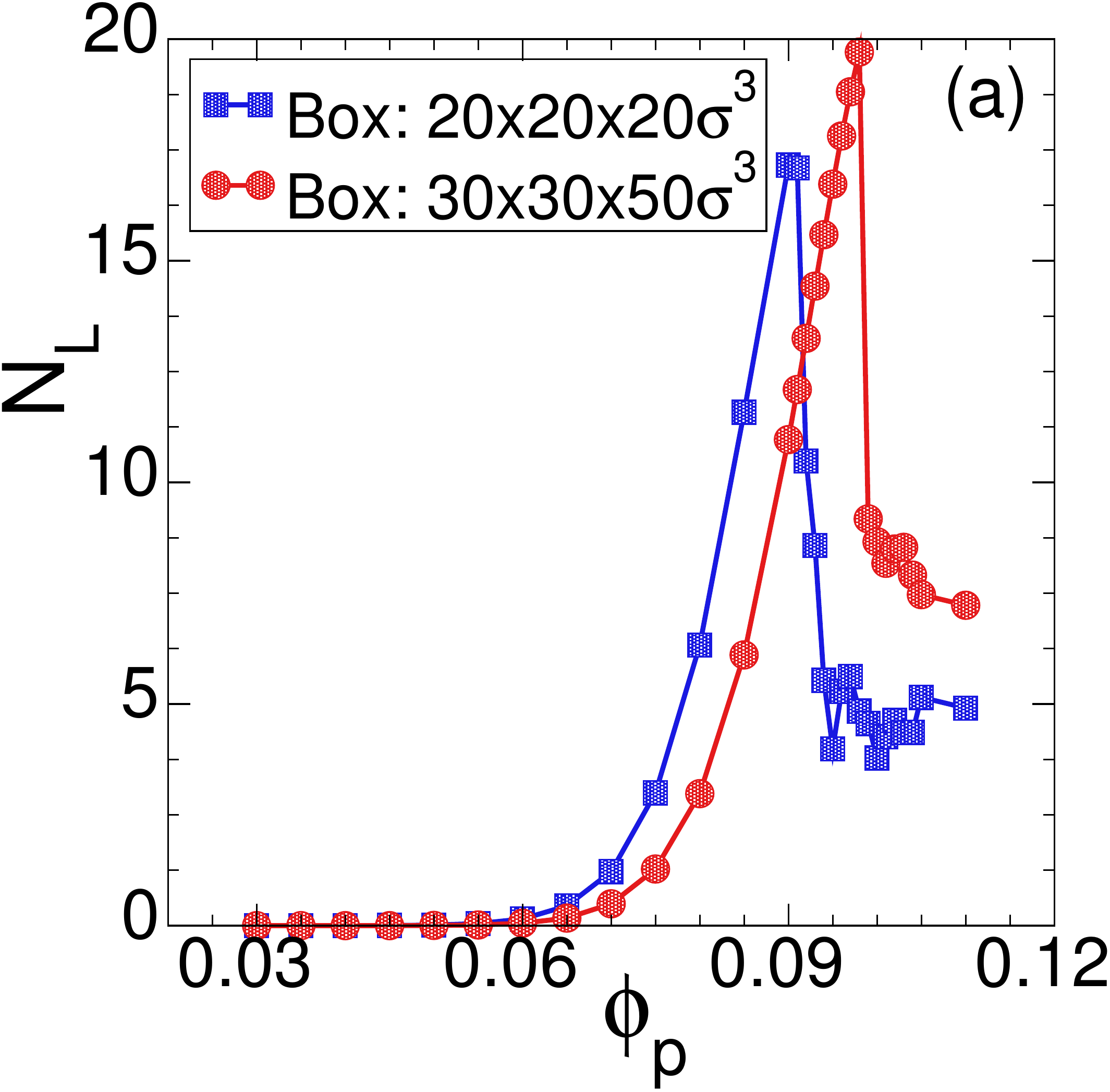}
\includegraphics[width=0.49\columnwidth]{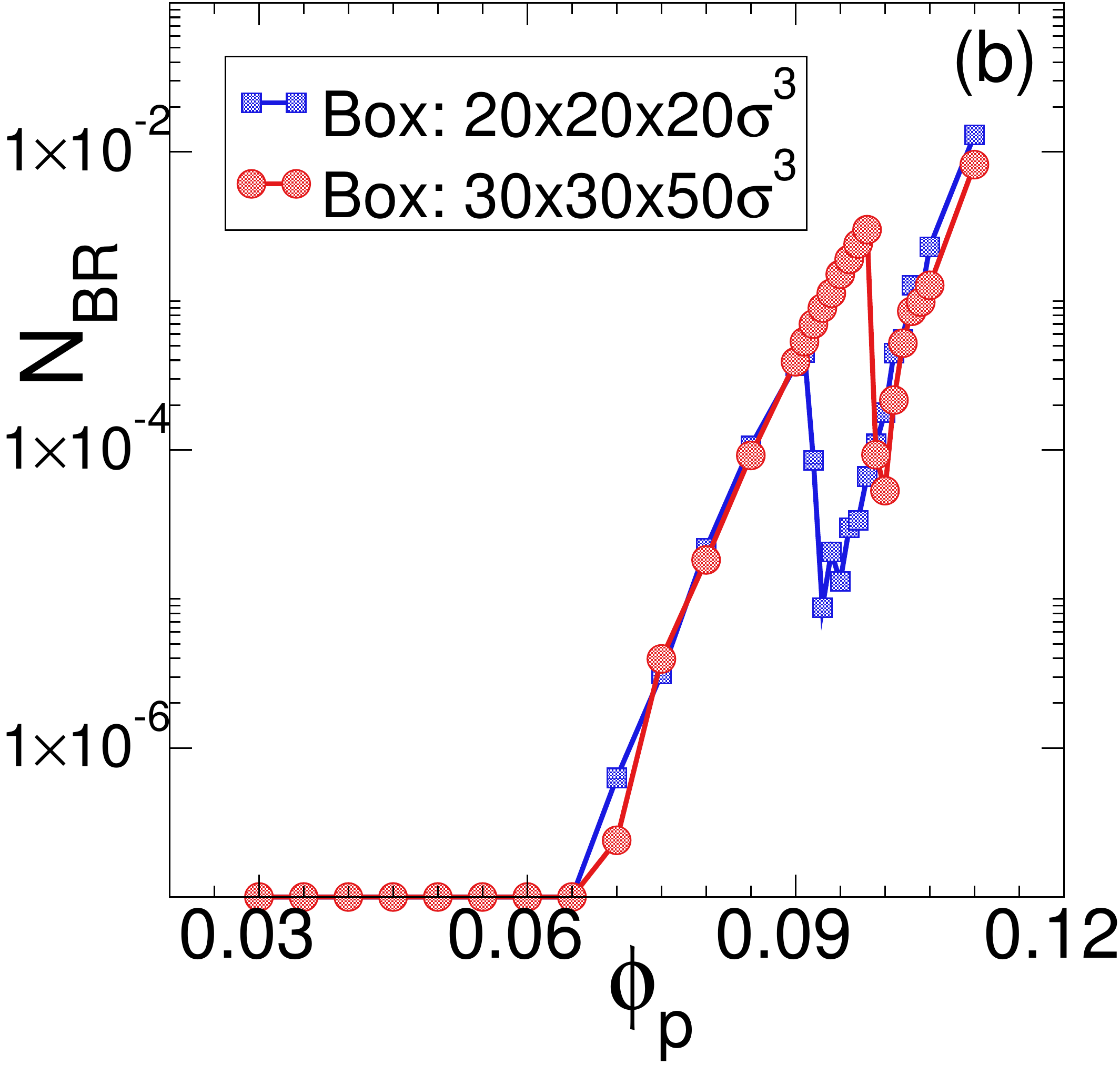}
\caption{\label{fig4} (a) The average number $N_L$  of self-assembled chains in a volume of  
$(10\sigma)^3$ ( obtained from simulations box sizes of $(20\sigma^3)$ and 
$30 \times 30 \times 50 \sigma^3$) is plotted versus the volume fraction of monomers $\phi_p$.
$N_L $ counts only those chains which number of monomers $L>3$ in a chain. (b) The plots shows the 
average number of branches $N_{BR}$ (suitably normalized for a box of size of $(10\sigma)^3$ as before)
 plotted against the monomer volume fraction $\phi_p$. 
Branched configurations need to have a minimum of $4$ or more monomers in a chain.
The values shown as $N_{BR} < 10^{-6}$ are actually zero, they have been by hand set to $10^{-7}$ to enable 
us to show them in a log-scale in the figure.
}
\end{figure}

Figure \ref{fig4}(a) shows that the number of chains $N_L$ with $L>3$ in simulation box, normalized by the volume
of a $(10 \sigma)^3$ simulation box (normalization reason discussed  previously, refer text of Fig. \ref{fig3}), 
though we present data for simulations performed in 
$30\times 30 \times 50 \sigma^3$ box, and in $(20 \sigma)^3$ box sizes. There are very few chains 
with length $L>3$ for $\phi_p<0.06$; however, $N_L$ increases with $\phi_p$ for $\phi_p>0.06$. For
$ \phi_p >0.091$, the chains straighten and become parallel to each other to form the line-hexagonal phase,
and the number of chains drops and remains nearly fixed at the number of parallel chains accommodated in the system.
The chain length distribution is no longer exponential, and large fluctuations are seen in the value of $N_L$ when chains break.
In a square cross section of $(10 \sigma)^2$, there are around $7$ to $9$ chains with area fraction of 
$8 \pi r^2/(10\sigma)^2 = 0.36$ considering $r=1.2 \sigma$ calculated from the second peak of $g_{AB}(r)$ data.
Furthermore, $\langle L \rangle$ is of the same order as the largest dimension of the box ($20 \sigma$ or  $50 \sigma$). 
Hence the numbers for $\langle L \rangle$ for $\phi_p>0.09$ can be expected to have strong finite size effects.
Fig.\ref{fig4}(b)  shows that the average number of branches in a particular microstate (snapshot) 
in the system is very small, compared to the number of chains $N_L (L>3)$ in the system. 
The number of branches increases as we change $\phi_p$ 
from $0.06$ to $0.09$, but still remains insignificant compared to the number of chains 
$n_L (L>3)$ in the system.  Since the number of branches per chain always has such a very low 
value, effectively there are no branches in our system for $\phi_p < 0.09$.

\begin{figure}[H]
\includegraphics[width=0.48\columnwidth]{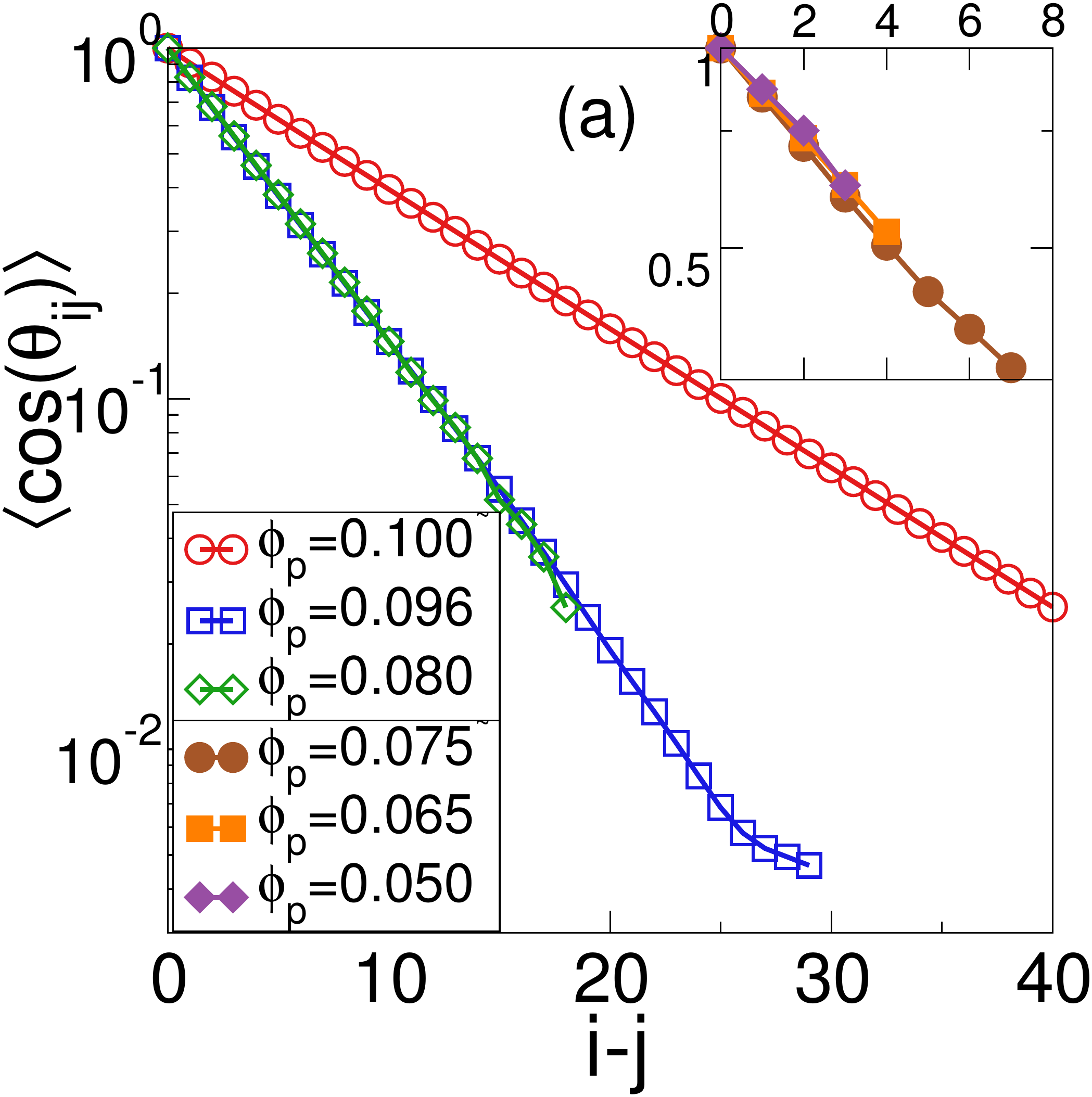}
\includegraphics[width=0.48\columnwidth]{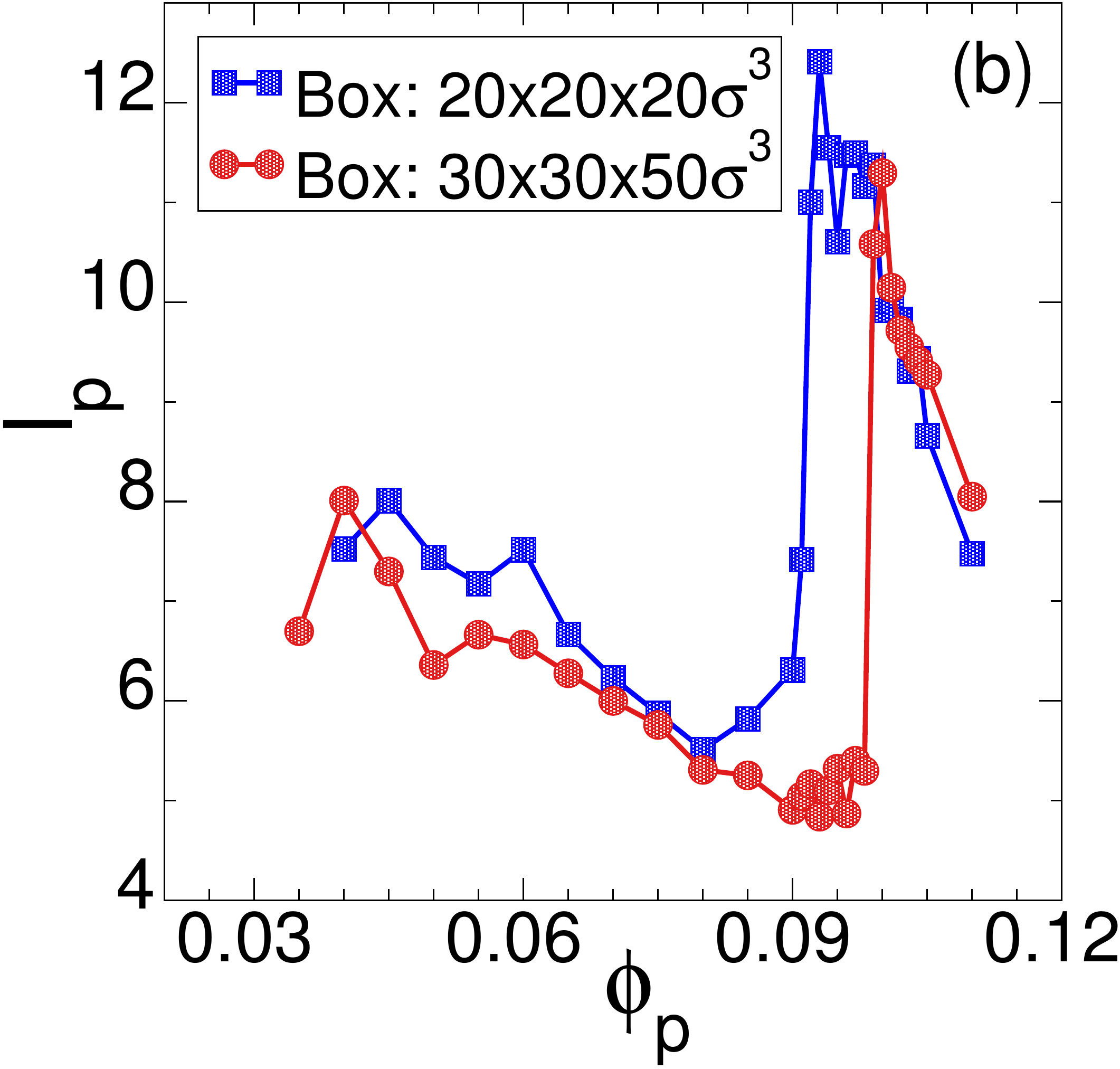}
\caption{\label{fig5} (a) The mean of the cosine of the angle $\theta_{ij}$  between two bond 
vectors $\vec{b_i}$ and $\vec{b_j}$ along the length of the chain as a function of the index $i-j$, 
where $i$ and $j$ are the bond vector indices along the contour length of the chain. 
The number of bonds in a chain with  $L$ monomers $L-1$. At low densities, since
the self assembled chains are very short the maximum value of $i-j$ is relatively low and 
the data is shown in the inset. The data from $30 \times 30 \times 50 \sigma^3$ box size
has been plotted in a semi-log plot to extract $l_p$ by using $\langle \cos (\theta_{ij}) \rangle \sim e^{- (i-j)\sigma/l_p}$.
(b) The persistence length $l_p$ versus the volume fraction $\phi_p$ of monomers have been shown for both box sizes. 
The standard deviation calculated for the data for smaller box is obtained from 10 
independent runs.   
}
\end{figure}

The tendency of the chains to line up parallel to each other at high densities ($\phi_p > 0.09$) also brings up the 
question of how semiflexible the chains are, and we quantify semiflexibily by
calculating the persistence length $l_p$ of the polymer chains as function of $\phi_p$.
To calculate $\ell_p$,  all chains with more than $3$ monomers are identified, and bond vectors $\vec{b_i}$ 
joining adjacent monomers along the chain contour are calculated. A chain with $L$ monomers will have $i$ 
running from $1$ to $L-1$. Then the quantity 
$\langle \cos (\theta_{ij}) \rangle = \langle \vec{b_j} \cdot \vec{b_i} \rangle$ is calculated 
for all values of $i-j$, where $i-j=1,2..,(L-2)$. The  plot of  $\langle \cos (\theta_{ij}) \rangle$ versus $i-j$ is shown
in Fig.\ref{fig5}(a) for different values of $\phi_p$ in a semilog plot. We do a exponential fit to calculate the values
of the peristence length $l_p(\phi_p)$ for each $\phi_p$, refer Fig.\ref{fig5}(b). The $l_p(\phi_p)$
(effective persistence length) shows a jump at $\phi_p=0.09$ as the chain get into the line-hexagonal 
ordered state, below that the $l_p$ is around 6 bond-lengths. Below $\phi_p <0.09$, the decrease
of $l_p$ with increasing $\phi_p$ could be due to the effect of self avoidance of many chains
trying to fit with each other in the box. Persistence length $l_p$ is a single chain property and should ideally 
be calculated at low densities of monomers. But at low densities of monomers, longer chains of self-assembling
monomers are relatively rare and values of $l_p$ around $\phi_p=0.05$ and $\phi_p=0.065$ should be considered reliable.

Such a simple model for self-assembled polymers can be used to study a range of properties of worm-like micellar systems
only if one can tune $l_p$ and branching of the chains. The persistence length  $l_p$ is a single chain property and should 
be calculate for dilute systems. To that end,  we changed parameters of the repulsive potential $V_{AA} (r)=V_{BB}(r)$.

\begin{figure}[H]
\includegraphics[width=0.49\columnwidth]{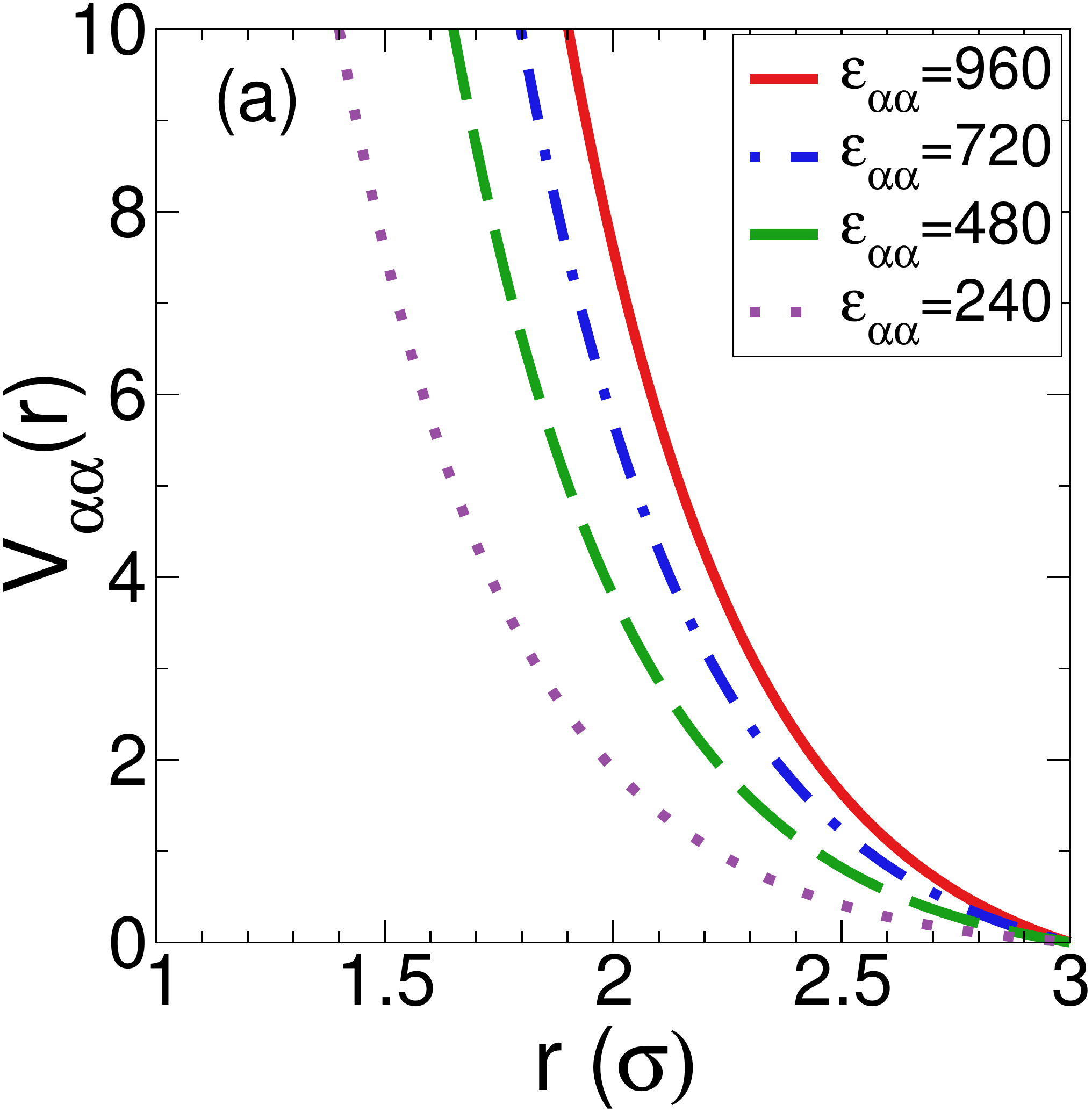}
\includegraphics[width=0.49\columnwidth]{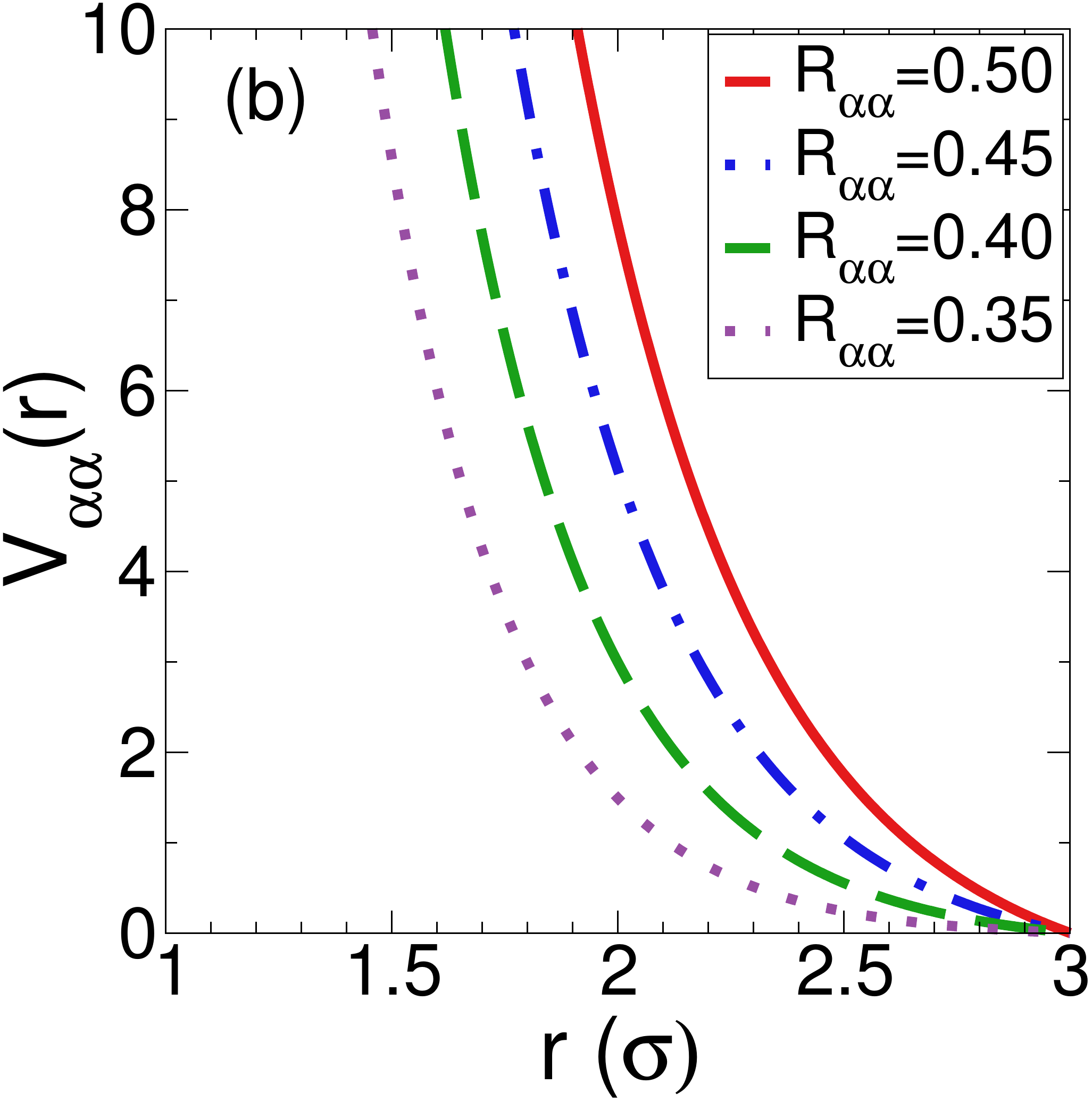}
\vskip0.3cm
\includegraphics[width=0.49\columnwidth]{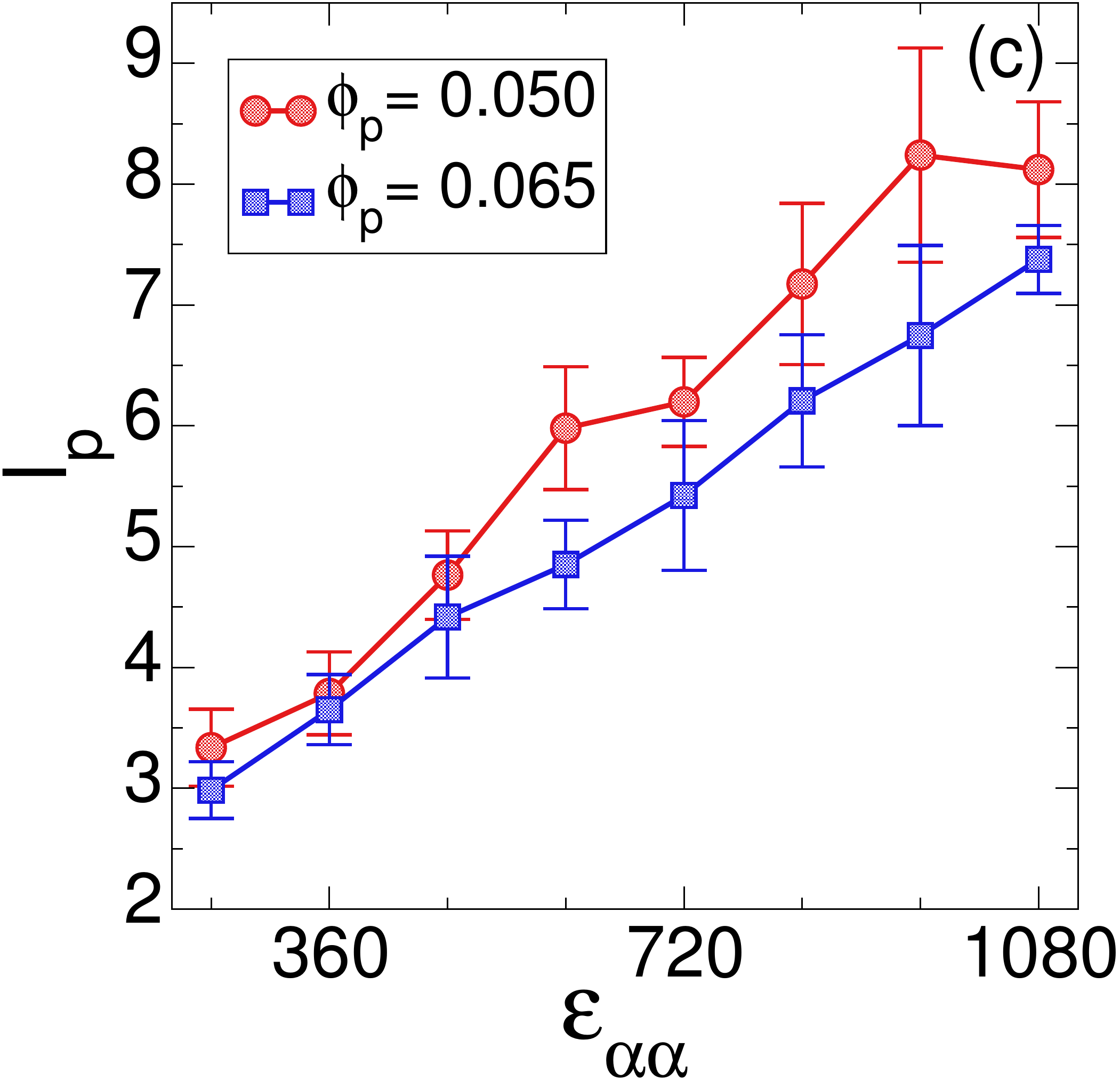}
\includegraphics[width=0.49\columnwidth]{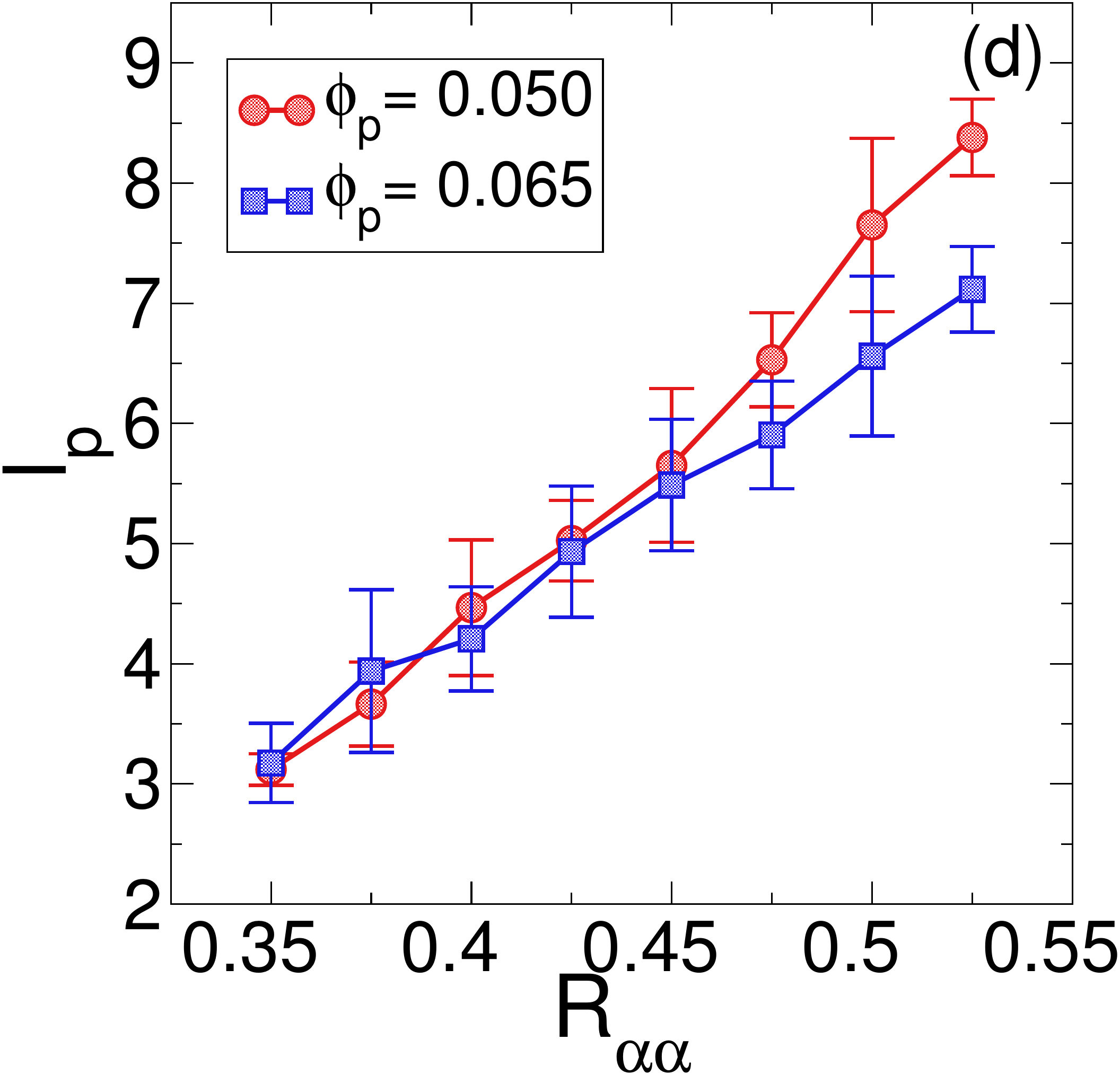}
\vskip0.3cm
\includegraphics[width=0.49\columnwidth]{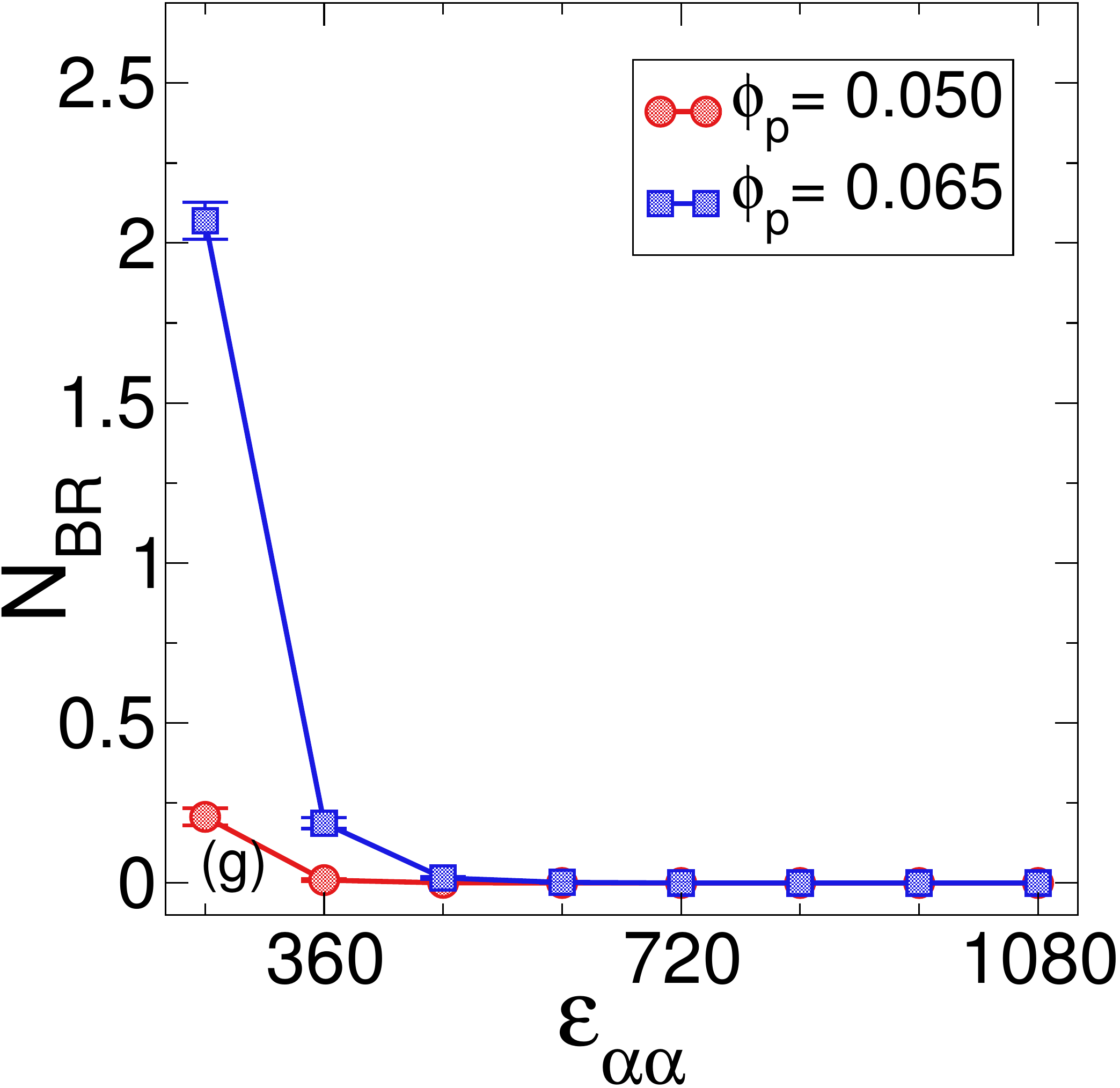}
\includegraphics[width=0.49\columnwidth]{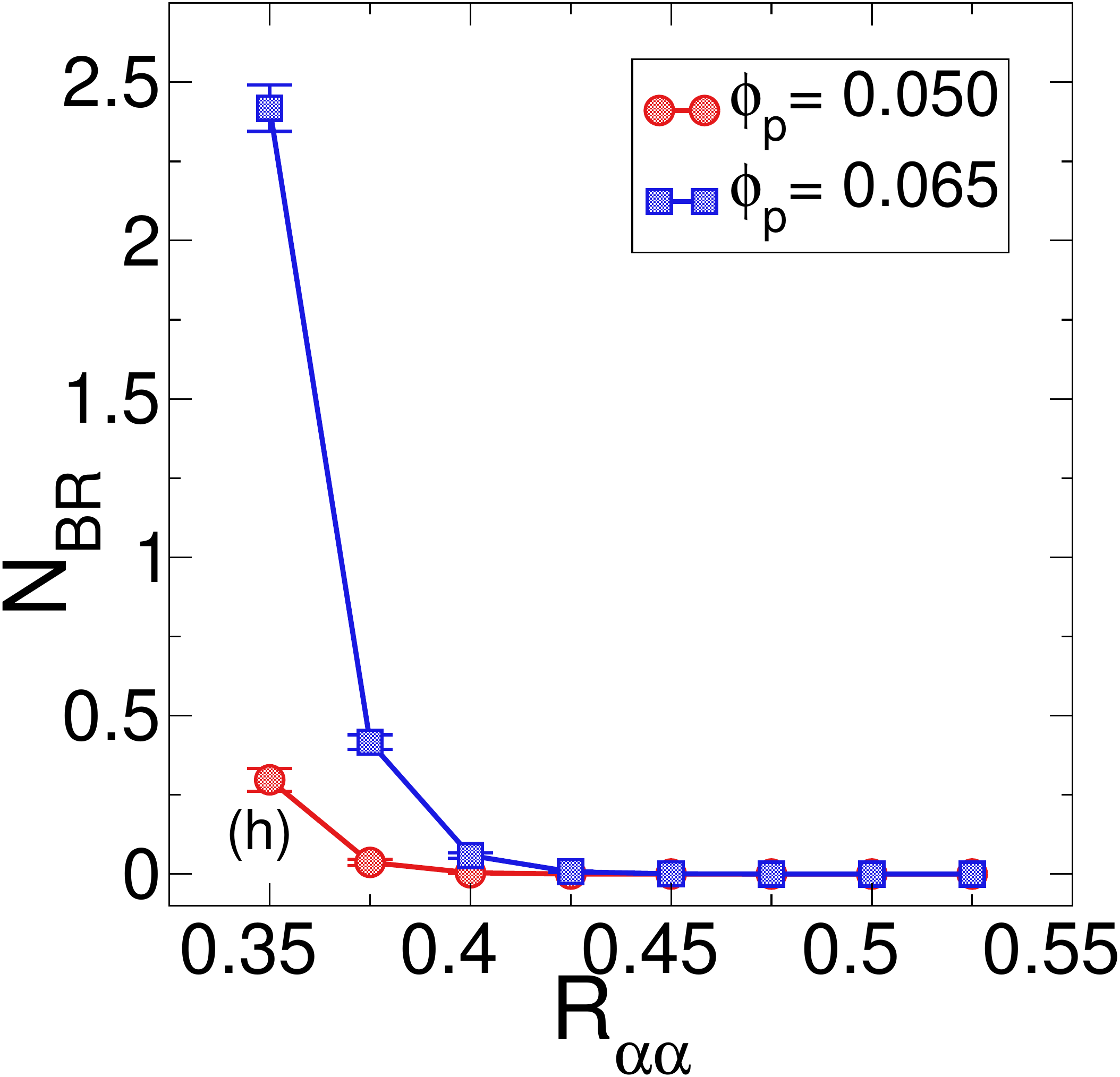}
\caption{\label{fig6}  Subplots (a) and (b) show the change in potential of interaction between
like particles $V_{\alpha \alpha} \,\, (V_{AA}, V_{BB})$ on changing parameters $\epsilon_{\alpha \alpha}$
(in units of $k_BT$) or $R_{\alpha \alpha}$ (in units of $\sigma$). 
Both $V_{AA}$ and $V_{BB}$ are changed simultaneously when we change the parameters.
Subplots (c) and (d) show corresponding changes in the persistence length $l_p$ (solid symbols) and
(e) and (f) show the number of branches (open symbols), normalized for a $(10 \sigma)^3$ box 
at two relatively low volume fractions $\phi_p$ of monomers.
}
\end{figure}

We change either the strength of the repulsive potential ($\epsilon_{\alpha \alpha}$) between like monomers
keeping $R_{AA}, R_{BB}$ constant. Alternatively, we change  the range of interactions ($R_{AA}$ and $R_{BB}$), 
keeping $\epsilon_{\alpha \alpha}$ fixed, refer Fig.\ref{fig6}(a,b).
We then investigate the corresponding change in the values of $l_p$ and $N_{br}$ on change of the above parameters. 
As one changes the parameters such that the range of the repulsive potential decreases,
the value of the persistence length decreases from the values of $ \approx 8.5 \sigma$ to lower 
values as seen in Figs.\ref{fig6}(c,d).  For these calculations we choose relatively lower values of 
$\phi_p=0.05$ or $\phi_p=0.06$ for reasons discussed previously.
The drop in values of $l_p$  occurs because decrease in $V_{AA}$ (and $V_{BB}$) potential allows two 
A particles in a A-B-A configuration to approach each other.  Consequently, increasing the repulsion 
between like-particles will keep $A-B-A$ configuration in straight line and increase $l_p$. 
Furthermore, there is a corresponding significant increase in branching with decrease in the repulsive
potential between like-particles, refer Figs.\ref{fig6}(e,f)
This is because lowering the repulsion between $A-A$ (or $B-B$) particles
allows a fourth particle (say) A to approach a (say) $-A-B-A-$ 
configuration by getting closer to (say) B in the triplet. 
Note that, the increase in $l_p$ is nearly the same on decrease of either $R_{\alpha \alpha}$ or
$\epsilon_{\alpha \alpha}$ as the relative decrease in the values of $V_{\alpha \alpha}$ is nearly the same
for the range of parameters considered. 

\section{Conclusions}

Spherically symmetric potentials with short range attraction and long range repulsion
have been suggested and used by previous studies \cite{mossa,zacarelli,shibananda,snigdha} but they obtained 
linear cluster of particles, or polymers with branches with additional three-body term to incorporate 
semi-flexibility.  There are two crucial ideas which led to our success in obtaining linear polymeric 
chains without branching using same class of potentials. 
Firstly, we used $ (1/r)^{24} - (1/r)^{12}$ instead of Lennard Jones term which led 
to a sharper minima of the attractive part of the potential. This enabled us to have a sharp  potential peak 
at rather short distances near $r=1.3 \sigma$. In the past we have tried to get a sharp peak at such distances 
with Lennard Jones potential but without success \cite{mubeena,alex_thesis}. Secondly, we used 2 kinds of 
particles A and B. The repulsion between like particles control  the persistence length 
and reduce the probability of branching. There have been previous studies which have obtained assemblies of particles 
with linear and other anisotropic morphologies \cite{geissler} starting out with isotropic potentials, 
but they have used 4 (or more) different kind of particles.
Our proposed model is much simpler. 
and could  be experimentally realizable with spherical charged colloidal particles with tunable attractive
potential \cite{virnau,virnau2,biben_frenkel,yodh,yethiraj} and screened Coulomb repulsions to form polymeric chains.  

We acknowledge funding and the use of a cluster bought using DST-SERB grant no.
EMR/2015/000018 to A. Chatterji. AC acknowledges funding support by DST Nanomission, India under
the Thematic Unit Program (grant no.:  SR/NM/TP-13/2016) and DBT project BT/PR16542/BID/7/654/2016.
AC acknowledges Rahul Pandit (IISc, Bangalore) who proposed this problem 20 years ago.

\bibliographystyle{apsrev4-1}
\bibliography{arxiv_alex}

\begin{thebibliography}{45}%
\makeatletter
\providecommand \@ifxundefined [1]{%
 \@ifx{#1\undefined}
}%
\providecommand \@ifnum [1]{%
 \ifnum #1\expandafter \@firstoftwo
 \else \expandafter \@secondoftwo
 \fi
}%
\providecommand \@ifx [1]{%
 \ifx #1\expandafter \@firstoftwo
 \else \expandafter \@secondoftwo
 \fi
}%
\providecommand \natexlab [1]{#1}%
\providecommand \enquote  [1]{``#1''}%
\providecommand \bibnamefont  [1]{#1}%
\providecommand \bibfnamefont [1]{#1}%
\providecommand \citenamefont [1]{#1}%
\providecommand \href@noop [0]{\@secondoftwo}%
\providecommand \href [0]{\begingroup \@sanitize@url \@href}%
\providecommand \@href[1]{\@@startlink{#1}\@@href}%
\providecommand \@@href[1]{\endgroup#1\@@endlink}%
\providecommand \@sanitize@url [0]{\catcode `\\12\catcode `\$12\catcode
  `\&12\catcode `\#12\catcode `\^12\catcode `\_12\catcode `\%12\relax}%
\providecommand \@@startlink[1]{}%
\providecommand \@@endlink[0]{}%
\providecommand \url  [0]{\begingroup\@sanitize@url \@url }%
\providecommand \@url [1]{\endgroup\@href {#1}{\urlprefix }}%
\providecommand \urlprefix  [0]{URL }%
\providecommand \Eprint [0]{\href }%
\providecommand \doibase [0]{http://dx.doi.org/}%
\providecommand \selectlanguage [0]{\@gobble}%
\providecommand \bibinfo  [0]{\@secondoftwo}%
\providecommand \bibfield  [0]{\@secondoftwo}%
\providecommand \translation [1]{[#1]}%
\providecommand \BibitemOpen [0]{}%
\providecommand \bibitemStop [0]{}%
\providecommand \bibitemNoStop [0]{.\EOS\space}%
\providecommand \EOS [0]{\spacefactor3000\relax}%
\providecommand \BibitemShut  [1]{\csname bibitem#1\endcsname}%
\let\auto@bib@innerbib\@empty
\bibitem [{\citenamefont {Gruenwald}\ and\ \citenamefont
  {Geissler}(2014)}]{geissler}%
  \BibitemOpen
  \bibfield  {author} {\bibinfo {author} {\bibfnamefont {M.}~\bibnamefont
  {Gruenwald}}\ and\ \bibinfo {author} {\bibfnamefont {P.~L.}\ \bibnamefont
  {Geissler}},\ }\href@noop {} {\bibfield  {journal} {\bibinfo  {journal} {{ACS
  Nano}}\ }\textbf {\bibinfo {volume} {8}},\ \bibinfo {pages} {5891} (\bibinfo
  {year} {2014})}\BibitemShut {NoStop}%
\bibitem [{\citenamefont {Jacobs}\ and\ \citenamefont
  {Frenkel}(2015)}]{frenkel}%
  \BibitemOpen
  \bibfield  {author} {\bibinfo {author} {\bibfnamefont {W.~M.}\ \bibnamefont
  {Jacobs}}\ and\ \bibinfo {author} {\bibfnamefont {D.}~\bibnamefont
  {Frenkel}},\ }\href@noop {} {\bibfield  {journal} {\bibinfo  {journal} {{Soft
  Matter}}\ }\textbf {\bibinfo {volume} {11}},\ \bibinfo {pages} {8930}
  (\bibinfo {year} {2015})}\BibitemShut {NoStop}%
\bibitem [{\citenamefont {Reinhardt}\ and\ \citenamefont
  {Frenkel}(2014)}]{frenkel1}%
  \BibitemOpen
  \bibfield  {author} {\bibinfo {author} {\bibfnamefont {A.}~\bibnamefont
  {Reinhardt}}\ and\ \bibinfo {author} {\bibfnamefont {D.}~\bibnamefont
  {Frenkel}},\ }\href@noop {} {\bibfield  {journal} {\bibinfo  {journal}
  {{Phys. Rev. Lett.}}\ }\textbf {\bibinfo {volume} {112}},\ \bibinfo {pages}
  {238103} (\bibinfo {year} {2014})}\BibitemShut {NoStop}%
\bibitem [{\citenamefont {Barros}\ and\ \citenamefont
  {Luijten}(2014)}]{luijten}%
  \BibitemOpen
  \bibfield  {author} {\bibinfo {author} {\bibfnamefont {K.}~\bibnamefont
  {Barros}}\ and\ \bibinfo {author} {\bibfnamefont {E.}~\bibnamefont
  {Luijten}},\ }\href@noop {} {\bibfield  {journal} {\bibinfo  {journal}
  {{Phys. Rev. Lett.}}\ }\textbf {\bibinfo {volume} {113}},\ \bibinfo {pages}
  {017801} (\bibinfo {year} {2014})}\BibitemShut {NoStop}%
\bibitem [{\citenamefont {Kegel}\ and\ \citenamefont {van~der
  Schoot}(2004)}]{schoot}%
  \BibitemOpen
  \bibfield  {author} {\bibinfo {author} {\bibfnamefont {W.~K.}\ \bibnamefont
  {Kegel}}\ and\ \bibinfo {author} {\bibfnamefont {P.}~\bibnamefont {van~der
  Schoot}},\ }\href@noop {} {\bibfield  {journal} {\bibinfo  {journal}
  {{Biophysical Journal}}\ }\textbf {\bibinfo {volume} {86}},\ \bibinfo {pages}
  {3905} (\bibinfo {year} {2004})}\BibitemShut {NoStop}%
\bibitem [{\citenamefont {Mubeena}\ and\ \citenamefont
  {Chatterji}(2015)}]{mubeena}%
  \BibitemOpen
  \bibfield  {author} {\bibinfo {author} {\bibfnamefont {S.}~\bibnamefont
  {Mubeena}}\ and\ \bibinfo {author} {\bibfnamefont {A.}~\bibnamefont
  {Chatterji}},\ }\href@noop {} {\bibfield  {journal} {\bibinfo  {journal}
  {Phys. Rev. E}\ }\textbf {\bibinfo {volume} {91}},\ \bibinfo {pages} {032602}
  (\bibinfo {year} {2015})}\BibitemShut {NoStop}%
\bibitem [{\citenamefont {Grason}(2016)}]{sa_grason}%
  \BibitemOpen
  \bibfield  {author} {\bibinfo {author} {\bibfnamefont {G.~M.}\ \bibnamefont
  {Grason}},\ }\href@noop {} {\bibfield  {journal} {\bibinfo  {journal} {The
  Journal of Chemical Physics}\ }\textbf {\bibinfo {volume} {145}},\ \bibinfo
  {pages} {110901} (\bibinfo {year} {2016})}\BibitemShut {NoStop}%
\bibitem [{\citenamefont {McManus}\ \emph {et~al.}(2016)\citenamefont
  {McManus}, \citenamefont {Charbonneau}, \citenamefont {Zaccarelli},\ and\
  \citenamefont {Asherie}}]{sa_mcmanus}%
  \BibitemOpen
  \bibfield  {author} {\bibinfo {author} {\bibfnamefont {J.~J.}\ \bibnamefont
  {McManus}}, \bibinfo {author} {\bibfnamefont {P.}~\bibnamefont
  {Charbonneau}}, \bibinfo {author} {\bibfnamefont {E.}~\bibnamefont
  {Zaccarelli}}, \ and\ \bibinfo {author} {\bibfnamefont {N.}~\bibnamefont
  {Asherie}},\ }\href@noop {} {\enquote {\bibinfo {title} {The physics of
  protein self-assembly},}\ } (\bibinfo {year} {2016}),\ \Eprint
  {http://arxiv.org/abs/arXiv:1602.00884v1 [cond-mat.soft]} {arXiv:1602.00884v1
  [cond-mat.soft]} \BibitemShut {NoStop}%
\bibitem [{\citenamefont {Angioletti-Uberti}\ \emph {et~al.}(2016)\citenamefont
  {Angioletti-Uberti}, \citenamefont {Mognetti},\ and\ \citenamefont
  {Frenkel}}]{sa_frenkel}%
  \BibitemOpen
  \bibfield  {author} {\bibinfo {author} {\bibfnamefont {S.}~\bibnamefont
  {Angioletti-Uberti}}, \bibinfo {author} {\bibfnamefont {B.~M.}\ \bibnamefont
  {Mognetti}}, \ and\ \bibinfo {author} {\bibfnamefont {D.}~\bibnamefont
  {Frenkel}},\ }\href@noop {} {\bibfield  {journal} {\bibinfo  {journal} {Phys.
  Chem. Chem. Phys.}\ }\textbf {\bibinfo {volume} {18}},\ \bibinfo {pages}
  {6373} (\bibinfo {year} {2016})}\BibitemShut {NoStop}%
\bibitem [{\citenamefont {Bruinsma}\ \emph {et~al.}(2016)\citenamefont
  {Bruinsma}, \citenamefont {Comas-Garcia}, \citenamefont {Garmann},\ and\
  \citenamefont {Grosberg}}]{sa_bruinsma}%
  \BibitemOpen
  \bibfield  {author} {\bibinfo {author} {\bibfnamefont {R.~F.}\ \bibnamefont
  {Bruinsma}}, \bibinfo {author} {\bibfnamefont {M.}~\bibnamefont
  {Comas-Garcia}}, \bibinfo {author} {\bibfnamefont {R.~F.}\ \bibnamefont
  {Garmann}}, \ and\ \bibinfo {author} {\bibfnamefont {A.~Y.}\ \bibnamefont
  {Grosberg}},\ }\href@noop {} {\bibfield  {journal} {\bibinfo  {journal}
  {Phys. Rev. E}\ }\textbf {\bibinfo {volume} {93}},\ \bibinfo {pages} {032405}
  (\bibinfo {year} {2016})}\BibitemShut {NoStop}%
\bibitem [{\citenamefont {Theodorakis}\ \emph {et~al.}(2015)\citenamefont
  {Theodorakis}, \citenamefont {Fytas}, \citenamefont {Kahl},\ and\
  \citenamefont {Dellago}}]{sa_dellago}%
  \BibitemOpen
  \bibfield  {author} {\bibinfo {author} {\bibfnamefont {P.~E.}\ \bibnamefont
  {Theodorakis}}, \bibinfo {author} {\bibfnamefont {N.~G.}\ \bibnamefont
  {Fytas}}, \bibinfo {author} {\bibfnamefont {G.}~\bibnamefont {Kahl}}, \ and\
  \bibinfo {author} {\bibfnamefont {C.}~\bibnamefont {Dellago}},\ }\href@noop
  {} {\enquote {\bibinfo {title} {Self-assembly of dna-functionalized
  colloids},}\ } (\bibinfo {year} {2015}),\ \Eprint
  {http://arxiv.org/abs/arXiv:1503.05384 [cond-mat.soft]} {arXiv:1503.05384
  [cond-mat.soft]} \BibitemShut {NoStop}%
\bibitem [{\citenamefont {Whitelam}\ and\ \citenamefont
  {Jack}(2015)}]{sa_whitelam}%
  \BibitemOpen
  \bibfield  {author} {\bibinfo {author} {\bibfnamefont {S.}~\bibnamefont
  {Whitelam}}\ and\ \bibinfo {author} {\bibfnamefont {R.~L.}\ \bibnamefont
  {Jack}},\ }\href@noop {} {\bibfield  {journal} {\bibinfo  {journal} {Annual
  Review of Physical Chemistry}\ }\textbf {\bibinfo {volume} {66}},\ \bibinfo
  {pages} {143} (\bibinfo {year} {2015})}\BibitemShut {NoStop}%
\bibitem [{\citenamefont {Glotzer}\ and\ \citenamefont
  {Solomon}(2007)}]{glotzer}%
  \BibitemOpen
  \bibfield  {author} {\bibinfo {author} {\bibfnamefont {S.~C.}\ \bibnamefont
  {Glotzer}}\ and\ \bibinfo {author} {\bibfnamefont {M.~J.}\ \bibnamefont
  {Solomon}},\ }\href@noop {} {\bibfield  {journal} {\bibinfo  {journal}
  {{Nature Materials}}\ }\textbf {\bibinfo {volume} {6}},\ \bibinfo {pages}
  {557} (\bibinfo {year} {2007})}\BibitemShut {NoStop}%
\bibitem [{\citenamefont {Akcora}\ \emph {et~al.}(2009)\citenamefont {Akcora},
  \citenamefont {Liu}, \citenamefont {Kumar}, \citenamefont {Moll},
  \citenamefont {Li}, \citenamefont {Benicewicz}, \citenamefont {Schadler},
  \citenamefont {Acehan}, \citenamefont {Panagiotopoulos}, \citenamefont
  {Pryamitsyn}, \citenamefont {Ganesan}, \citenamefont {Ilavsky}, \citenamefont
  {Thiyagarajan}, \citenamefont {Colby},\ and\ \citenamefont
  {Douglas}}]{sanat}%
  \BibitemOpen
  \bibfield  {author} {\bibinfo {author} {\bibfnamefont {P.}~\bibnamefont
  {Akcora}}, \bibinfo {author} {\bibfnamefont {H.}~\bibnamefont {Liu}},
  \bibinfo {author} {\bibfnamefont {S.~K.}\ \bibnamefont {Kumar}}, \bibinfo
  {author} {\bibfnamefont {J.}~\bibnamefont {Moll}}, \bibinfo {author}
  {\bibfnamefont {Y.}~\bibnamefont {Li}}, \bibinfo {author} {\bibfnamefont
  {B.~C.}\ \bibnamefont {Benicewicz}}, \bibinfo {author} {\bibfnamefont
  {L.~S.}\ \bibnamefont {Schadler}}, \bibinfo {author} {\bibfnamefont
  {D.}~\bibnamefont {Acehan}}, \bibinfo {author} {\bibfnamefont {A.~Z.}\
  \bibnamefont {Panagiotopoulos}}, \bibinfo {author} {\bibfnamefont
  {V.}~\bibnamefont {Pryamitsyn}}, \bibinfo {author} {\bibfnamefont
  {V.}~\bibnamefont {Ganesan}}, \bibinfo {author} {\bibfnamefont
  {J.}~\bibnamefont {Ilavsky}}, \bibinfo {author} {\bibfnamefont
  {P.}~\bibnamefont {Thiyagarajan}}, \bibinfo {author} {\bibfnamefont {R.~H.}\
  \bibnamefont {Colby}}, \ and\ \bibinfo {author} {\bibfnamefont {J.~F.}\
  \bibnamefont {Douglas}},\ }\href@noop {} {\bibfield  {journal} {\bibinfo
  {journal} {{Nature Materials}}\ }\textbf {\bibinfo {volume} {8}},\ \bibinfo
  {pages} {354} (\bibinfo {year} {2009})}\BibitemShut {NoStop}%
\bibitem [{\citenamefont {Mossa}\ \emph {et~al.}(2004)\citenamefont {Mossa},
  \citenamefont {Sciortino}, \citenamefont {Tartaglia},\ and\ \citenamefont
  {Zaccarelli}}]{mossa}%
  \BibitemOpen
  \bibfield  {author} {\bibinfo {author} {\bibfnamefont {S.}~\bibnamefont
  {Mossa}}, \bibinfo {author} {\bibfnamefont {F.}~\bibnamefont {Sciortino}},
  \bibinfo {author} {\bibfnamefont {P.}~\bibnamefont {Tartaglia}}, \ and\
  \bibinfo {author} {\bibfnamefont {E.}~\bibnamefont {Zaccarelli}},\
  }\href@noop {} {\bibfield  {journal} {\bibinfo  {journal} {{Langmuir}}\
  }\textbf {\bibinfo {volume} {20}},\ \bibinfo {pages} {10756} (\bibinfo {year}
  {2004})}\BibitemShut {NoStop}%
\bibitem [{\citenamefont {Sciortino}\ \emph {et~al.}(2004)\citenamefont
  {Sciortino}, \citenamefont {Mossa}, \citenamefont {Zaccarelli},\ and\
  \citenamefont {Tartaglia}}]{zacarelli}%
  \BibitemOpen
  \bibfield  {author} {\bibinfo {author} {\bibfnamefont {F.}~\bibnamefont
  {Sciortino}}, \bibinfo {author} {\bibfnamefont {S.}~\bibnamefont {Mossa}},
  \bibinfo {author} {\bibfnamefont {E.}~\bibnamefont {Zaccarelli}}, \ and\
  \bibinfo {author} {\bibfnamefont {P.}~\bibnamefont {Tartaglia}},\ }\href@noop
  {} {\bibfield  {journal} {\bibinfo  {journal} {{Phys. Rev. Lett.}}\ }\textbf
  {\bibinfo {volume} {93}},\ \bibinfo {pages} {055701} (\bibinfo {year}
  {2004})}\BibitemShut {NoStop}%
\bibitem [{\citenamefont {Das}\ \emph {et~al.}(2017)\citenamefont {Das},
  \citenamefont {Riest}, \citenamefont {Winkler}, \citenamefont {Gompper},
  \citenamefont {Dhont},\ and\ \citenamefont {Naegele}}]{shibananda}%
  \BibitemOpen
  \bibfield  {author} {\bibinfo {author} {\bibfnamefont {S.}~\bibnamefont
  {Das}}, \bibinfo {author} {\bibfnamefont {J.}~\bibnamefont {Riest}}, \bibinfo
  {author} {\bibfnamefont {R.~G.}\ \bibnamefont {Winkler}}, \bibinfo {author}
  {\bibfnamefont {G.}~\bibnamefont {Gompper}}, \bibinfo {author} {\bibfnamefont
  {J.~K.~G.}\ \bibnamefont {Dhont}}, \ and\ \bibinfo {author} {\bibfnamefont
  {G.}~\bibnamefont {Naegele}},\ }\href@noop {} {\bibfield  {journal} {\bibinfo
   {journal} {Soft Matter}\ ,\ } (\bibinfo {year} {2017})}\BibitemShut
  {NoStop}%
\bibitem [{\citenamefont {Chen}\ \emph {et~al.}(2011)\citenamefont {Chen},
  \citenamefont {Mao}, \citenamefont {Thakur}, \citenamefont {Xu},\ and\
  \citenamefont {Liu}}]{snigdha}%
  \BibitemOpen
  \bibfield  {author} {\bibinfo {author} {\bibfnamefont {J.-X.}\ \bibnamefont
  {Chen}}, \bibinfo {author} {\bibfnamefont {J.-W.}\ \bibnamefont {Mao}},
  \bibinfo {author} {\bibfnamefont {S.}~\bibnamefont {Thakur}}, \bibinfo
  {author} {\bibfnamefont {J.-R.}\ \bibnamefont {Xu}}, \ and\ \bibinfo {author}
  {\bibfnamefont {F.-y.}\ \bibnamefont {Liu}},\ }\href@noop {} {\bibfield
  {journal} {\bibinfo  {journal} {The Journal of Chemical Physics}\ }\textbf
  {\bibinfo {volume} {135}},\ \bibinfo {pages} {094504} (\bibinfo {year}
  {2011})}\BibitemShut {NoStop}%
\bibitem [{\citenamefont {Vliegenthart}\ \emph {et~al.}(1999)\citenamefont
  {Vliegenthart}, \citenamefont {Lodge},\ and\ \citenamefont
  {Lekkerkerker}}]{vliegenthart}%
  \BibitemOpen
  \bibfield  {author} {\bibinfo {author} {\bibfnamefont {G.}~\bibnamefont
  {Vliegenthart}}, \bibinfo {author} {\bibfnamefont {J.}~\bibnamefont {Lodge}},
  \ and\ \bibinfo {author} {\bibfnamefont {H.}~\bibnamefont {Lekkerkerker}},\
  }\href@noop {} {\bibfield  {journal} {\bibinfo  {journal} {{Physica A:
  Statistical Mechanics and its Applications}}\ }\textbf {\bibinfo {volume}
  {263}},\ \bibinfo {pages} {378 } (\bibinfo {year} {1999})},\ \bibinfo {note}
  {{Proceedings of the 20th IUPAP International Conference on Statistical
  Physics}}\BibitemShut {NoStop}%
\bibitem [{\citenamefont {{Berret}}(2004)}]{berret}%
  \BibitemOpen
  \bibfield  {author} {\bibinfo {author} {\bibfnamefont {J.-F.}\ \bibnamefont
  {{Berret}}},\ }\href@noop {} {\enquote {\bibinfo {title} {Rheology of
  wormlike micelles : Equilibrium properties and shear banding transition},}\ }
  (\bibinfo {year} {2004}),\ \Eprint {http://arxiv.org/abs/eprint
  arXiv:cond-mat/0406681} {eprint arXiv:cond-mat/0406681} \BibitemShut
  {NoStop}%
\bibitem [{\citenamefont {Dreiss}(2007)}]{dreiss}%
  \BibitemOpen
  \bibfield  {author} {\bibinfo {author} {\bibfnamefont {C.~A.}\ \bibnamefont
  {Dreiss}},\ }\href@noop {} {\bibfield  {journal} {\bibinfo  {journal} {Soft
  Matter}\ }\textbf {\bibinfo {volume} {3}},\ \bibinfo {pages} {956} (\bibinfo
  {year} {2007})}\BibitemShut {NoStop}%
\bibitem [{\citenamefont {Dreiss}(2017)}]{dreiss1}%
  \BibitemOpen
  \bibfield  {author} {\bibinfo {author} {\bibfnamefont {C.~A.}\ \bibnamefont
  {Dreiss}},\ }in\ \href@noop {} {\emph {\bibinfo {booktitle} {Wormlike
  Micelles: Advances in Systems{,} Characterisation and Applications}}}\
  (\bibinfo  {publisher} {The Royal Society of Chemistry},\ \bibinfo {year}
  {2017})\ pp.\ \bibinfo {pages} {1--8}\BibitemShut {NoStop}%
\bibitem [{\citenamefont {Cates}\ and\ \citenamefont
  {Candau}(1990)}]{cates_candau}%
  \BibitemOpen
  \bibfield  {author} {\bibinfo {author} {\bibfnamefont {M.~E.}\ \bibnamefont
  {Cates}}\ and\ \bibinfo {author} {\bibfnamefont {S.~J.}\ \bibnamefont
  {Candau}},\ }\href@noop {} {\bibfield  {journal} {\bibinfo  {journal}
  {Journal of Physics: Condensed Matter}\ }\textbf {\bibinfo {volume} {2}},\
  \bibinfo {pages} {6869} (\bibinfo {year} {1990})}\BibitemShut {NoStop}%
\bibitem [{\citenamefont {Milchev}\ and\ \citenamefont
  {Landau}(1995)}]{milchev}%
  \BibitemOpen
  \bibfield  {author} {\bibinfo {author} {\bibfnamefont {A.}~\bibnamefont
  {Milchev}}\ and\ \bibinfo {author} {\bibfnamefont {D.~P.}\ \bibnamefont
  {Landau}},\ }\href@noop {} {\bibfield  {journal} {\bibinfo  {journal} {Phys.
  Rev. E}\ }\textbf {\bibinfo {volume} {52}},\ \bibinfo {pages} {6431}
  (\bibinfo {year} {1995})}\BibitemShut {NoStop}%
\bibitem [{\citenamefont {Huang}\ \emph {et~al.}(2006)\citenamefont {Huang},
  \citenamefont {Xu},\ and\ \citenamefont {Ryckaert}}]{ryck1}%
  \BibitemOpen
  \bibfield  {author} {\bibinfo {author} {\bibfnamefont {C.}~\bibnamefont
  {Huang}}, \bibinfo {author} {\bibfnamefont {H.}~\bibnamefont {Xu}}, \ and\
  \bibinfo {author} {\bibfnamefont {J.-P.}\ \bibnamefont {Ryckaert}},\
  }\href@noop {} {\bibfield  {journal} {\bibinfo  {journal} {The Journal of
  Chemical Physics}\ }\textbf {\bibinfo {volume} {125}},\ \bibinfo {pages}
  {094901} (\bibinfo {year} {2006})}\BibitemShut {NoStop}%
\bibitem [{\citenamefont {Huang}\ \emph {et~al.}(2008)\citenamefont {Huang},
  \citenamefont {Xu},\ and\ \citenamefont {Ryckaert}}]{ryck2}%
  \BibitemOpen
  \bibfield  {author} {\bibinfo {author} {\bibfnamefont {C.-C.}\ \bibnamefont
  {Huang}}, \bibinfo {author} {\bibfnamefont {H.}~\bibnamefont {Xu}}, \ and\
  \bibinfo {author} {\bibfnamefont {J.~P.}\ \bibnamefont {Ryckaert}},\
  }\href@noop {} {\bibfield  {journal} {\bibinfo  {journal} {EPL (Europhysics
  Letters)}\ }\textbf {\bibinfo {volume} {81}},\ \bibinfo {pages} {58002}
  (\bibinfo {year} {2008})}\BibitemShut {NoStop}%
\bibitem [{\citenamefont {Huang}\ \emph {et~al.}(2009)\citenamefont {Huang},
  \citenamefont {Ryckaert},\ and\ \citenamefont {Xu}}]{ryck3}%
  \BibitemOpen
  \bibfield  {author} {\bibinfo {author} {\bibfnamefont {C.-C.}\ \bibnamefont
  {Huang}}, \bibinfo {author} {\bibfnamefont {J.-P.}\ \bibnamefont {Ryckaert}},
  \ and\ \bibinfo {author} {\bibfnamefont {H.}~\bibnamefont {Xu}},\ }\href@noop
  {} {\bibfield  {journal} {\bibinfo  {journal} {Phys. Rev. E}\ }\textbf
  {\bibinfo {volume} {79}},\ \bibinfo {pages} {041501} (\bibinfo {year}
  {2009})}\BibitemShut {NoStop}%
\bibitem [{\citenamefont {Padding}\ \emph {et~al.}(2009)\citenamefont
  {Padding}, \citenamefont {Briels}, \citenamefont {Stukan},\ and\
  \citenamefont {Boek}}]{padding}%
  \BibitemOpen
  \bibfield  {author} {\bibinfo {author} {\bibfnamefont {J.~T.}\ \bibnamefont
  {Padding}}, \bibinfo {author} {\bibfnamefont {W.~J.}\ \bibnamefont {Briels}},
  \bibinfo {author} {\bibfnamefont {M.~R.}\ \bibnamefont {Stukan}}, \ and\
  \bibinfo {author} {\bibfnamefont {E.~S.}\ \bibnamefont {Boek}},\ }\href@noop
  {} {\bibfield  {journal} {\bibinfo  {journal} {Soft Matter}\ }\textbf
  {\bibinfo {volume} {5}},\ \bibinfo {pages} {4367} (\bibinfo {year}
  {2009})}\BibitemShut {NoStop}%
\bibitem [{\citenamefont {Padding}\ and\ \citenamefont
  {Boek}(2004{\natexlab{a}})}]{pa1}%
  \BibitemOpen
  \bibfield  {author} {\bibinfo {author} {\bibfnamefont {J.~T.}\ \bibnamefont
  {Padding}}\ and\ \bibinfo {author} {\bibfnamefont {E.~S.}\ \bibnamefont
  {Boek}},\ }\href@noop {} {\bibfield  {journal} {\bibinfo  {journal} {EPL
  (Europhysics Letters)}\ }\textbf {\bibinfo {volume} {66}},\ \bibinfo {pages}
  {756} (\bibinfo {year} {2004}{\natexlab{a}})}\BibitemShut {NoStop}%
\bibitem [{\citenamefont {Padding}\ and\ \citenamefont
  {Boek}(2004{\natexlab{b}})}]{pa2}%
  \BibitemOpen
  \bibfield  {author} {\bibinfo {author} {\bibfnamefont {J.~T.}\ \bibnamefont
  {Padding}}\ and\ \bibinfo {author} {\bibfnamefont {E.~S.}\ \bibnamefont
  {Boek}},\ }\href@noop {} {\bibfield  {journal} {\bibinfo  {journal} {Phys.
  Rev. E}\ }\textbf {\bibinfo {volume} {70}},\ \bibinfo {pages} {031502}
  (\bibinfo {year} {2004}{\natexlab{b}})}\BibitemShut {NoStop}%
\bibitem [{\citenamefont {Padding}\ \emph {et~al.}(2005)\citenamefont
  {Padding}, \citenamefont {Boek},\ and\ \citenamefont {Briels}}]{pa3}%
  \BibitemOpen
  \bibfield  {author} {\bibinfo {author} {\bibfnamefont {J.~T.}\ \bibnamefont
  {Padding}}, \bibinfo {author} {\bibfnamefont {E.~S.}\ \bibnamefont {Boek}}, \
  and\ \bibinfo {author} {\bibfnamefont {W.~J.}\ \bibnamefont {Briels}},\
  }\href@noop {} {\bibfield  {journal} {\bibinfo  {journal} {Journal of
  Physics: Condensed Matter}\ }\textbf {\bibinfo {volume} {17}},\ \bibinfo
  {pages} {S3347} (\bibinfo {year} {2005})}\BibitemShut {NoStop}%
\bibitem [{\citenamefont {Chatterji}\ and\ \citenamefont
  {Pandit}(2001)}]{pandit}%
  \BibitemOpen
  \bibfield  {author} {\bibinfo {author} {\bibfnamefont {A.}~\bibnamefont
  {Chatterji}}\ and\ \bibinfo {author} {\bibfnamefont {R.}~\bibnamefont
  {Pandit}},\ }\href@noop {} {\bibfield  {journal} {\bibinfo  {journal} {EPL
  (Europhysics Letters)}\ }\textbf {\bibinfo {volume} {54}},\ \bibinfo {pages}
  {213} (\bibinfo {year} {2001})}\BibitemShut {NoStop}%
\bibitem [{\citenamefont {Thakur}\ \emph {et~al.}(2010)\citenamefont {Thakur},
  \citenamefont {Prathyusha}, \citenamefont {Deshpande}, \citenamefont
  {Laradji},\ and\ \citenamefont {Kumar}}]{sunilkumar}%
  \BibitemOpen
  \bibfield  {author} {\bibinfo {author} {\bibfnamefont {S.}~\bibnamefont
  {Thakur}}, \bibinfo {author} {\bibfnamefont {K.~R.}\ \bibnamefont
  {Prathyusha}}, \bibinfo {author} {\bibfnamefont {A.~P.}\ \bibnamefont
  {Deshpande}}, \bibinfo {author} {\bibfnamefont {M.}~\bibnamefont {Laradji}},
  \ and\ \bibinfo {author} {\bibfnamefont {P.~B.~S.}\ \bibnamefont {Kumar}},\
  }\href@noop {} {\bibfield  {journal} {\bibinfo  {journal} {Soft Matter}\
  }\textbf {\bibinfo {volume} {6}},\ \bibinfo {pages} {489} (\bibinfo {year}
  {2010})}\BibitemShut {NoStop}%
\bibitem [{\citenamefont {Prathyusha}\ \emph {et~al.}(2013)\citenamefont
  {Prathyusha}, \citenamefont {Deshpande}, \citenamefont {Laradji},\ and\
  \citenamefont {Sunil~Kumar}}]{sunilkumar1}%
  \BibitemOpen
  \bibfield  {author} {\bibinfo {author} {\bibfnamefont {K.~R.}\ \bibnamefont
  {Prathyusha}}, \bibinfo {author} {\bibfnamefont {A.~P.}\ \bibnamefont
  {Deshpande}}, \bibinfo {author} {\bibfnamefont {M.}~\bibnamefont {Laradji}},
  \ and\ \bibinfo {author} {\bibfnamefont {P.~B.}\ \bibnamefont
  {Sunil~Kumar}},\ }\href@noop {} {\bibfield  {journal} {\bibinfo  {journal}
  {Soft Matter}\ }\textbf {\bibinfo {volume} {9}},\ \bibinfo {pages} {9983}
  (\bibinfo {year} {2013})}\BibitemShut {NoStop}%
\bibitem [{\citenamefont {Bandyopadhyay}\ \emph {et~al.}(2000)\citenamefont
  {Bandyopadhyay}, \citenamefont {Basappa},\ and\ \citenamefont {Sood}}]{sood}%
  \BibitemOpen
  \bibfield  {author} {\bibinfo {author} {\bibfnamefont {R.}~\bibnamefont
  {Bandyopadhyay}}, \bibinfo {author} {\bibfnamefont {G.}~\bibnamefont
  {Basappa}}, \ and\ \bibinfo {author} {\bibfnamefont {A.~K.}\ \bibnamefont
  {Sood}},\ }\href@noop {} {\bibfield  {journal} {\bibinfo  {journal} {Phys.
  Rev. Lett.}\ }\textbf {\bibinfo {volume} {84}},\ \bibinfo {pages} {2022}
  (\bibinfo {year} {2000})}\BibitemShut {NoStop}%
\bibitem [{\citenamefont {Bandyopadhyay}\ and\ \citenamefont
  {Sood}(2001)}]{sood1}%
  \BibitemOpen
  \bibfield  {author} {\bibinfo {author} {\bibfnamefont {R.}~\bibnamefont
  {Bandyopadhyay}}\ and\ \bibinfo {author} {\bibfnamefont {A.~K.}\ \bibnamefont
  {Sood}},\ }\href@noop {} {\bibfield  {journal} {\bibinfo  {journal} {EPL
  (Europhysics Letters)}\ }\textbf {\bibinfo {volume} {56}},\ \bibinfo {pages}
  {447} (\bibinfo {year} {2001})}\BibitemShut {NoStop}%
\bibitem [{\citenamefont {Ganapathy}\ \emph {et~al.}(2008)\citenamefont
  {Ganapathy}, \citenamefont {Majumdar},\ and\ \citenamefont
  {Sood}}]{ganapathy}%
  \BibitemOpen
  \bibfield  {author} {\bibinfo {author} {\bibfnamefont {R.}~\bibnamefont
  {Ganapathy}}, \bibinfo {author} {\bibfnamefont {S.}~\bibnamefont {Majumdar}},
  \ and\ \bibinfo {author} {\bibfnamefont {A.~K.}\ \bibnamefont {Sood}},\
  }\href@noop {} {\bibfield  {journal} {\bibinfo  {journal} {Phys. Rev. E}\
  }\textbf {\bibinfo {volume} {78}},\ \bibinfo {pages} {021504} (\bibinfo
  {year} {2008})}\BibitemShut {NoStop}%
\bibitem [{\citenamefont {Das}\ \emph {et~al.}(2005)\citenamefont {Das},
  \citenamefont {Chakrabarti}, \citenamefont {Dasgupta}, \citenamefont
  {Ramaswamy},\ and\ \citenamefont {Sood}}]{chakrabarti}%
  \BibitemOpen
  \bibfield  {author} {\bibinfo {author} {\bibfnamefont {M.}~\bibnamefont
  {Das}}, \bibinfo {author} {\bibfnamefont {B.}~\bibnamefont {Chakrabarti}},
  \bibinfo {author} {\bibfnamefont {C.}~\bibnamefont {Dasgupta}}, \bibinfo
  {author} {\bibfnamefont {S.}~\bibnamefont {Ramaswamy}}, \ and\ \bibinfo
  {author} {\bibfnamefont {A.~K.}\ \bibnamefont {Sood}},\ }\href@noop {}
  {\bibfield  {journal} {\bibinfo  {journal} {Phys. Rev. E}\ }\textbf {\bibinfo
  {volume} {71}},\ \bibinfo {pages} {021707} (\bibinfo {year}
  {2005})}\BibitemShut {NoStop}%
\bibitem [{\citenamefont {Angioletti-Uberti}\ \emph {et~al.}(2014)\citenamefont
  {Angioletti-Uberti}, \citenamefont {Varilly}, \citenamefont {Mognetti},\ and\
  \citenamefont {Frenkel}}]{sa_uberti_frenkel}%
  \BibitemOpen
  \bibfield  {author} {\bibinfo {author} {\bibfnamefont {S.}~\bibnamefont
  {Angioletti-Uberti}}, \bibinfo {author} {\bibfnamefont {P.}~\bibnamefont
  {Varilly}}, \bibinfo {author} {\bibfnamefont {B.~M.}\ \bibnamefont
  {Mognetti}}, \ and\ \bibinfo {author} {\bibfnamefont {D.}~\bibnamefont
  {Frenkel}},\ }\href@noop {} {\bibfield  {journal} {\bibinfo  {journal} {Phys.
  Rev. Lett.}\ }\textbf {\bibinfo {volume} {113}},\ \bibinfo {pages} {128303}
  (\bibinfo {year} {2014})}\BibitemShut {NoStop}%
\bibitem [{\citenamefont {Abraham}(2016)}]{alex_thesis}%
  \BibitemOpen
  \bibfield  {author} {\bibinfo {author} {\bibfnamefont {A.}~\bibnamefont
  {Abraham}},\ }\href@noop {} {\enquote {\bibinfo {title} {Self-assembly of
  polymeric chains under a new radially symmetric potential},}\ }\bibinfo
  {howpublished} {MS Thesis} (\bibinfo {year} {2016})\BibitemShut {NoStop}%
\bibitem [{\citenamefont {Binder}\ \emph {et~al.}(2014)\citenamefont {Binder},
  \citenamefont {Virnau},\ and\ \citenamefont {Statt}}]{virnau}%
  \BibitemOpen
  \bibfield  {author} {\bibinfo {author} {\bibfnamefont {K.}~\bibnamefont
  {Binder}}, \bibinfo {author} {\bibfnamefont {P.}~\bibnamefont {Virnau}}, \
  and\ \bibinfo {author} {\bibfnamefont {A.}~\bibnamefont {Statt}},\
  }\href@noop {} {\bibfield  {journal} {\bibinfo  {journal} {The Journal of
  Chemical Physics}\ }\textbf {\bibinfo {volume} {141}},\ \bibinfo {pages}
  {140901} (\bibinfo {year} {2014})}\BibitemShut {NoStop}%
\bibitem [{\citenamefont {Zausch}\ \emph {et~al.}(2009)\citenamefont {Zausch},
  \citenamefont {Virnau}, \citenamefont {Binder}, \citenamefont {Horbach},\
  and\ \citenamefont {Vink}}]{virnau2}%
  \BibitemOpen
  \bibfield  {author} {\bibinfo {author} {\bibfnamefont {J.}~\bibnamefont
  {Zausch}}, \bibinfo {author} {\bibfnamefont {P.}~\bibnamefont {Virnau}},
  \bibinfo {author} {\bibfnamefont {K.}~\bibnamefont {Binder}}, \bibinfo
  {author} {\bibfnamefont {J.}~\bibnamefont {Horbach}}, \ and\ \bibinfo
  {author} {\bibfnamefont {R.~L.}\ \bibnamefont {Vink}},\ }\href {\doibase
  10.1063} {\bibfield  {journal} {\bibinfo  {journal} {The Journal of Chemical
  Physics}\ }\textbf {\bibinfo {volume} {130}},\ \bibinfo {pages} {064906}
  (\bibinfo {year} {2009})}\BibitemShut {NoStop}%
\bibitem [{\citenamefont {Biben}\ \emph {et~al.}(1996)\citenamefont {Biben},
  \citenamefont {Bladon},\ and\ \citenamefont {Frenkel}}]{biben_frenkel}%
  \BibitemOpen
  \bibfield  {author} {\bibinfo {author} {\bibfnamefont {T.}~\bibnamefont
  {Biben}}, \bibinfo {author} {\bibfnamefont {P.}~\bibnamefont {Bladon}}, \
  and\ \bibinfo {author} {\bibfnamefont {D.}~\bibnamefont {Frenkel}},\
  }\href@noop {} {\bibfield  {journal} {\bibinfo  {journal} {Journal of
  Physics: Condensed Matter}\ }\textbf {\bibinfo {volume} {8}},\ \bibinfo
  {pages} {10799} (\bibinfo {year} {1996})}\BibitemShut {NoStop}%
\bibitem [{\citenamefont {Crocker}\ \emph {et~al.}(1999)\citenamefont
  {Crocker}, \citenamefont {Matteo}, \citenamefont {Dinsmore},\ and\
  \citenamefont {Yodh}}]{yodh}%
  \BibitemOpen
  \bibfield  {author} {\bibinfo {author} {\bibfnamefont {J.~C.}\ \bibnamefont
  {Crocker}}, \bibinfo {author} {\bibfnamefont {J.~A.}\ \bibnamefont {Matteo}},
  \bibinfo {author} {\bibfnamefont {A.~D.}\ \bibnamefont {Dinsmore}}, \ and\
  \bibinfo {author} {\bibfnamefont {A.~G.}\ \bibnamefont {Yodh}},\ }\href@noop
  {} {\bibfield  {journal} {\bibinfo  {journal} {Phys. Rev. Lett.}\ }\textbf
  {\bibinfo {volume} {82}},\ \bibinfo {pages} {4352} (\bibinfo {year}
  {1999})}\BibitemShut {NoStop}%
\bibitem [{\citenamefont {Yethiraj}(2007)}]{yethiraj}%
  \BibitemOpen
  \bibfield  {author} {\bibinfo {author} {\bibfnamefont {A.}~\bibnamefont
  {Yethiraj}},\ }\href@noop {} {\bibfield  {journal} {\bibinfo  {journal} {Soft
  Matter}\ }\textbf {\bibinfo {volume} {3}},\ \bibinfo {pages} {1099} (\bibinfo
  {year} {2007})}\BibitemShut {NoStop}%
\end{thebibliography}%

\clearpage
\begin{widetext}

\chapter{\large\bf Supplementary Materials}
\label{supplementary}


\renewcommand\thefigure{S\arabic{figure}}
\setcounter{figure}{0}

\vskip1.0cm

\begin{figure*}[hbt]
\includegraphics[width=0.8\textwidth]{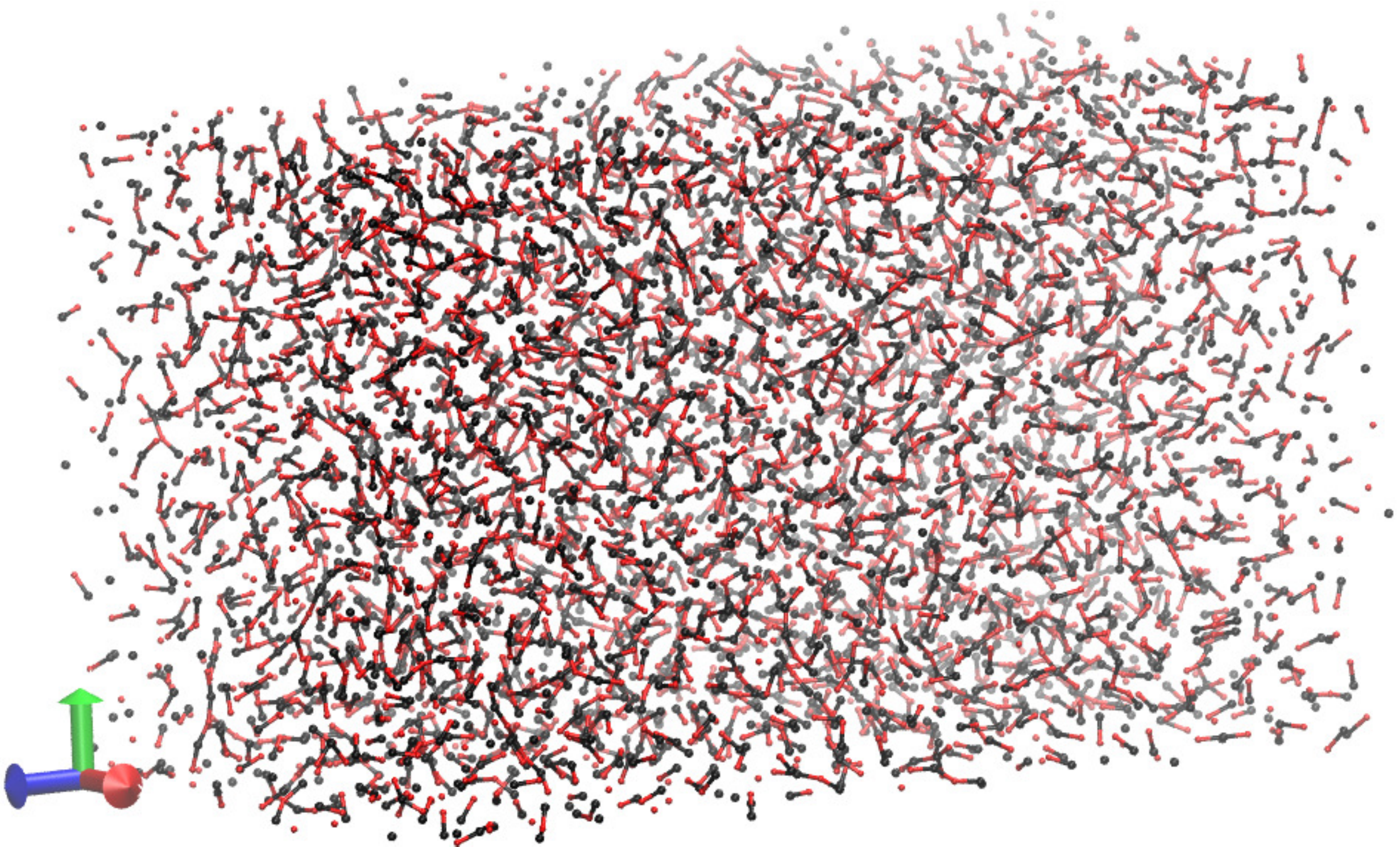}
\vskip0.3cm
\includegraphics[width=0.8\textwidth]{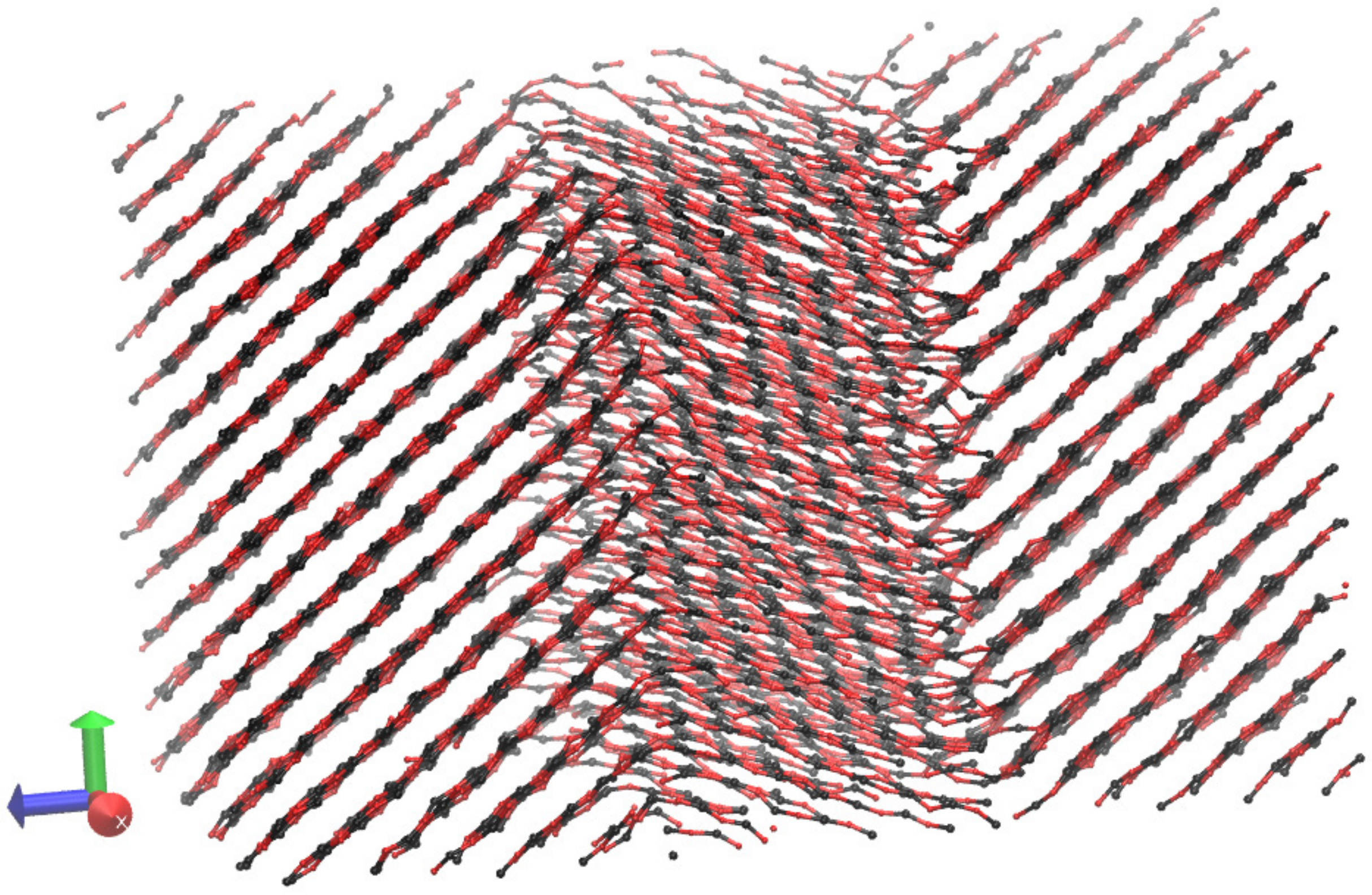}
\caption{\label{sup1} 
Respresentative snapshots of the self assembled polymers
at volume fraction $\phi_p=0.08$ (top) and $\phi_p=0.102$ (bottom) in box size $30 \times 30 \times 50 \sigma^3$. The figure at the bottom shows ordered polymers
but there are 2 different domains where the direction or orietnation is different. 
}
\end{figure*}

\begin{figure*}[hbt]
\includegraphics[width=0.8\textwidth]{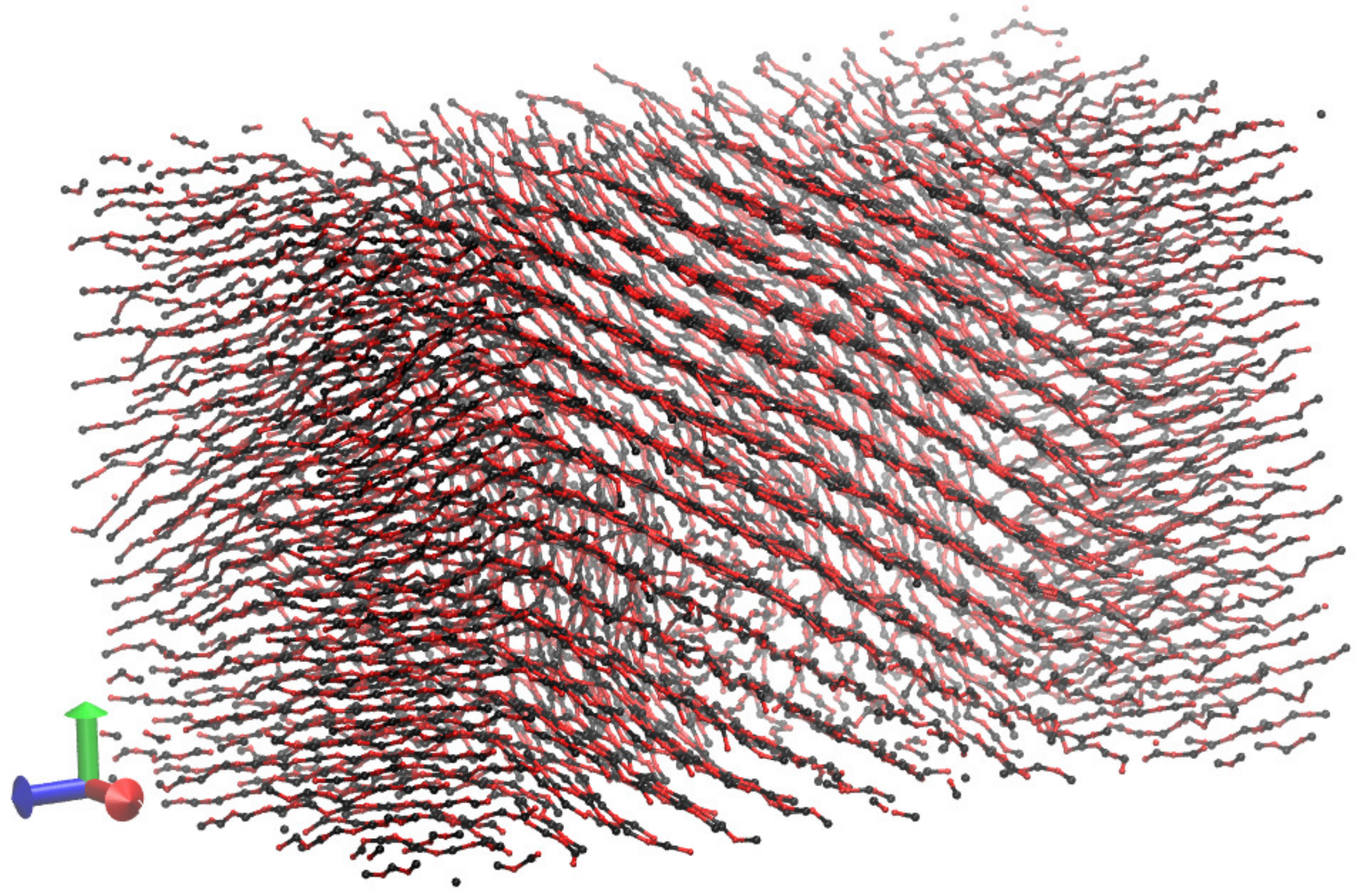}
\vskip0.3cm
\caption{\label{sup2}
Respresentative snapshots of the self assembled polymers
at volume fraction $\phi_p=0.11$  in box size $30 \times 30 \times 50 \sigma^3$.
One observes that as we increase density to $\phi_p=0.11$,
 there is a domain of ordered polymers (line hexagonal order) but in the other region the 
organization of chains is lost: the chains look disordered. 
}
\end{figure*}

\begin{figure*}[hbt]
\center
\includegraphics[width=0.329\textwidth]{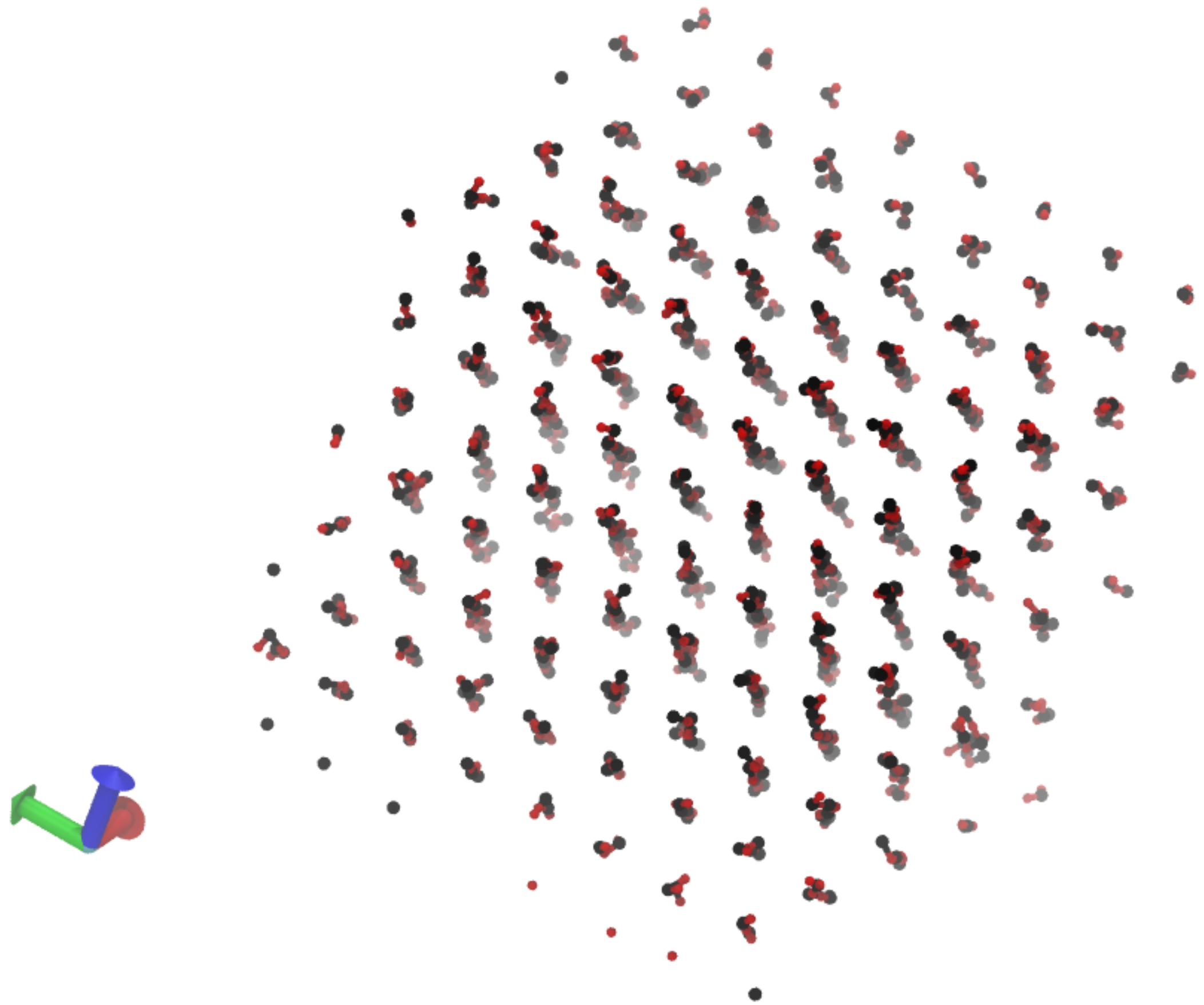}
\includegraphics[width=0.329\textwidth]{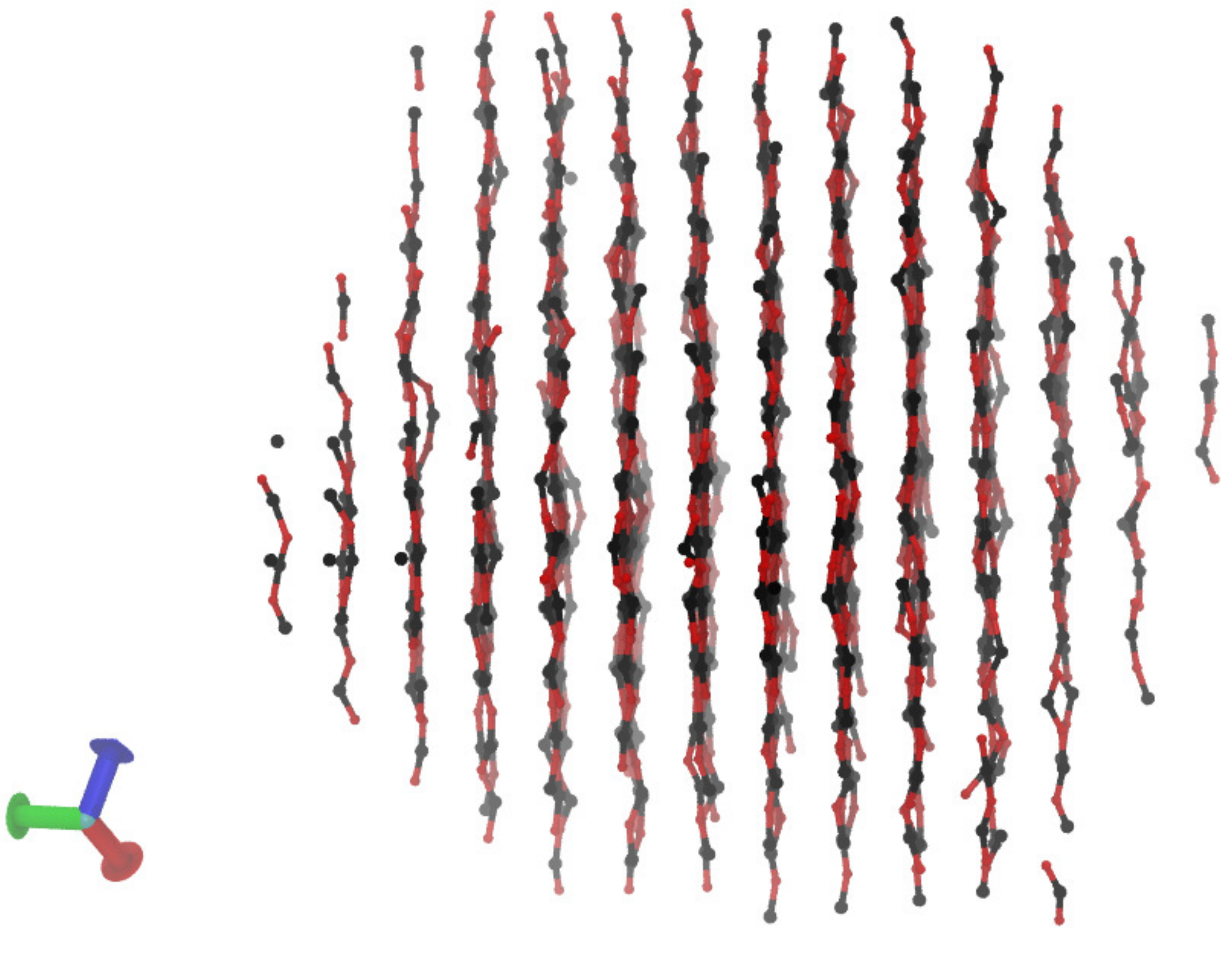}
\includegraphics[width=0.329\textwidth]{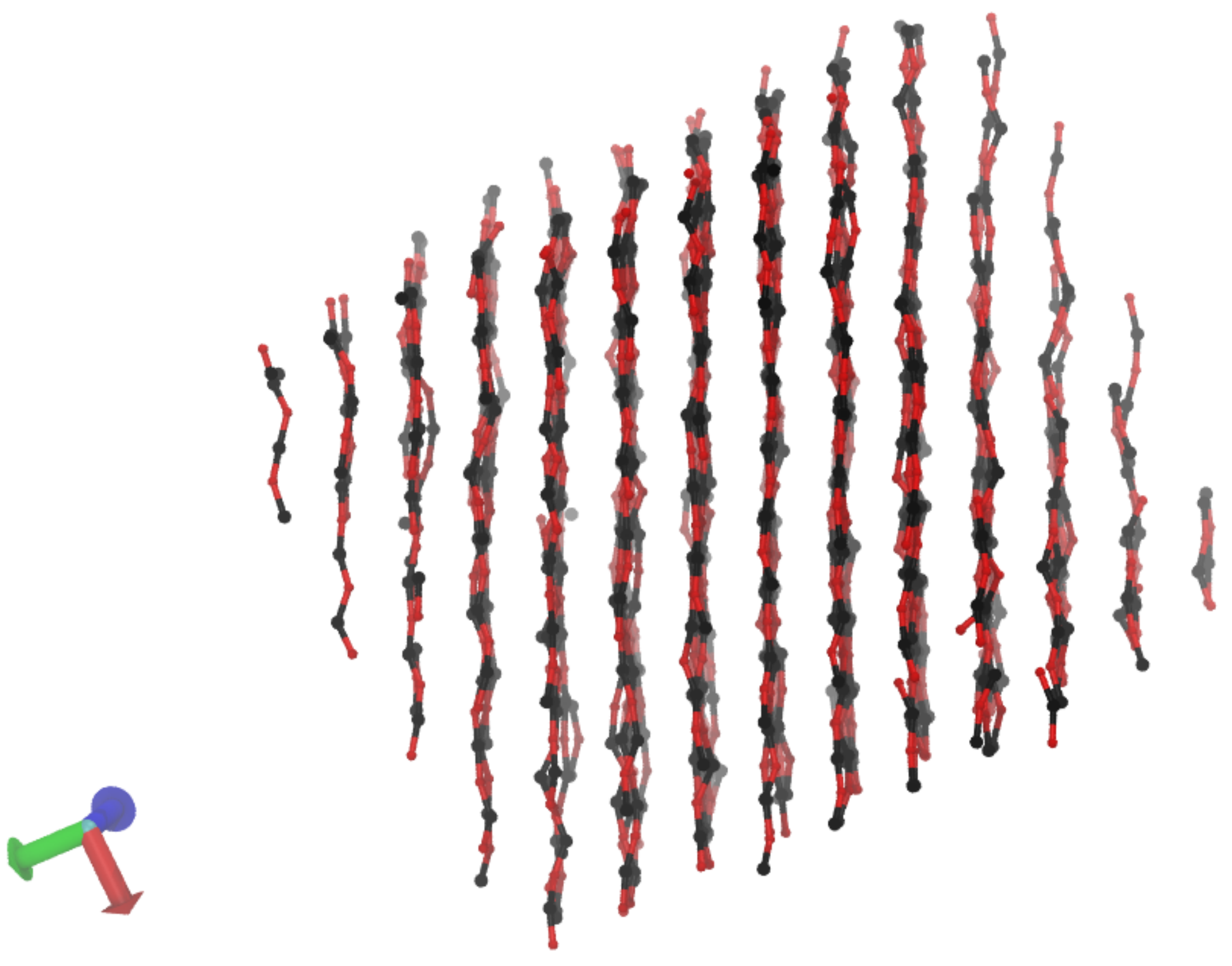}
\caption{\label{sup3}
Snapshots of system at $\phi_p=0.095$ with boxsize $20\times20\times20$, viewed from different 
directions, as it is continuously rotated to enable to reader to deciper the inter-chain and inter-particle ordering.}
\end{figure*}

\begin{figure*}[hbt]
\center
\includegraphics[width=0.8\textwidth]{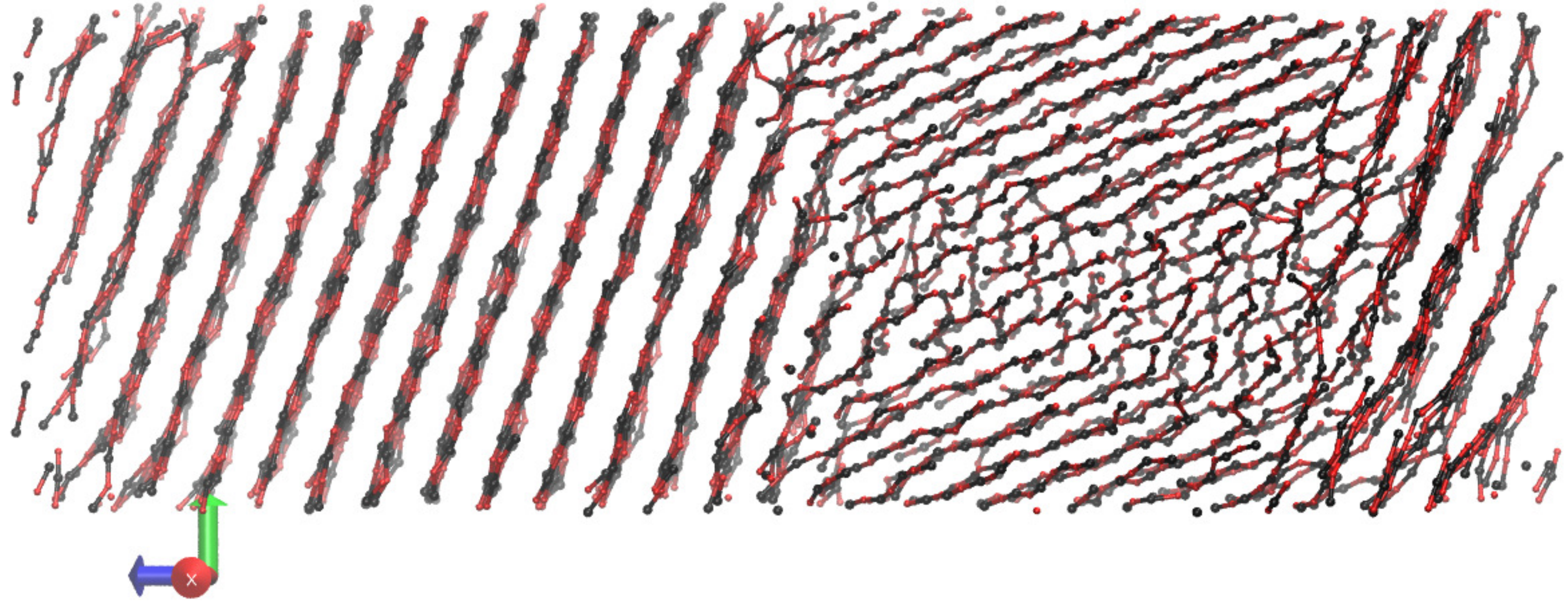}
\includegraphics[width=0.8\textwidth]{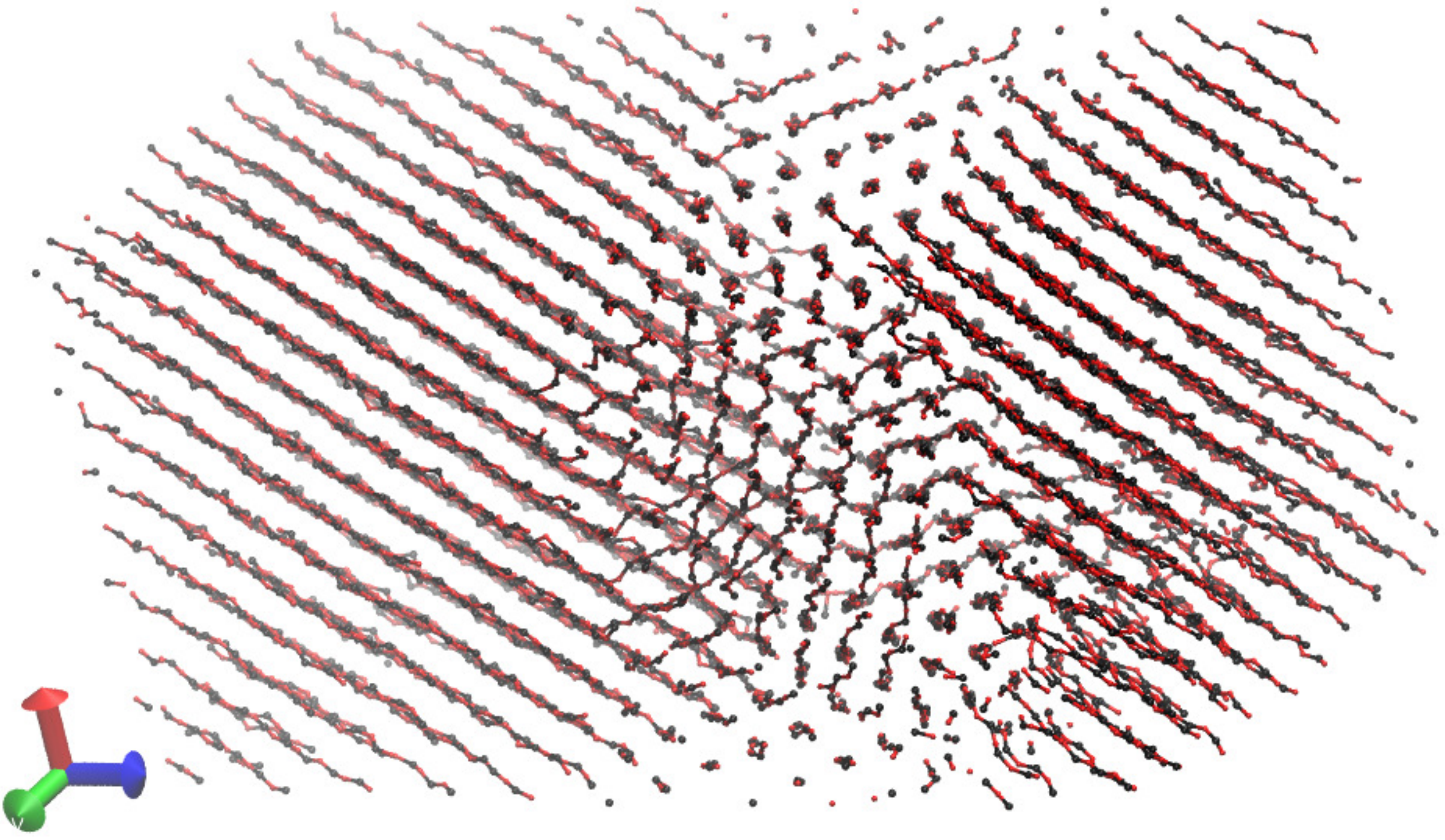}
\caption{\label{sup4}
Snapshots of system at $\phi_p=0.095$ with boxsizes (a) $20\times20\times60$ and (b) $30\times30\times60$.}
\end{figure*}

\begin{figure*}[hbt]
\center
\includegraphics[width=0.55\textwidth]{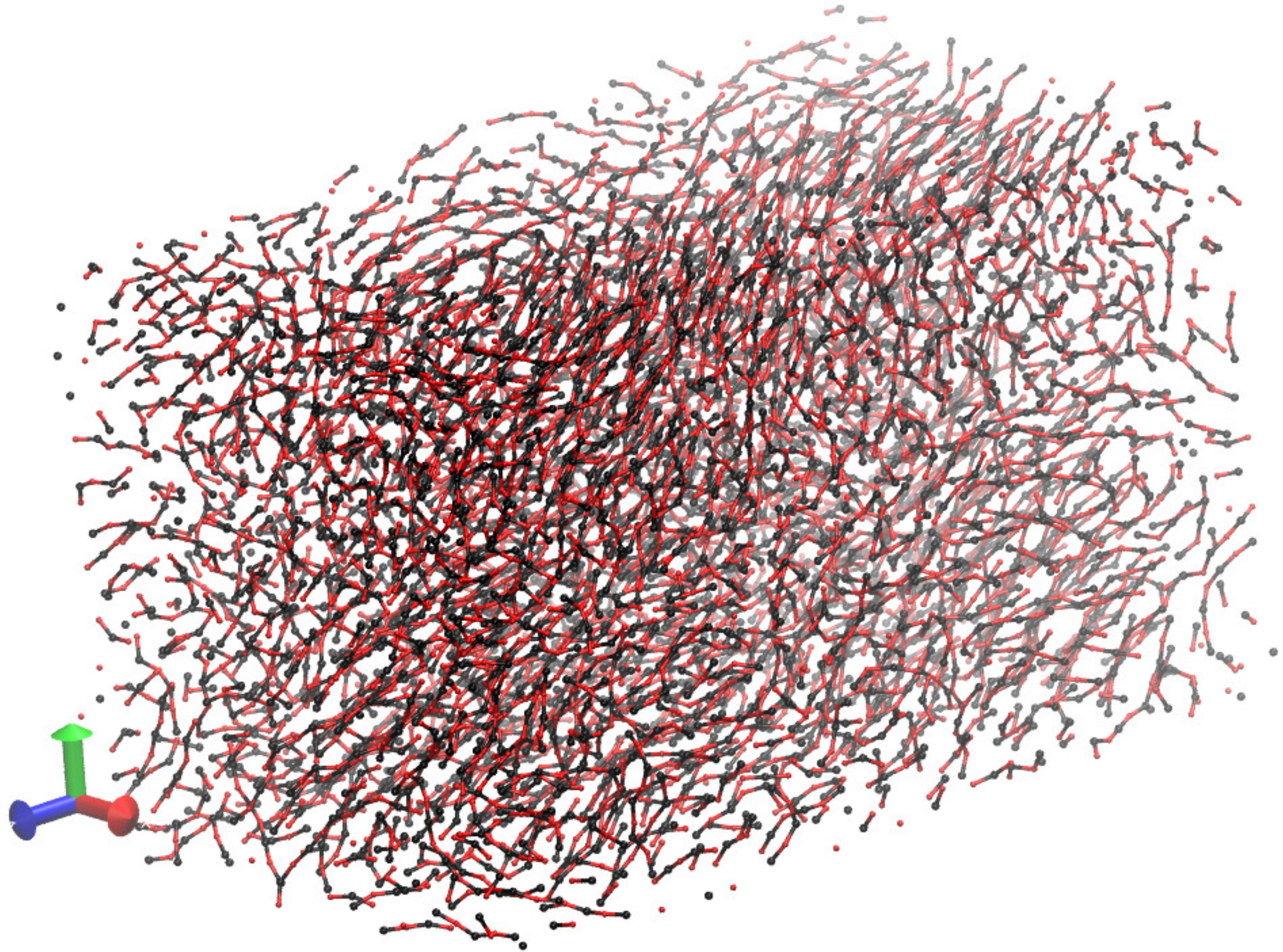}
\includegraphics[width=0.55\textwidth]{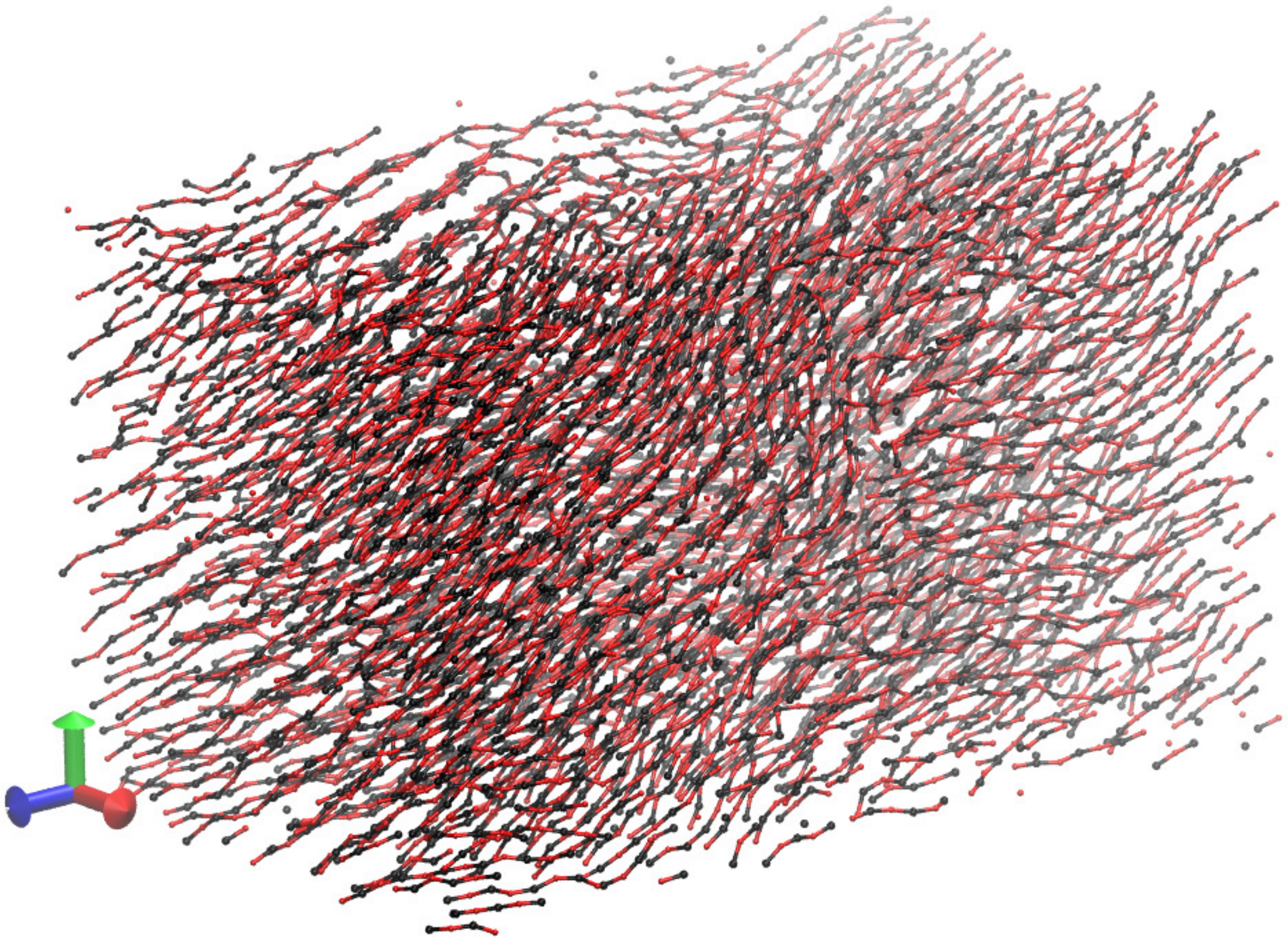}
\includegraphics[width=0.55\textwidth]{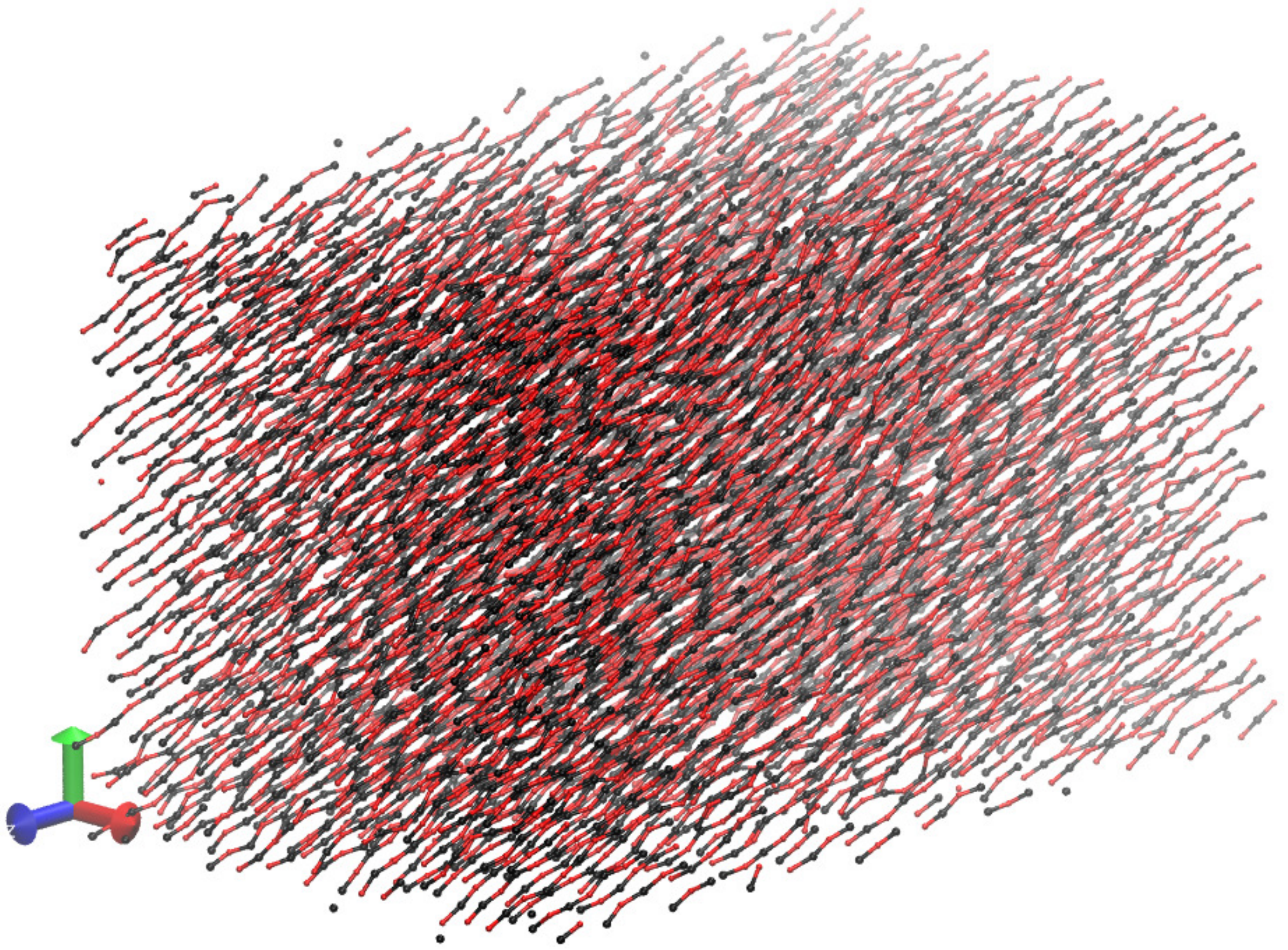}
\caption{\label{sup5}
Snapshots of system at different simulation times for $\phi_p=0.101$. 
The top panel is at $1.6 \times 10^5$ MCS after the beginning of the run, the middle panel is
at $2.4 \times 10^5$ MCS and the bottom is at $10 \times 10^5$ MCS.  Data is for 
bigger box size.}
\end{figure*}

\begin{figure*}[hbt]
\includegraphics[width=0.8\textwidth]{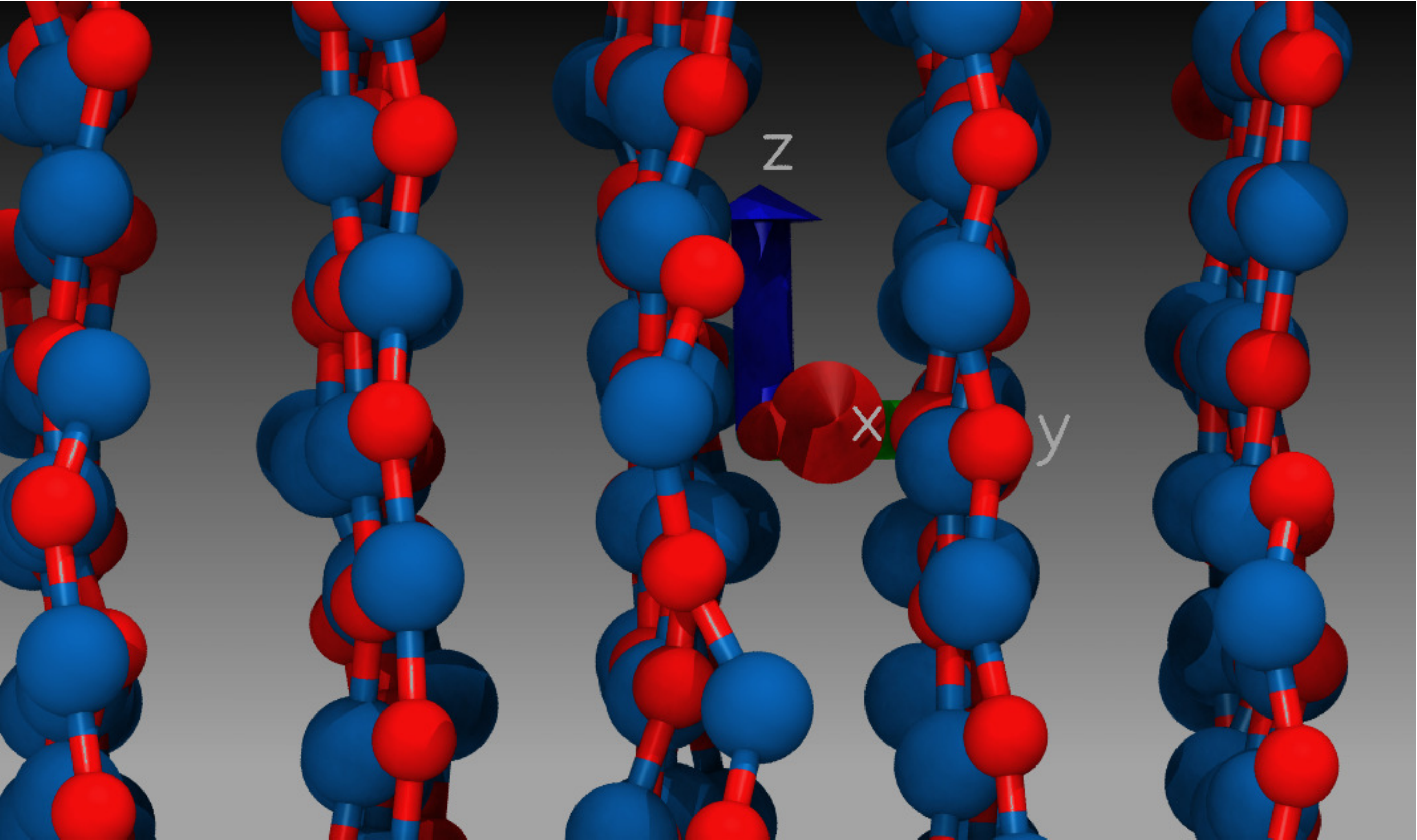}
\caption{\label{sup6} A closer look at chains, which are aligned in the z-direction. One can see that, nearly in all cases, the closest neighbours to an A-particle (red) are B-particles (blue) and vice versa.}
\end{figure*}

\end{widetext}

\end{document}